\documentclass[11pt,twoside]{article}

\usepackage[utf8]{inputenc}
\usepackage[english]{babel}
\usepackage{textcomp}
\usepackage{amsmath}
\usepackage{gensymb}
\usepackage{multirow}
\usepackage{multicol}
\usepackage{amsbsy}
\usepackage{amssymb}
\usepackage{capt-of}
\usepackage[T1]{fontenc} 

\usepackage[hmarginratio=1:1,top=20mm,bottom=20mm,columnsep=15pt]{geometry}

\usepackage{multicol} 
\usepackage{multirow}
\usepackage{lscape}
\usepackage[hang, small,labelfont=bf,up,textfont=it,up]{caption} 
\usepackage{booktabs} 
\usepackage{float} 

\usepackage{hyperref} 
\usepackage{graphicx} 
\usepackage{epstopdf} 
\usepackage{subfig}
\usepackage{paralist} 

\usepackage{abstract} 

\usepackage{titlesec} 

\titleformat{\section}[block]{\large\bfseries\scshape}{\thesection.}{0.5em}{} 
\titleformat{\subsection}[block]{\large\bfseries}{\thesubsection.}{0.5em}{}

\usepackage{fancyhdr} 
\usepackage{siunitx}
\usepackage[usenames,dvipsnames]{color}

\usepackage{verbatim}

\usepackage{bm}

\usepackage{isomath}
\usepackage{upgreek}

\usepackage{physics}

\newcommand{\te}{\tensorsym}

\newcommand{\dotp}[2]{\left(#1\vdot#2\right)}

\usepackage{bbold}
\numberwithin{equation}{section}
\usepackage[nottoc]{tocbibind}
\usepackage{amsthm}

\newtheorem*{claim}{Claim}
\newtheorem*{theorem}{Theorem}

\newcommand{\ii}{\mathrm{i}}
\newcommand{\TT}{\mathrm{T}}
\newcommand{\ee}{\mathrm{e}}
\newcommand{\Var}{\mathrm{\Delta}}

\usepackage{tikz-feynman}
\tikzfeynmanset{compat=1.1.0}

\usepackage{import}

\title{\vspace{50mm}
\noindent\rule{15cm}{0.4pt}
\huge\scshape{\textbf{Introduction to Renormalisation}}
\noindent\rule{15cm}{0.4pt}
} 

\author{
\Large\textsc{Jo\~ao F. Melo}\\[2mm] 
\large Department of Applied Mathematics and Theoretical Physics\\
\large University of Cambridge
}
\date{}

\begin{document}

\maketitle 
\thispagestyle{empty}

\pagebreak
\setcounter{page}{1}

\begin{abstract}
These are introductory lecture notes aimed at beginning graduate students covering fundamental concepts and ideas behind the renormalisation group. Our main goal is to motivate it and then explore its consequences, in the context of quantum field theory. We shall reproduce the standard results without needing any mention of cancellation of infinities. To keep the story simple and compact, we'll ignore the role of symmetry in this game. We'll also restrict to scalar fields, and not mention neither fermions nor gauge fields. Moreover, we'll only briefly mention the connection with statistical field theory.

Knowledge of quantum mechanics and special relativity is essential. Prior knowledge of quantum field theory is not required, however, this is not designed to be an introduction to QFT, so it may not be the most pedagogical way to introduce the topic. Previous familiarity with path integrals is also not assumed.
\end{abstract}

\noindent\rule{15cm}{0.4pt}

\tableofcontents

\newpage
\section*{Acknowledgements}
I have to thank a number of people for the making of these notes for several different reasons. Firstly, David Tong and David Skinner for teaching me the ways of RG the correct way, not by cancelling infinities. Not only that but there were several important discussions with my friends throughout the past years that were crucial for me to be able to tame this beast, I couldn't name (or remember) them all, but definitely the most important one (and the one which, in some sense, still hasn't ended) was with Paolo Pichini. I also have to thank to everyone who actually came to my (somewhat experimental) lectures which were vastly important to improving these notes. And finally, to Bruno Bento, for endless encouragement, outstanding typo-spotting skills, actually making the lectures happen with his great marketing skills, and for producing every gorgeous illustration present in the final version. I also thank the Cambridge Trust for my Vice-Chancellor's award to support my studies.

\newpage
\section*{Useful References}

Throughout these notes I haven't been very careful about citing the original references. To ameliorate this fact here is a list of the references I found useful in preparing them:

\begin{itemize}
    \item Michael E. Peskin, Daniel V. Schroeder: \textit{An Introduction to Quantum Field Theory}\\
        This is one of the standard references in QFT. Covers all the way from the basics to renormalisation and gauge theories. It has a very clear physical insight and covers many of the technical details explicitly. With respect to RG it follows quite a different path from these lectures, more historically inclined.
    \item David Tong: \textit{Statistical Field Theory}\\
        In many aspects this was the main inspiration for these lecture notes and was my first exposure to the wonders of renormalisation. It is all about renormalisation in the context of phase transitions and condensed matter with only a hint of the quantum world. Cannot recommend it too highly to learn the mysteries of RG.
    \item David Skinner: \textit{Quantum Field Theory II}\\
        These lectures cover some more advanced topics of quantum field theory in the path integral perspective. They are more on the mathematical side than what we'll cover, but offer a very clear view of what is going on. The first chapter on the path integral perspective of quantum nechanics and later on the section on the local potential approximation are very heavily based on David Skinner's notes.
    \item Hugh Osborn: \textit{Advanced Quantum Field Theory}\\
        Another very useful reference for path integrals and quantum field yheory in general, following a more classical perspective on the subject. The appendix on the derivation of Feynman diagrams is based heavily on these notes.
    \item Timo Weigand: \textit{Quantum Field Theory I+II}\\
        These are generic lecture notes covering two courses worth of QFT from the very basics to more advanced topics. They are quite clear and with a lot of physical insight. Highly recommended as an introduction to QFT. Although I didn't take anything in particular from them they were a constant companion to clear up some misconceptions.
    \item A. Zee: \textit{Quantum Field Theory in a Nutshell}\\
        This is a charming book, almost purely conceptual and with very little mathematical rigour. It won't teach you how to compute 2-loop diagrams but it is the place to go when the physical interpretation isn't clear.
    \item Yuji Iragashi, Katsumi Itoh, Hidenori Sonoda: \textit{Realization of Symmetry in the ERG Approach to Quantum Field Theory}\\
        Although this is a more generic review on the role of symmetry in Exact RG its first chapter has the clearest explanation of the Exact RG programme I ever came accross. It was the basis for the last section of these lecture notes.
    \item Timothy J. Hollowood: \textit{6 Lectures on QFT, RG and SUSY}\\
        This is probably one of the best explanations of renormalisation in the QFT perspective on the market. Highly recommend the first few lectures to get a hold of what this is all about.
\end{itemize}

\newpage

\section{Introduction}

In school (and in undergraduate physics degrees) we're taught many facts about the world, and it's a curious thing how stuff that seem quite relevant at that level soon become obsolete and other very simple ideas only grow stronger in importance as we continue on learning. An example of the latter will be the main focus of these lecture notes, the concept of scale. But there is one key difference between the concept of scale we meet as undergraduates and the concept of scale we'll delve into during this course. The former is a useful trick, the latter is probably one the deepest discoveries ever made in theoretical physics.

In classical physics, we're taught that physics is fundamentally scale invariant. We can always rescale our variables so that they are phrased in terms of dimensionless parameters, and, as long as those are fixed, everything should remain the same. Nonetheless, there are several hints to the contrary. First of all, it is very often easier to deal with an averaged, effective description rather than follow all details. Like when you use a Lagrange\footnote{Joseph-Louis Lagrange: born Giuseppe Lodovico Lagrangia in 1736 in Turin, Piedmont-Sardinia; died in 1813 in Paris, France; Academic Advisors: Leonhard Euler, Giovanni Battista Beccaria} multiplier to constrain a system, rather than calculating all forces. We're also taught about relativity and quantum mechanics and how they really only come about at scales different from our ordinary lives. So is physics fundamentally scale invariant or not?

Quite remarkably, we have a very definite answer to this question. Physics is, at its core, \textit{not} scale invariant. The world looks different at different scales. Even more surprisingly, phrasing things in terms of dimensionless parameters does not help. Fundamental constants, even dimensionless ones, are not constant! They depend on the scale at which we perform our experiment. When people say "The fine structure constant has the value $\alpha\approx1/137$" they are not correct in general, they are only correct at our ordinary energy scale.

If you think about it, this is quite bizarre. It seems to defy all our intuition built on dimensional analysis, and when even dimensional analysis doesn't work, what \textit{can} we trust?? We have dimensionless couplings, but, somehow, we can trade them by functions of some esoteric scale that seems to appear out of nowhere. And even condensed matter physicists aren't safe! Near certain phase transitions they found some quantities that seemed to scale with \textit{irrational} exponents! And they weren't known numbers like $\ee$ or $\pi$, they were more like new fundamental \textit{mathematical} constants. Even worse, how things scale is usually dictated by dimensional analysis, but all naive attempts failed massively.  And, to add to the confusion, these exponents were the same for wildly different physical systems like magnets and gases. They called this mysterious phenomenon \textit{universality}.

In due time we shall explain all these phenomena using the framework of the renormalisation group. But here is a hint of what's to come, it all boils down to a simple statement: field theories don't necessarily make sense to arbitrarily small length scales. If we try to make calculations in the continuum we simply get garbage. This "garbage" is the famous infinities that surround quantum field theory. However, if we introduce some fundamental scale in the game to define our theory, some minimal length beyond which our description breaks down, suddenly everything makes sense again. And using this framework we have been able to recover all those weird experimental facts to an astounding precision. And I do mean astounding. Renormalisation has the best prediction in the history of science. The $g$-factor of the the electron is the quantity that has the best agreement between data and theoretical prediction. Period. No caveats\footnote{To be fair renormalisation also has arguably the worst prediction in the history of science, which is the cosmological constant problem. However, whether that famous $10^{120}$ discrepancy really comes from renormalisation or by unjustified assumptions is a matter of heated debate}.

There is one issue with this. Renormalisation isn't simple. It isn't one of those ideas that you see it and think "how did I not see this before?". It is incredibly subtle. It took the better part of the 20$^\text{th}$ century to fully unfold, and, to some extent, we're still in the dark about many of its aspects. For this reason, there are lots of misconceptions and confusions surrounding renormalisation, which mostly stem from historical confusions. But all evidence points that it is the language of nature and we ought to understand it properly. It isn't a "hocus pocus" procedure like Feymman\footnote{Richard Phillips Feynman: born in 1918 in Queens, USA; died in 1988 in Los Angeles, USA; Doctoral Advisor: John Archibald Wheeler; Nobel Prize in Physics 1965 together with Sin-Itiro Tomonaga and Julian Schwinger "for their fundamental work in quantum electrodynamics, with deep-ploughing consequences for the physics of elementary particles"} once said, it isn't about sweeping infinities under a very large rug. It is about scale, and about phrasing predictions in terms of measurable quantities.

The point of these lectures is to convey precisely that. Stripping away all unnecessary technical details, we'll go to the simplest possible example, just a single real scalar field in flat spacetime. No discussion of symmetries, or of fermions\footnote{Enrico Fermi: born in 1901 in Rome, Italy; died in 1954 in Chicago, USA; Academic Advisors: Luigi Puccianti, Max Born, Paul Ehrenfest; Nobel Prize in Physics 1938 "for his demonstrations of the existence of new radioactive elements produced by neutron irradiation, and for his related discovery of nuclear reactions brought about by slow neutrons"}. Even at this level there is a lot to unpack, and lots of subtleties to confuse us. And the hope is that, when you go and learn about the more complex aspects of renormalisation, if ever you're confused about the meaning of it all, you can come back here, to the simple case, and then use that to understand what is actually happening in the more complicated one.

It won't be an easy journey, there are lots of dangers on the way, lot's of things one can get lost in. But, if you persevere, you'll come out with a new tool in the box, one that is incredibly powerful. So I hope you join me in this journey. See you on the other side.

\newpage
\section{Path integrals in quantum mechanics}\label{pathqm}

We start our journey of explaining the origin of the renormalisation group by delving into the path integral formulation of quantum mechanics. This is a very simple setting where the physics is clear and we have another very robust framework to compare with. We'll begin with deriving the path integral starting from the usual Hilbert\footnote{David Hilbert: born in 1862 in Königsberg or Wehlau, Kingdom of Prussia; died in 1943 in Göttingen, Germany; Doctoral Advisor: Ferdinand von Lindemann} space formulation, and then we'll discuss how to recover basic quantum facts like the non-commutativity of position and momentum.

\subsection{The quantum mechanical amplitude as a functional integral}

We'll consider the simplest case. A single particle, with mass $m$, position $\vb{x}(t)\in \mathbb{R}^3$ for $t\in\mathbb{R}$, and subject to a potential $V(\vb{x})$ which does not depend on the momenta. The Hamiltonian\footnote{William Rowan Hamilton: born in 1805 in Dublin, Ireland; died in 1865 in Dublin, Ireland; Academic Advisor: John Brinkley} is

\begin{equation}
    H(\vb{x},\vb{p})=\frac{\vb{p}^2}{2m}+V(\vb{x})
\end{equation}
where $\vb{p}=m\dot{\vb{x}}$.

In this case, there are no operator ordering ambiguities when going to the quantum theory. When more particles are present, or for momenta-dependent potentials, we have the usual ambiguities. This is a very hard problem to solve in general, but it has no bearing in the following discussion, hence, I'll ignore it. 

We now do the usual thing, promote the canonical variables $\vb{x}$ and $\vb{p}$ to operators $\hat{\vb{x}}$ and $\hat{\vb{p}}$, and impose the canonical commutation relations:

\begin{equation}
    [\hat{x}_i,\hat{p}_j]=\ii\delta_{ij}
\end{equation}

The Schrödinger\footnote{Erwin Rudolf Josef Alexander Schrödinger: born in 1887 in Vienna, Austria-Hungary; died in 1961 in Vienna, Austria; Doctoral Advisor: Friedrich Hasenöhrl; Nobel Prize in Physics 1933 together with Paul Adrien Maurice Dirac "for the discovery of new productive forms of atomic theory"} equation is then,

\begin{equation}
    \ii\pdv{\ket{\psi,t}}{t}=\hat{H}\ket{\psi,t}
\end{equation}
which is solved by (since the Hamiltonian does not depend explicitly on time)
\begin{equation}
    \ket{\psi,t}=\ee^{-\ii\hat{H}t}\ket{\psi,0}
\end{equation}
or in the position basis, i.e. written in terms of the wave-function
\begin{equation}
    \psi(\vb{x},t)=\bra{\vb{x}}\ee^{-\ii\hat{H}t}\ket{\psi,0}
\end{equation}
Inserting the identity operator in the form $\int\dd[3]x\dyad{\vb{x}}{\vb{x}}$ we get
\begin{equation}
    \psi(\vb{x},t)=\int\dd[3]y\bra{\vb{x}}\ee^{-\ii\hat{H}t}\ket{\vb{y}}\bra{\vb{y}}\ket{\psi,0}=\int\dd[3]y\bra{\vb{x}}\ee^{-\ii\hat{H}t}\ket{\vb{y}}\psi(\vb{y},0)
\end{equation}

Therefore, the dynamics is entirely specified by the transition amplitude associated to a particle being prepared at $\vb{x}_\text{i}$ at $t=0$ and then found at $\vb{x}_\text{f}$ at $t=T$:

\begin{equation}
    \mathcal{A}_T(\vb{x}_\text{f},\vb{x}_\text{i})=\bra{\vb{x}_\text{f}}\ee^{-\ii\hat{H}T}\ket{\vb{x}_\text{i}}
\end{equation}

The trick to calculate this is to divide the interval $[0,T]$ in $N$ slices of size $\Var t=\frac{T}{N}$ and at the end take the limit $N\to\infty$. We then have

\begin{align}
    \mathcal{A}_T(\vb{x}_\text{f},\vb{x}_\text{i})&=\bra{\vb{x}_\text{f}}\prod_{j=1}^N\ee^{-\ii\hat{H}\Var t}\ket{\vb{x}_\text{i}}=\nonumber\\
    &=\int\qty(\prod_{j=1}^{N-1}\dd[3]{x_j})\bra{\vb{x}_\text{f}}\ee^{-\ii\hat{H}\Var t}\ket{\vb{x}_{N-1}}\bra{\vb{x}_{N-1}}\ee^{-\ii\hat{H}\Var t}\ket{\vb{x}_{N-2}}\dots\bra{\vb{x}_1}\ee^{-\ii\hat{H}\Var t}\ket{\vb{x}_\text{i}}
\end{align}

Where, in going to the last line, we once more inserted $\int\dd[3]x\dyad{\vb{x}}{\vb{x}}$ between each Hamiltonian. Now we only need to evaluate expressions of the form

\begin{equation}
    \bra{\vb{x}_{i+1}}\ee^{-\ii\hat{H}\Var t}\ket{\vb{x}_i}
\end{equation}
for small $\Var t$. To do that, we use two tricks. The first one is that $\Var t$ is small, hence, at leading order in $\Var t$, the Baker\footnote{Henry Frederick Baker: born in 1866 in Cambridge, England; died in 1956 in Cambridge England; Doctoral Advisor: Arthut Cayley}-Campbell\footnote{John Edward Campbell: born in 1862 in Lisburn, Ireland; died in 1924 in Oxford, England}-Hausdorff\footnote{Felix Hausdorff: born on 1868 in Breslau, Kindgom of Prussia; died in 1942 in Bonn, Germany; Doctoral Advisors: Heinrich Bruns, Adolph Mayer} (BCH) formula gives

\begin{equation}
    \exp(-\ii\hat{H}\Var t)\approx\exp(-\ii\Var t\frac{\hat{\vb{p}}^2}{2m}) \exp(-\ii\Var tV(\hat{\vb{x}}))
\end{equation}

The second trick is inserting the identity in the form $\int\frac{\dd[3]p}{(2\pi)^3}\dyad{\vb{p}}{\vb{p}}$, which gives

\begin{align}
    \bra{\vb{x}_{i+1}}\exp(-\ii\hat{H}\Var t)\ket{\vb{x}_i}&=\int\frac{\dd[3]p_i}{(2\pi)^3}\bra{\vb{x}_{i+1}}\ket{\vb{p}_i}\bra{\vb{p}_i}\exp(-\ii\hat{H}\Var t)\ket{\vb{x}_i}\approx\nonumber\\
    &\approx\int\frac{\dd[3]p_i}{(2\pi)^3}\bra{\vb{x}_{i+1}}\ket{\vb{p}_i}\bra{\vb{p}_i}\ket{\vb{x}_i}\exp(-\ii\Var t\qty(\frac{\vb{p}_i^2}{2m}+V(\vb{x}_i)))
\end{align}
where we went from operators to eigenvalues by acting with $\exp(-\ii\Var tV(\hat{\vb{x}}))$ on the right and $\exp(-\ii\Var t\frac{\hat{\vb{p}}^2}{2m})$ on the left.

Using $\bra{\vb{p}}\ket{\vb{x}}=\exp(-\ii\dotp{\vb{p}}{\vb{x}})$ we get

\begin{align}
    \bra{\vb{x}_{i+1}}\exp(-\ii\hat{H}\Var t)\ket{\vb{x}_i}\approx\int\frac{\dd[3]p_i}{(2\pi)^3}\exp\Big(\ii\dotp{\vb{p}_i}{\qty(\vb{x}_{i+1}-\vb{x}_{i})}\Big)\exp(-\ii\Var t\qty(\frac{\vb{p}_i^2}{2m}+V(\vb{x}_i)))
\end{align}

This integral is a Gaussian\footnote{Johann Carl Friedrich Gauss: born in 1777 in Brunswick, Principality of Brunswick-Wolfenbüttel; died in 1855 in Göttingen, German Confederation; Doctoral Advisor: Johann Friedrich Pfaff} integral, hence, in principle, we know how to perform it. However, the coefficient of the quadratic term does not have a negative real part, it's pure imaginary. Therefore this integral is not convergent. To solve this issue we use two tricks: Wick\footnote{Gian Carlo Wick: born in 1909 in Turin, Kingdom of Italy; died in 1992 in Turin, Italy; Doctoral Advisor: Gleb Wataghin} rotation and analytic continuation. 

The vague idea behind this is to think of the integral as a function of \textit{complex} $\Var t$ instead of real (and positive). If we know the value of this function in a given open region, there is a unique procedure to assign values to this function in other points, so long as we don't cross any poles/branch cuts. This is the idea of \textit{analytic continuation}.

In our case, we know the value of the integral so long as $\Re{\ii\Var t}>0$, since, in this case, the integral is damped for $\vb{p}_i^2\to\infty$. We then define $\Var\tau=\ii\Var t$ and rotate to real (and positive) $\Var\tau$. The variable $\tau$ is called \textit{Euclidean\footnote{Euclid: born c. 325 BCE in an unknown place; died c. 270 BCE in (possibly) Alexandria, Hellenistic Egypt} time}, and this substitution is called \textit{Wick rotation}. We can then perform this integral because it is actually convergent. Because the final answer will make sense in an open region in the complex plane (the region $\Re{\ii\Var t}>0$ is open) the mathematics of analytic continuation will ensure that, if we rotate back to real physical time $\Var t$, the answer we get is unique. 

I know this may feel a bit hand-wavy and not very rigorous, but there is actually serious mathematics behind this and, in the end, it works, i.e. it agrees with experiment. For the more complex case of quantum field theory, the status of Wick rotation is a bit shakier and it is still not entirely clear to what extent we are actually loosing any important physics by doing this. We shall mostly ignore all these issues and fully work with Euclidean time objects. Our integral then becomes\footnote{If you need some reminding of how to do Gaussian integrals refer to Appendix \ref{pathqft}}


\begin{align}
    \bra{\vb{x}_{i+1}}\exp(-\Var\tau\hat{H})\ket{\vb{x}_i}&\approx\int\frac{\dd[3]p_i}{(2\pi)^3}\exp(-\Var\tau\qty(\frac{\vb{p}_i^2}{2m}+V(\vb{x}_i))+\ii\dotp{\vb{p}_i}{\qty(\vb{x}_{i+1}-\vb{x}_{i})})=\nonumber\\
    &=\qty(\frac{m}{2\pi\Var\tau})^\frac{3}{2}\exp(-\Var\tau\qty(\frac{1}{2}m\frac{\abs{\vb{x}_{i+1}-\vb{x}_i}^2}{\Var\tau^2}+V(\vb{x}_i)))
\end{align}

The full amplitude is then (in Euclidean time $T_\text{E}=\ii T$)

\begin{align}
    K_{T_\text{E}}(\vb{x}_\text{f},\vb{x}_\text{i})&= \bra{\vb{x}_\text{f}}\exp(-T_\text{E}\hat{H})\ket{\vb{x}_\text{i}}=\nonumber\\
    &=\lim_{N\to\infty}\int\prod_{i=1}^{N-1}\qty[\dd[3]x_i\qty(\frac{m}{2\pi\Var\tau})^\frac{3}{2}]\exp(-\sum\limits_{i=0}^{N-1}\Var\tau\qty(\frac{1}{2}m\frac{\abs{\vb{x}_{i+1}-\vb{x}_i}^2}{\Var\tau^2}+V(\vb{x}_i)))
    \label{discrete}
\end{align}
where we renamed the amplitude in Euclidean time as $K_{T_\text{E}}$ to emphasise the change to Euclidean time\footnote{Also, in Euclidean time we are essentially doing statistical physics and in that context this object would be the heat kernel.}, and defined $\vb{x}_\text{i}=\vb{x}_0$, $\vb{x}_\text{f}=\vb{x}_N$. 

This form of the amplitude is really suggestive. It would be very convenient if we could make the following identifications

\begin{align}
    \lim_{N\to\infty}\sum_{i=0}^{N-1}\Var\tau\qty(\frac{1}{2}m\frac{\abs{\vb{x}_{i+1}-\vb{x}_i}^2}{\Var\tau^2}+V(\vb{x}_i))&\overset{?}{=}\int_0^{T_\text{E}}\dd\tau\qty(\frac{1}{2}m\dot{\vb{x}}(\tau)+V(\vb{x}(\tau)))=S_\text{E}[\vb{x}]\label{action}\\
    \lim_{N\to\infty}\prod_{i=1}^{N-1}\qty[\dd[3]x_i\qty(\frac{m}{2\pi\Var\tau})^\frac{3}{2}]&\overset{?}{=}\mathcal{D}x\label{measure}
\end{align}
which we would identify as a Euclidean time version of the classical action and a functional integration measure. We'll attest the validity of these limits in the next section. Assuming this to be a legitimate procedure we write

\begin{equation}
    K_{T_\text{E}}(\vb{x}_\text{f},\vb{x}_\text{i})=\int_{\vb{x}_\text{i},0}^{\vb{x}_\text{f},T_\text{E}}\mathcal{D}x~\ee^{-S_\text{E}[\vb{x}]}
\end{equation}

This expression can be interpreted as stating that to calculate the amplitude for some process we should sum over all possible paths that the particle can take, consistent with the measured initial and final positions, weighed with a factor given by the value of the classical action evaluated on the given path, with exponential suppression. This integral is dominated by the regions where the action is least, which means we recover classical mechanics by taking the saddle point approximation of this integral.\footnote{Roughly, this approximation means assuming that the integral will just equal (the exponential of) the action evaluated at its extremum. Therefore the particle takes a single trajectory, the one that extremizes the action.}

\subsection{Non-commutativity and the continuum limit}

In the last chapter, we solved the Schrödinger equation by calculating the transition amplitude to go from one position to another. By writing this amplitude in terms of an integral over all possible paths, we fully determined the dynamics of our system. Given some initial state we can use the path integral to compute it's evolution, in principle at least\footnote{Emphasis on \textit{in principle}, path integrals are usually really hard to compute, as in, most problems in current theoretical physics can be traced back to a path integral that no one knows how to compute}. But this is not all there is to it in quantum mechanics, we're missing operators. 

If we're just asking about expectation values (or more generally correlation functions) at a particular time, then we're done. We can calculate the state hence we have the required probability distribution to calculate the expectation values, it's the whole point of calculating the state. However, we can imagine more complicated arrangements where the path integral comes in handy. Imagine that we know both the initial and the final states in some process (because we've measured them, or a little bird told us in advance), and we want to calculate the expectation value of an observable which is measured at some intermediate time $t$. This may seem like a ridiculously complicated arrangement, but when we move to quantum field theory, these kinds of statements are actually the only ones we can make, so bear with me.\footnote{In that context we usually choose the initial and final states to be the vacuum, and the operator to be just a simple product of the fundamental fields}

We'll consider the operator to be a function of $\hat{\vb{x}}$, and we'll denote it by $\hat{O}(\hat{\vb{x}})$, then this very complicated expectation value will be given by\footnote{We'll want to express this in terms of path integrals, where we don't have operators, only usual functions. In that context, the position is always a function of time, $\vb{x}(t)$, hence the notation on the LHS. On the RHS we're dealing with operators which can or cannot depend on time depending on the picture, hence the notation on the RHS, c.f. Appendix \ref{notation}}

\begin{align}
    \expval{O(\vb{x}(t))}&=\bra{\vb{x}_\text{f}}\ee^{-\ii\hat{H}(T-t)}\hat{O}(\hat{\vb{x}})\ee^{-\ii\hat{H}t}\ket{\vb{x}_\text{i}}=\nonumber\\
    &=\int\dd[3]{x}\bra{\vb{x}_\text{f}}\ee^{-\ii\hat{H}(T-t)}\hat{O}(\hat{\vb{x}})\dyad{\vb{x}}{\vb{x}}\ee^{-\ii\hat{H}t}\ket{\vb{x}_\text{i}}=\nonumber\\
    &=\int\dd[3]{x}O(\vb{x})\bra{\vb{x}_\text{f}}\ee^{-\ii\hat{H}(T-t)}\dyad{\vb{x}}{\vb{x}}\ee^{-\ii\hat{H}t}\ket{\vb{x}_\text{i}}=\nonumber\\
    &=\int\dd[3]{x}O(\vb{x})\mathcal{A}_{T-t}(\vb{x}_\text{f},\vb{x})\mathcal{A}_{t}(\vb{x},\vb{x}_\text{i})
    \label{opexp}
\end{align}
The complicated exponential factors in the first line are there because we need to evolve the initial state until time $t$ before we can act with the operator in order to perform the measurement, and then, to compare it with the known final state, we must evolve it the rest of the time interval from $t$ to $T$. In the remaining steps we merely wrote it in terms of the transition amplitudes, which we know how to express in terms of path integrals. We then rotate to Euclidean time and write it as
\begin{align}
    \expval{O(\vb{x}(\tau))}=&\int\dd[3]{x}O(\vb{x})K_{T_\text{E}-\tau}(\vb{x}_\text{f},\vb{x})K_{\tau}(\vb{x},\vb{x}_\text{i})=\nonumber\\
    =&\int\dd[3]{x}O(\vb{x})\int_{\vb{x},\tau}^{\vb{x}_\text{f},T_\text{E}}\mathcal{D}y~\ee^{-S_\text{E}[\vb{y}]}\int_{\vb{x}_\text{i},0}^{\vb{x},\tau}\mathcal{D}z~\ee^{-S_\text{E}[\vb{z}]}=\nonumber\\
    =&\int_{\vb{x}_\text{i},0}^{\vb{x}_\text{f},T_\text{E}}\mathcal{D}y~O(\vb{x}(\tau))\ee^{-S_\text{E}[\vb{y}]}
\end{align}
where in the last line we used the fact that integrating over all paths from from $\vb{x}_\text{i}$ to $\vb{x}$, then integrating over all paths from $\vb{x}$ to $\vb{x}_\text{f}$, and finally integrating over all possible intermediate points $\vb{x}$ is exactly the same thing as integrating over all paths from $\vb{x}_\text{i}$ to $\vb{x}_\text{f}$, provided the operator (which is now just a function) is evaluated at the correct time. This is called the \textit{concatenation property}.

This is a bit surprising, the somewhat complicated arrangement we concocted, is actually a very natural thing to calculate with path integral methods. We're just calculating the expectation value of a function of position, which, in this context, is a function of time.

Now, what happens if we try to calculate the expectation value of a composite operator, one which is a product of several operators evaluated at different times? Said another way, what happens if we make multiple successive measurements at different times? The derivation proceeds exactly as in (\ref{opexp}) inserting the identity in terms of the position basis in between every operator, but we must be careful. We make measurements in order, so we must insert operators such that operators evaluated at earlier times are to the right. Otherwise we get factors similar to

\begin{equation}
    \bra{\vb{x}}\ee^{\hat{H}T_\text{E}}\ket{\vb{x}'}
\end{equation}
for $T_\text{E}>0$, which do not lead to convergent integrals. Defining $\mathcal{T}\qty(\prod_iO_i(\tau_i))$ to be such \textit{time-ordering} we get

\begin{align}
    \expval{\mathcal{T}\qty(\prod_iO_i(\vb{x}(\tau_i)))}&=
    \bra{\vb{x}_\text{f}}\ee^{-\hat{H}T_\text{E}}\mathcal{T}\qty(\prod_i\ee^{\hat{H}\tau_i}\hat{O}_i(\hat{\vb{x}})\ee^{-\Hat{H}\tau_i})\ket{\vb{x}_\text{i}}=\nonumber\\
    &=\int_{\vb{x}_\text{i}}^{\vb{x}_\text{f}}\mathcal{D}x~\ee^{-S_\text{E}[\vb{x}]}\prod_iO_i(\vb{x}(\tau_i))
\end{align}

This is a very important point, the path integral naturally picks a preferred ordering of operators inside expectation values, the \textit{time-ordering}. Since, inside the path integral, the operators became mere functions, everything commutes, there's no notion of ordering. However, a random product of operators in the path integral perspective becomes a time-ordered product of operators in the canonical quantisation perspective. If operators are evaluated at the same time they won't have a preferred ordering, however, since we're considering only functions of position, they commute even when viewed as operators acting on a Hilbert space, so there's no issue there. But what if we consider functions of momentum? Well, there are a few issues when considering functions of momentum that complicate matters

\begin{enumerate}
    \item Hermiticity\footnote{Charles Hermite: born in 1822 in Dieuze, France; died in 1901 in Paris, France; Doctoral Advisor: Eugène Charles Catalan} is a bit bizarre once we go to Euclidean time. There are analogues, but it's a bit fiddly if you do the calculations in certain ways. We'll gently sweep this one under the rug, by doing the calculations in a way this isn't an issue.
    \item For arbitrary potentials momentum is not conserved. This can show up in some calculations using path integrals where, if you do your calculations in some ways, it appears that there are additional contributions to the momentum proportional to derivatives of the potential. This is very subtle and has to do with the fact that, at least classically, the potential generates an impulse and contributes to the future momentum. I've just mentioned this here so that you don't go mad trying to understand where you went wrong if you do the calculation yourself. However, it's not very relevant for quantum field theory and we'll set $V=0$
\end{enumerate}

This illustrates what is simultaneously the greatest strength and the greatest weakness of the path integral. If you try to hack the path integral, you'll quickly get non-sense, you need to be very careful about what you mean from the get go in order to proceed. With operators, you can very easily be naughty and ignore some subtleties, and still get more or less the correct answer. What usually happens is that simple calculations and proofs are way easier with operators than path integrals, however, for genuinely subtle situations (like gauge theories), the path integral makes dead obvious from the start where you can go wrong, and doesn't let you go through until you truly understand what's going on. Whereas with operators it's very hard to see what assumption went wrong most of the times. It's a trade off that depends on the situation, for ordinary quantum mechanics, operators are the way to go, for quantum field theory, usually path integrals are easier, but in any given problem you should analyse it carefully.

So, because we have a lot of subtleties going on, I'll restrict to showing those that carry through to quantum field theory and simplify the rest. This means I'll set $V=0$ and consider only one spatial dimension. Increasing the number of dimensions is easy, turning on the potential is not so easy. We'll proceed nevertheless by calculating the expectation value of the momentum. 

First just a note on how to write momentum in Euclidean time
\begin{equation}
    p=m\dv{x}{t}=\ii m\dv{x}{\tau}=\ii m \dot{x}
\end{equation}
and also,
\begin{equation}
    \bra{x}\hat{p}\ket{\psi}=-\ii\pdv{x}\bra{x}\ket{\psi}
\end{equation}

The expectation value of the momentum is
\begin{align}
    \expval{p(\tau)}&=\bra{x_\text{f}}\ee^{-\hat{H}(T_\text{E}-\tau)}\hat{p}\ee^{-\hat{H}\tau}\ket{x_\text{i}}=\nonumber\\
    &=\int\dd{x}\bra{x_\text{f}}\ee^{-\hat{H}(T_\text{E}-\tau)}\dyad{x}{x}\hat{p}\ee^{-\hat{H}\tau}\ket{x_\text{i}}=\nonumber\\
    &=\int\dd{x}\bra{x_\text{f}}\ee^{-\hat{H}(T_\text{E}-\tau)}\ket{x}\qty(-\ii\pdv{x}\bra{x}\ee^{-\hat{H}\tau}\ket{x_\text{i}})=\nonumber\\
    &=\int\dd{x}K_{T_\text{E}-\tau}(x_\text{f},x)\qty(-\ii\pdv{K_{\tau}(x,x_\text{i})}{x})
    \label{expvalp}
\end{align}

To calculate this derivative let's go to the discrete version of the heat kernel given by (\ref{discrete}). Note that only the last term depends on the end point, hence we just need to calculate

\begin{equation}
    \pdv{x_\tau}K_{\Var\tau}(x_\tau,x_{\tau-\Var\tau})
\end{equation}
for $\Var\tau\ll1$. Plugging (\ref{discrete}) in here we get
\begin{align}
    \pdv{x_\tau}K_{\Var\tau}(x_\tau,x_{\tau-\Var\tau})&=\pdv{x_\tau}\qty(\sqrt{\frac{m}{2\pi\Var\tau}}\exp(-\frac{m}{2}\frac{(x_\tau-x_{\tau-\Var\tau})^2}{\Var\tau}))=\nonumber\\
    &=-m\frac{x_\tau-x_{\tau-\Var\tau}}{\Var\tau}K_{\Var\tau}(x_\tau,x_{\tau-\Var\tau})
    \label{derivative}
\end{align}

Therefore,
\begin{equation}
    \pdv{K_{\tau}(x,x_\text{i})}{x}=-m\eval{\dv{x}{\tau}}_\tau K_{\tau}(x,x_\text{i})=\ii p(\tau) K_{\tau}(x,x_\text{i})
\end{equation}

Plugging back into (\ref{expvalp}) we finally get
\begin{align}
    \expval{p(\tau)}&=\int\dd{x}K_{T_\text{E}-\tau}(x_\text{f},x)p(\tau)K_{\tau}(x,x_\text{i})=\nonumber\\
    &=\int\dd{x}p(\tau)\int_{x,\tau}^{x_\text{f},T_\text{E}}\mathcal{D}y~\ee^{-S_\text{E}[y]}\int_{x_\text{i},0}^{x,\tau}\mathcal{D}z~\ee^{-S_\text{E}[z]}=\nonumber\\
    &=\int_{x_\text{i},0}^{x_\text{f},T_\text{E}}\mathcal{D}y~p(\tau)\ee^{-S_\text{E}[y]}
\end{align}

It was a bit harder than before, but we ended up with a nice result, which is reassuring. It appears there is no difference whether it's position or momentum, you just insert it inside the path integral to calculate the expectation vale. However, there is an issue if we insert a function of both position and momentum. Let's say we want to insert the product of the two. If they're evaluated at different times there's no issue, everything goes exactly as expected, time ordering prevails. But, what if we insert them at the same time? Then we have a problem, in the operator perspective the result depends on the order in which we insert them, while in the path integral perspective they're ordinary functions and hence order cannot matter! How can we solve this?

We need to be clever, remember that the path integral picks an ordering, time-ordering. We'll exploit that by considering two different times $\tau_-$ and $\tau_+$ such that $0<\tau_-<\tau<\tau_+<T_\text{E}$. In this way we can force an ordering of the two operators by inserting them at different times. After that we make $\tau_-$ approach $\tau$ from below, and $\tau_+$ approach $\tau$ from above; and calculate the difference between the two approaches. First let's examine what the canonical point of view has to say about this,

\begin{align}
    \expval{x(\tau)p(\tau_-)}&=\bra{x_\text{f}}\ee^{-\hat{H}(T_\text{E}-\tau)}\hat{x}\ee^{-\hat{H}(\tau-\tau_-)}\hat{p}\ee^{-\hat{H}\tau_-}\ket{x_\text{i}}\to\bra{x_\text{f}}\ee^{-\hat{H}(T_\text{E}-\tau)}\hat{x}\hat{p}\ee^{-\hat{H}\tau}\ket{x_\text{i}}\\
    \expval{p(\tau_+)x(\tau)}&=\bra{x_\text{f}}\ee^{-\hat{H}(T_\text{E}-\tau_+)}\hat{p}\ee^{-\hat{H}(\tau_+-\tau)}\hat{x}\ee^{-\hat{H}\tau}\ket{x_\text{i}}\to\bra{x_\text{f}}\ee^{-\hat{H}(T_\text{E}-\tau)}\hat{p}\hat{x}\ee^{-\hat{H}\tau}\ket{x_\text{i}}
\end{align}
where the limits taken are $\tau_-\nearrow\tau$ and $\tau_+\searrow\tau$. Subtracting the two we get
\begin{equation}
    \bra{x_\text{f}}\ee^{-\hat{H}(T_\text{E}-\tau)}[\hat{x},\hat{p}]\ee^{-\hat{H}\tau}\ket{x_\text{i}}=\ii\bra{x_\text{f}}\ee^{-\hat{H}T_\text{E}}\ket{x_\text{i}}=\ii K_{T_\text{E}}(x_\text{f},x_\text{i})
\end{equation}
using the fact that
\begin{equation}
    [\hat{x},\hat{p}]=\ii
\end{equation}

This is one of the most fundamental results in the quantum theory, it was our first step in setting up the problem. The path integral better be able to recover this. To do just that we'll move to the discrete version of the path integral given by (\ref{discrete}). We'll focus on the integration around $\tau$, taking $\tau_-=\tau-\Var\tau$ and $\tau_+=\tau+\Var\tau$,

\begin{align}
    &\int\dd{x_\tau}K_{\Var\tau}(x_{\tau+\Var\tau},x_\tau)\qty(x_\tau m\ii\frac{x_\tau-x_{\tau-\Var\tau}}{\Var\tau}-x_\tau m\ii\frac{x_{\tau+\Var\tau}-x_\tau}{\Var\tau})K_{\Var\tau}(x_\tau,x_{\tau-\Var\tau})=\nonumber\\
    =&-\ii\int\dd{x_\tau}x_\tau\pdv{x_\tau} \Big(K_{\Var\tau}(x_{\tau+\Var\tau},x_\tau)K_{\Var\tau}(x_\tau,x_{\tau-\Var\tau})\Big)=\nonumber\\
    =&\ii\int\dd{x_\tau}K_{\Var\tau}(x_{\tau+\Var\tau},x_\tau)K_{\Var\tau}(x_\tau,x_{\tau-\Var\tau})=\ii K_{2\Var\tau}(x_{\tau+\Var\tau},x_{\tau-\Var\tau})
\end{align}
where in going to the second line we used (\ref{derivative}), then we integrated by parts, and finally we used the concatenation property. The remaining integrals are easy to do, we're left with

\begin{equation}
    \expval{x(\tau)p(\tau_-)-p(\tau_+)x(\tau)}\xrightarrow{\tau_{+/-}\to\tau}\ii K_{T_\text{E}}(x_\text{f},x_\text{i})
    \label{limit}
\end{equation}
which agrees with the operator approach.

Wait, what? How was this possible? For differentiable functions that limit is exactly zero, no matter what we do. However, when computing the path integral using the discrete version we definitely don't obtain zero as an answer. The only possible solution is that the function $x(\tau)$ is a continuous but non-differentiable function, it actually must not be differentiable anywhere for (\ref{limit}) to make sense for all $\tau$! This is against our common intuition, but then again, this is quantum mechanics, what isn't against our intuition in this field? But this means the limit in (\ref{action}) doesn't make sense, there is no continuum action.

If (\ref{action}) doesn't make sense, does (\ref{measure}) make sense? It turns out it also doesn't make sense. The reason for this is rather formal and beyond the main scope of these lectures but suffice to say that there exists no standard Lebesgue\footnote{Henri Léon Lebesgue: born in 1875 in Beauvais, France; died in 1941 in Paris, France; Doctoral Advisor: Émile Borel} measure for an infinite dimensional space. Does anything make sense in the continuum? Or do we need to always work with the discrete version of the path integral? It turns out that, although the measure and the action don't exist separately, the product between the measure and the kinetic part of the action has a well defined continuum limit. It constitutes a Wiener\footnote{Norbert Wiener: born in 1894 in Columbia, USA; died in 1964 in Stockholm, Sweden; Doctoral Advisor: Karl Schmidt} measure, which is not a Lebesgue measure in the usual sense. This works so long as we use paths that are nowhere-differentiable. This is the reason why we were always capable of taking the continuum limit after we're done calculating and get sensible answers. And this is also connected to why standard quantum mechanics works without the need to discretise time. 

However, this only works for non-relativistic quantum mechanics. When we move to quantum field theory things are not so simple, not all theories have a well defined continuum limit, and even for those that do, no one has been able to construct the analogue of the Wiener measure. We'll spend the rest of these lectures exploring how to deal with this and what does this tell us about our world.
\newpage
\section{The Wilsonian renormalisation group}\label{wilsonrg}

In the last chapter, we've seen that in order to make sense of path integrals, we must make time a discrete variable and only at the end take the continuum limit. In the case of quantum field theory, our dynamical variables will depend on space as well as time so it makes sense to discretise space too. However, there is no Wiener measure to save us now, we must take this discretisation into account and fully understand its role in our description of Nature. There is no guarantee that some continuum limit exists for quantum field theory. This means there is a new scale in the game, the size of the spacing, call it $a$.

This raises many questions. First of all, what is this scale and what does it mean to say that we're discretising spacetime? Secondly, why this discretisation in particular, why do we choose $a$ and not $a/2$ or $2a$, or why don't we use a smoother procedure? Thirdly, how to make contact with experiment, can we actually measure the effects of this scale or is it some theoretical trick? And finally, when can we remove this discretisation, i.e. when does our theory make sense in the continuum?

The rest of these notes will try to answer these questions. This first chapter will deal with the condensed matter or statistical point of view of effective field theories, which is also very relevant for the high energy/particle physics realm 

\subsection{Coarse graining}

To be concrete, let's consider the most general theory with a single real scalar field $\phi(\vb{x})$ in $d$ dimensions\footnote{We assume that we have Wick rotated to Euclidean signature therefore $\vb{x}=(x^1,x^2,x^3,\dots,x^d)$, where $x^d=\tau$ is the Euclidean time.}. To do that, we write every possible term in the action compatible with the symmetries we impose. In particular we'll impose two symmetries, Euclidean group $E(d)\equiv ISO(d)$ composed of rotations and translations\footnote{The Lorentzian signature counterpart would be the Poincaré group}, and the discrete $\mathbb{Z}_2$ symmetry $\phi\to-\phi$. And I do mean every possible term: $\phi^{892}$ is there, and so is $\phi^7\nabla^4\phi^{2075}$ all with their respective couplings. However, $\partial_1\phi$ isn't there because it isn't rotationally invariant, $\phi^3$ is also forbidden because it is odd in $\phi$, and $1/\phi$ is also not allowed because it blows up at $\phi=0$. The first few terms in the action are,

\begin{equation}
    S[\phi]=\int\dd[d]x\qty(\frac{1}{2}(\grad{\phi})^2+\frac{1}{2}m^2\phi^2+\frac{\lambda_4}{4!}\phi^4+\dots)
\end{equation}
where we fixed the coefficient in from of the kinetic term to be $1/2$ thereby fixing the overall scale of our field.

We could proceed to work directly in real space and use a lattice, however, it will be simpler to work in momentum space, which is entirely equivalent. Therefore, we'll regularise our theory by imposing that the fields have finite support in momentum space, i.e.

\begin{equation}
    \Tilde{\phi}(\vb{p})=\begin{cases}
    0 &  \abs{\vb{p}}>\Lambda_0 \\
    \Tilde{\phi}(\vb{p}) &  \abs{\vb{p}}<\Lambda_0
    \end{cases}
\end{equation}

We write the partition function as

\begin{equation}
    Z_{\Lambda_0}=\int\mathcal{D}\phi~\ee^{-S_{\Lambda_0}[\phi]}
\end{equation}
leaving explicit the dependence of both the partition function and the action in $\Lambda_0$.

Note that we do not know how to define our theory without the regularisation, the cutoff is part of the \textit{definition} of our theory. This is a very important point, the theory is defined \textit{at this scale}, we cannot answer questions about what happens above that scale, because our theory knows nothing of it. This is the clue to interpret the cutoff, it is the maximum scale to which we give credit to our theory, you can think of it as the resolution of our measurement apparatus. Above that scale we have absolutely no clue about what happens, this theory may continue to be valid, or it may be complete bollocks. But we can't access it with experiment so, who cares? I don't.

Now imagine that we have tested the theory with some experiments up to the scale $\Lambda_0$, so we know it works all the way up to that energy scale. But, we are only interested in describing phenomena at some scale way below that initial cutoff, we may not have the money to build an equally big accelerator. If this is the case we should be able to use a different theory, an \textit{effective} theory with a different cutoff $\Lambda<\Lambda_0$ as long as we are only interested in energies $E\ll\Lambda$. Let's build that theory.

We'll split our field into two components, the useless/high energy/UV modes, $\phi^+$, that have support between $\Lambda$ and $\Lambda_0$ i.e.

\begin{equation}
    \Tilde{\phi}^+(\vb{p})=\begin{cases}
    0 & \abs{\vb{p}}>\Lambda_0 \\
    \Tilde{\phi}(\vb{p}) & \Lambda<\abs{\vb{p}}<\Lambda_0 \\
    0 & \abs{\vb{p}}<\Lambda
    \end{cases}
\end{equation}
and the useful/low energy/IR modes, $\phi^-$, that have support below $\Lambda$,

\begin{equation}
    \Tilde{\phi}^-(\vb{p})=\begin{cases}
    0 & \abs{\vb{p}}>\Lambda \\
    \Tilde{\phi}(\vb{p}) & \abs{\vb{p}}<\Lambda
    \end{cases}
\end{equation}

The partition function becomes

\begin{equation}
    Z_{\Lambda_0}=\int\mathcal{D}\phi^- \mathcal{D}\phi^+ ~\ee^{-S_{\Lambda_0}[\phi^+ +\phi^-]}\overset{!}{=}\int\mathcal{D}\phi^- ~\ee^{-S_{\Lambda}[\phi^-]}
\end{equation}
where in the last line we expressed our wish of writing the partition function solely as an integration over the low energy modes. To do this we define the \textit{effective action} as,

\begin{equation}
    \ee^{-S_\Lambda[\phi^-]}=\int\mathcal{D}\phi^+\ee^{-S_{\Lambda_0}[\phi^++\phi^-]}
\end{equation}

Note the similarity with the way we defined the Wilsonian\footnote{Kenneth Geddes "Ken" Wilson: born in 1936 in Waltham, USA; died in 2013 in Saco, USA; Doctoral Advisor: Murray Gell-Mann; Nobel Prize in Physics 1982 "for his theory for critical phenomena in connection with phase transitions"} effective action in (\ref{wilsoneff}), this is the Wilsonian effective action where the low-energy fields are considered as external sources in the path integral.

What is the form of this effective action? We started with the most general possible action, therefore we can only end up with an action that has exactly the same form but different coefficients

\begin{equation}
    S_\Lambda[\phi^-]=\int\dd[d]x\qty(\frac{1}{2}Z'_\phi(\grad{\phi^-})^2+\frac{1}{2}{m'}^2(\phi^-)^2+\frac{\lambda'_4}{4!}(\phi^-)^4+\dots)
\end{equation}

Let's stop for a moment and consider the implications. If we can only access lower energy scales, we'll "see"\footnote{We'll give a more precise meaning to this in the next chapter} different effective couplings in our theory. The masses and charges/couplings of the fundamental fields that make up our universe aren't fixed, they depend on the scales which we can access! If I didn't blow your mind just now, I don't know why do you study physics.

As we're physicists we'd like to be able to compare the two theories, however, at this moment, that is impossible for two reasons. Firstly, the theory is defined \textit{with} a cutoff, therefore if we just compared the two actions straight away we'd be comparing apples to oranges. Said another way, there is an implicit dependence on the cutoff inside the spacetime integration which would make our comparison meaningless. To get rid of this we rescale our spacetime variables as

\begin{equation}
    \vb{x}\to \vb{x}'=\frac{\Lambda}{\Lambda_0}\vb{x},\quad \vb{p}\to \vb{p}'=\frac{\Lambda_0}{\Lambda}\vb{p}
\end{equation}
which ensures that $\abs{\vb{p}}=\Lambda\Rightarrow\abs{\vb{p}'}=\Lambda_0$.

Secondly, rescaling the fields by an overall constant factor has no physical significance, therefore we must fix the overall scale of the field in some way between the two theories. The best way to do this depends on the particular theory you're considering, in our case, we'll fix the coefficient of the kinetic term to be exactly $1/2$ by defining

\begin{equation}
    \phi'(\vb{x}')=\sqrt{Z_\phi}\phi^-(\vb{x})
\end{equation}
where $Z_\phi=\qty(\frac{\Lambda_0}{\Lambda})^{d-2}Z'_\phi$.

The final action is

\begin{equation}
    S_\Lambda[\phi']=\int\dd[d]x'\qty(\frac{1}{2}(\grad '\phi')^2+\frac{1}{2}m^2(\Lambda){\phi'}^2+\frac{\lambda_4(\Lambda)}{4!}{\phi'}^4+\dots)
\end{equation}

This procedure is the famous \textit{renormalisation group} (RG). Which is a very bad name for it, since nothing is normalised more than once, and it is not a group since we cannot invert this procedure. But the name stuck and it would be criminal of me to call it any other thing.

To recap, the steps of the renormalisation group are the following:

\begin{enumerate}
    \item Coarse graining: this means integrating out the degrees of freedom which are not accessible to yield an effective theory at some lower resolution
    \item Fix the scale of the spacetime: this means rescaling the spacetime variables to make the dependence on the new cutoff explicit
    \item Fix the scale of the field: this means removing the ambiguity in the overall scale factor of the field variables by some independent manner
\end{enumerate}

\begin{center}
    \includegraphics[scale=0.45]{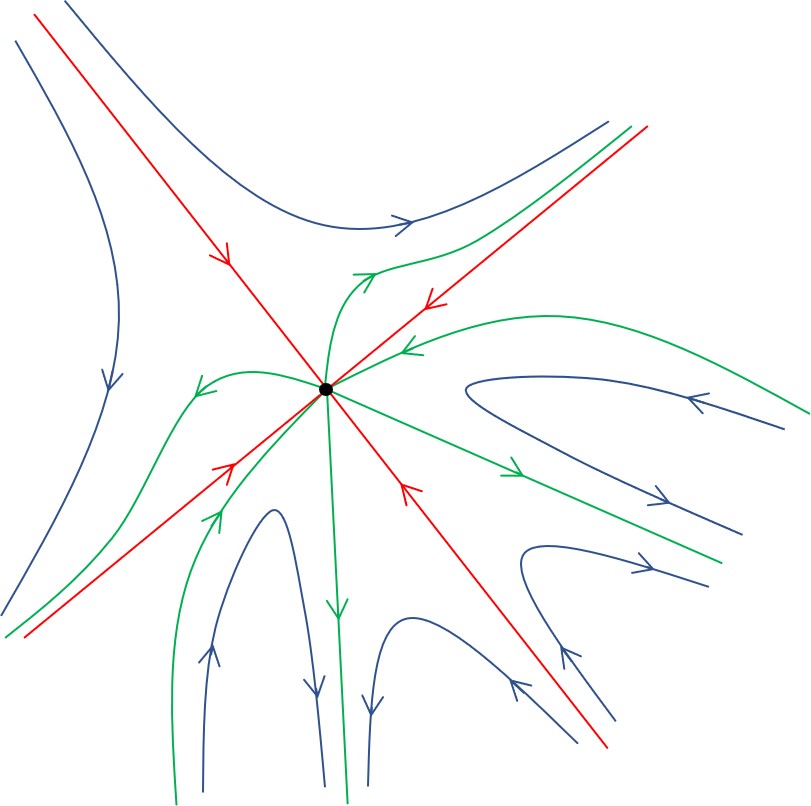}
    \captionsetup{type=figure}
    \caption{RG flows in Theory Space}
\end{center}

To picture what's going on, imagine a space such that each point corresponds to a given value of all couplings, i.e. each point corresponds to a different theory, this is the \textit{theory space}. Now, imagine we start at some point then we lower the scale of our theory by performing the three steps of RG, the end result is that we flow to some other point in theory space. Therefore this picture really has lots of flow lines that represent what happens under RG.

What are the possibilities for these flows? In general, this is an infinite dimensional space, and the flows can be rather complicated, understanding the shape of RG flows is something that occupies large swathes of modern theoretical physics. But there will be some special points for which the RG flow does nothing. These are \textit{fixed points} and correspond to conformal field theories (CFT). So let us plant ourselves in such a point. Now imagine we deform our theory by changing a coupling by hand, moving away from the fixed point. What is the RG flow then? There are three possibilities:

\begin{enumerate}
    \item We flow back towards the fixed point. In this case it is called an \textit{irrelevant deformation} since those deformations don't matter for the low-energy physics. It will turn out that most of the deformations you can think of are of this type, which vastly simplifies our analysis. The region in theory space around the fixed point that corresponds to irrelevant deformations is called the \textit{critical surface}, it is the set of points that under RG flow converge to our fixed point
    \item We flow away from the fixed point. In this case it is called a \textit{relevant deformation} since this is the kind of deformation that matters in the IR. There are typically only a few of these. The flow that emanates from a fixed point by turning on a relevant deformation is called the \textit{renormalised trajectory}
    \item Nothing happens. In this case they are called \textit{marginal deformations} and signal the fact that the fixed point is not truly a point but rather a line or some more complicated object. Exactly marginal deformations are very rare, usually they are only approximately marginal, we'll see examples of this later on, in those cases they are either \textit{marginally irrelevant} or \textit{marginally relevant} deformations
\end{enumerate}

\begin{center}
    \includegraphics[scale=0.5]{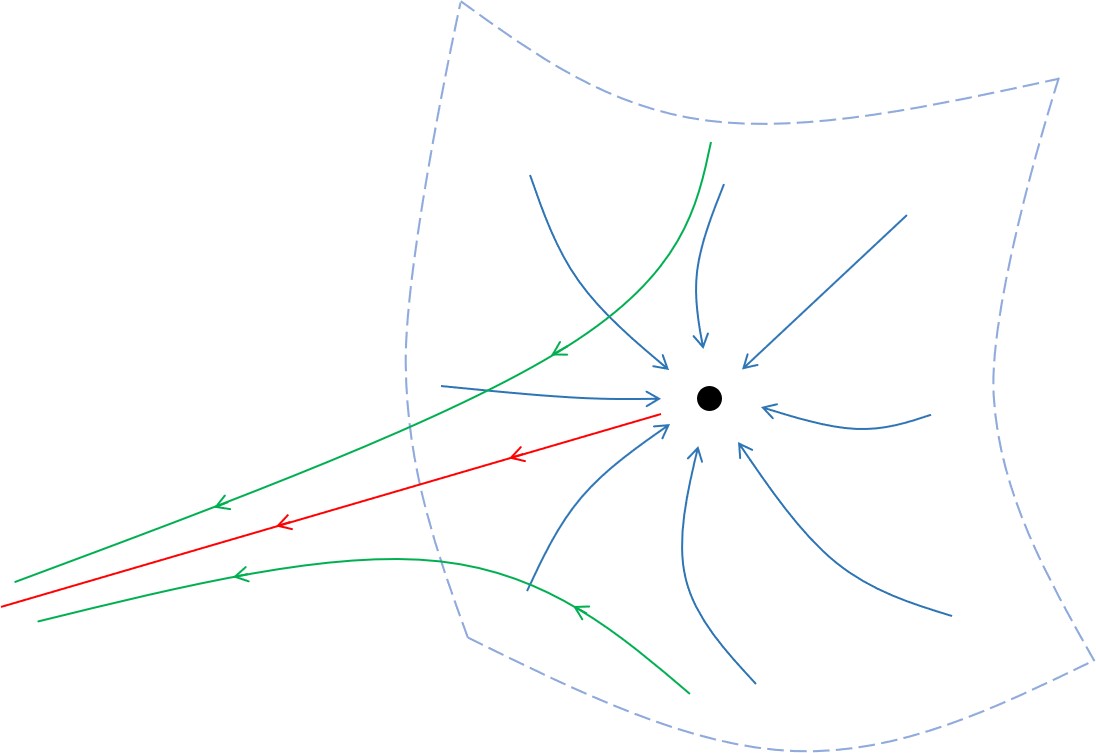}
    \captionsetup{type=figure}
    \caption{Flow near a fixed point: in blue, irrelevant directions along the critical surface; in red, renormalised trajectory; in greed, generic direction}
\end{center}

If you start in some arbitrary point in theory space more complicated things can happen like flowing to a limit cycle, these kinds of exotic RG flows can be ruled out in most cases, but our grasp away from fixed points is rather limited.

The fact that this flow is highly converging on a renormalised trajectory is the reason for the phenomenon of universality we mentioned in the introduction, where distinct physical systems like gases and magnets behave in similar ways, they start from different points in theory space but end up flowing to the same trajectory. It is also what makes high energy physics really challenging because there is very little hope to extract the details of the UV if you can only do experiments deep in the IR. 

This is all very pretty, but it's hard to get a feeling for that this procedure is actually doing. What always helps is a simple example. So we'll do the simplest possible thing. It's too simple to see a lot of the features we discussed, but we must start somewhere.

Consider just the first two terms in our action, i.e. a massive Klein\footnote{Oskar Benjamin Klein: born in 1894 in Mörby, Sweden; died in 1977 in Stockholm, Sweden; Doctoral Advisor: Svante Arrhenius}-Gordon\footnote{Walter Gordon: born in 1893 in Apolda, Germany; died in 1939 in Stockholm, Sweden; Doctoral Advisor: Max Planck} field:

\begin{align}
    S_{\Lambda_0}[\phi]&=\int\dd[d]x\frac{1}{2}\qty((\grad{\phi})^2+m_0^2\phi^2)=\nonumber\\
    &=\int_{\abs{\vb{p}}<\Lambda_0}\frac{\dd[d]p}{(2\pi)^d}\frac{1}{2}\Tilde{\phi}(-\vb{p})\qty(\vb{p}^2+m_0^2)\Tilde{\phi}(\vb{p})
\end{align}

In this case, since $\phi^+$ and $\phi^-$ have supports in momentum space that do not intersect, the action factors out completely,

\begin{align}
    S_{\Lambda_0}[\phi^++\phi^-]=&\int_{\abs{\vb{p}}<\Lambda}\frac{\dd[d]p}{(2\pi)^d}\frac{1}{2}\Tilde{\phi}^-(-\vb{p})\qty(\vb{p}^2+m_0^2)\Tilde{\phi}^-(\vb{p})+\nonumber\\
    +&\int_{\Lambda<\abs{\vb{p}}<\Lambda_0}\frac{\dd[d]p}{(2\pi)^d}\frac{1}{2}\Tilde{\phi}^+(-\vb{p})\qty(\vb{p}^2+m_0^2)\Tilde{\phi}^+(\vb{p})
\end{align}
therefore,

\begin{align}
    \ee^{-S_\Lambda[\phi^-]}=&\int\mathcal{D}\phi^+\ee^{-S_{\Lambda_0}[\phi^++\phi^-]}=\nonumber\\
    =&\mathcal{N}\exp(-\int_{\abs{\vb{p}}<\Lambda}\frac{\dd[d]p}{(2\pi)^d}\frac{1}{2}\Tilde{\phi}^-(-\vb{p})\qty(\vb{p}^2+m_0^2)\Tilde{\phi}^-(\vb{p}))
\end{align}
where $\mathcal{N}$ is some unimportant constant we'll ignore in what follows. This means the first step is trivial. Now we rescale the momenta to yield,

\begin{equation}
    S_\Lambda[\phi^-]=\int_{\abs{\vb{p}}<\Lambda_0}\frac{\dd[d]p'}{(2\pi)^d}\qty(\frac{\Lambda}{\Lambda_0})^d\frac{1}{2}\Tilde{\phi}^-(-\vb{p})\qty(\qty(\frac{\Lambda}{\Lambda_0})^2\vb{p}'^2+m_0^2)\Tilde{\phi}^-(\vb{p})
\end{equation}
hence,
\begin{equation}
    Z_\phi=\qty(\frac{\Lambda}{\Lambda_0})^{d+2},\text{ and }\Tilde{\phi}'(\vb{p}')=\qty(\frac{\Lambda}{\Lambda_0})^{\frac{d+2}{2}}\Tilde{\phi}^-(\vb{p})
\end{equation}
the final action is (dropping primes)
\begin{equation}
    S_\Lambda[\phi]=\int_{\abs{\vb{p}}<\Lambda_0}\frac{\dd[d]p}{(2\pi)^d}\frac{1}{2}\Tilde{\phi}(-\vb{p})\qty(\vb{p}^2+\qty(\frac{\Lambda}{\Lambda_0})^2m_0^2)\Tilde{\phi}(\vb{p})
\end{equation}

We don't generate any new terms, but the mass gets rescaled as
\begin{equation}
    m^2(\Lambda)=\qty(\frac{\Lambda_0}{\Lambda})^2m_0^2
    \label{massflow}
\end{equation}

If $m^2=0$ the RG procedure has no effect, therefore this is a fixed point, the \textit{Gaussian fixed point}. In these notes we'll never venture too far away from this point in theory space.

A common way to capture equations like (\ref{massflow}) is to differentiate wrt to the logarithm of the cutoff, this is the \textit{beta-function}

\begin{equation}
    \beta_{m^2}(\Lambda)=\Lambda\dv{m^2(\Lambda)}{\Lambda}=-2m^2(\Lambda)
\end{equation}

The coefficient on the RHS is negative, which means $m^2$ \textit{increases} in the IR, therefore it is a relevant deformation. 

So far so boring right? Let's spice things up a little bit. Consider a generic small deformation around this fixed point $\sim \phi^n$, more concretely we add the following term to the action in momentum space (in this calculation we'll only consider powers of $\phi$, but it's not very hard to extend this analysis to the case where we also add derivatives of $\phi$, the conclusions are pretty much unchanged)

\begin{equation}
    \int\prod_{i=1}^n\qty(\frac{\dd[d]p_i}{(2\pi)^d}\Tilde{\phi}(\vb{p}_i))\lambda_{0,n}\delta^{(d)}\qty(\sum_{i=1}^n\vb{p}_i)
\end{equation}

We'll still assume the first step to be trivial (we'll correct this assumption in the next section), what is the effect of the second and third steps? The measure brings out a factor of $\qty(\frac{\Lambda}{\Lambda_0})^{nd}$ and the delta-function gives a factor of $\qty(\frac{\Lambda_0}{\Lambda})^d$ during the second step. The third step contributes $\qty(\frac{\Lambda_0}{\Lambda})^{n\frac{d+2}{2}}$. The end result is

\begin{equation}
    \lambda_n(\Lambda)=\qty(\frac{\Lambda_0}{\Lambda})^{\qty(1-\frac{n}{2})d+n}\lambda_{0,n}\Rightarrow\beta_{\lambda_n}(\Lambda)=-\qty(\qty(1-\frac{n}{2})d+n)\lambda_n
\end{equation}

It's clear that the behaviour of the beta function is dependent not only on the number of fields $n$ but also on the number of dimensions $d$. The following table summarises the possible behaviours

\begin{center}
\begin{tabular}{ c|c|c|c|c|c|c }
          & $d=1$ & $d=2$ & $d=3$ & $d=4$ & $d=5$ & $d=6$ \\ 
 \hline
 $\phi^2$ & $-2$  & $-2$  & $-2$  & $-2$  & $-2$  & $-2$  \\ 
 \hline
 $\phi^4$ & $-3$  & $-2$  & $-1$  & $0$   & $1$   & $2$   \\ 
 \hline
 $\phi^6$ & $-4$  & $-2$  & $0$   & $2$   & $4$   & $6$   \\ 
 \hline
 $\phi^8$ & $-5$  & $-2$  & $1$   & $4$   & $7$   & $10$
\end{tabular}
\end{center}

For $d=1,2$ the coefficient is always negative, this means all operators\footnote{Although we're in the path integral, and these are just functions, it's commonplace to give the name of operators to these deformations of the action, reminiscent of the canonical perspective} of this type are relevant and if we turn any of them on we quickly flow away from the Gaussian fixed point and our analysis breaks down. For $d=3,4$ we have some relevant operators, one marginal, and then an infinite number of irrelevant operators. Because the irrelevant operators die off quite quickly, we can generally not turn them on at all since they won't matter for low energy physics, this means we really only need to deal with a handful of interactions, which is a massive simplification. For $d=3$ however, the quartic term is relevant so we quickly flow away from the fixed point, for $d=4$ we have some hope of staying near the fixed point and use some sort of perturbation theory, we'll analyse this in more detail in the next section. In all higher dimensions the mass term is the only relevant operator hence the low energy theory is always free.

Even from this very simple analysis we start to get some physics out of this, for example, the real reason you spent so much time studying $\phi^4$ theory in your QFT course is because it is the most general theory you can write down, in 4 dimensions, at low energies. All other interactions are literally irrelevant. However, the structure behind all this is still a bit obscure. What if I told you all this was only dimensional analysis? To see this, let's rephrase the three steps of RG in a slightly different way. We shall have the need to do this many times throughout this course in order to fully get our heads around this very subtle field, and, as is always the case in physics, every approach will have it's advantages and disadvantages, what it enlightens and what it obscures, but the only way to really understand a topic is to tackle it in as many ways as possible. So let's do it.

You may have noticed not all steps of RG are born equal. The first step sounds ridiculously harder than the other two. In comparison they seem like mere technicalities, whilst all the physics is in the first step, the coarse graining. And, although there is some truth in this, we have just seen how they can actually already lead to some interesting physics. Let's bring them to the centre stage. To do this, we must first realise how the third step has contributions both from integrating out fields, the $Z_\phi'$, and from scaling the spacetime variables, the factor of $\qty(\frac{\Lambda}{\Lambda_0})^{d+2}$. In the following I'll split these contributions apart, calling the contribution from the coarse graining the \textit{non-trivial part}, and the contribution from scaling the spacetime the \textit{trivial part}\footnote{I should point out, these names are not standard, it's merely for my convenience}. 

The next trick is to express every coupling in terms of a dimensionless coupling using the cutoff to absorb the mass dimension of the coupling. We'll denote mass dimension by square brackets, i.e. the mass dimension of $\lambda_n$ is denoted by $[\lambda_n]$ (this is sometimes also called the engineering dimension). Since $[x]=-1$, looking at the gradient term of the action we conclude that

\begin{equation}
    [\phi]=\frac{d-2}{2}
    \label{dimphi}
\end{equation}
and therefore, the mass dimension of the coupling to $\phi^n$ is
\begin{equation}
    [\lambda_n]=\qty(1-\frac{n}{2})d+n
\end{equation}
and we write
\begin{equation}
    \lambda_n=\Lambda^{[\lambda_n]}g_n
\end{equation}
where $g_n$ is a dimensionless coupling and $\Lambda$ is the current cutoff of the theory.

Now we perform RG, for convenience I'll go back to real space and write,

\begin{equation}
    \int\dd[d]{x}\Lambda_0^{[\lambda_n]}g_{0,n}\phi^n(\vb{x}),~~~~~\lambda_{0,n}=\Lambda_0^{[\lambda_n]}g_{0,n}
\end{equation}
where we made explicit the connection with our previous notation in terms of dimensionful couplings. After this we do step 1 and non-trivial part of step 3, to get

\begin{equation}
    \int\dd[d]{x}\Lambda^{[\lambda_n]}g_n(\Lambda)(\phi^-)^n(\vb{x}),~~~~~\lambda'_n=\Lambda^{[\lambda_n]}g_n(\Lambda)
\end{equation}
notice we had to keep the current cutoff out front for consistency. Now we can do step 2 and trivial part of step 3 to get

\begin{equation}
    \int\dd[d]{x'}\Lambda_0^{[\lambda_n]}g_n(\Lambda)\phi'^n,~~~~~\lambda_n(\Lambda)=\Lambda_0^{[\lambda_n]}g_n(\Lambda)
\end{equation}

Wait, this last step did nothing to the dimensionless coupling, it merely changed the cutoff factor. This means we only need to track the change in $g_n$ to get the full beta function. Tracking the change in the dimensionless coupling effectively does the trivial bits automatically. More explicitly, the beta function is

\begin{equation}
    \beta_{\lambda_n}(\Lambda)=\Lambda\dv{\lambda_n}{\Lambda}=\Lambda_0^{[\lambda_n]}\Lambda\dv{g_n}{\Lambda}=\Lambda_0^{[\lambda_n]}\beta_n(\Lambda)
\end{equation}

In this way of parametrising the flow we only need to do steps 1 and non-trivial 3 to get the full beta functions. This may seem like a big simplification but it can be a bit tricky to figure out exactly where all the factors go with this method, doing the full thing may be less prone to mistakes.

Let's look at our previous calculations once more. By neglecting steps 1 and non-trivial 3, we get $\lambda_{0,n}=\lambda'_n$, which, in terms of the dimensionless couplings means

\begin{equation}
    \Lambda^{[\lambda_n]}g_n=\Lambda_0^{[\lambda_n]}g_{0,n}\Leftrightarrow g_n(\Lambda)=\qty(\frac{\Lambda_0}{\Lambda})^{[\lambda_n]}g_{0,n}
\end{equation}
making it obvious that
\begin{equation}
    \beta_n=-[\lambda_n]g_n
\end{equation}
which is precisely the result we had previously.

This means the behaviour of RG near the Gaussian fixed point is in part determined by the mass dimension of the coupling. In particular, it also sheds some light on the meaning of steps 2 and 3. They're not mere technicalities, they are capturing the flow due to the intrinsic mass dimension of the coupling. Even if there was no functional integration (i.e. if we were dealing with a classical, rather than quantum, field theory) RG wouldn't stay silent, this contribution would still be there, albeit with a slightly different interpretation: how do the couplings change as we scale spacetime, paying due attention to the scale of the field in order to be able to make comparisons. For all these reasons, this contribution is called the \textit{classical} or \textit{scaling} contribution.

However, this analysis seems to only make sense near the Gaussian fixed point. In fact, this is not true, it is actually generic behaviour near any fixed point. To see why let's imagine we're at a fixed point, not necessarily the Gaussian fixed point, this may be a highly non-trivial interacting fixed point. How do we know we're at a fixed point? Well, if we perform RG and the couplings don't change, this means the beta functions are zero, $\beta_n^*=0$. Away from this point in theory space, they will generically be non-zero, and very complicated functions that mix all the couplings in a highly non-linear fashion. However, since they vanish at that special point, we can linearise around it, and then diagonalise the resulting matrix transformation. We end up with

\begin{equation}
    \beta_{g_a}=-\Delta_{g_a} g_a
\end{equation}
where we used a different index $a$ for the couplings to emphasise that the operators that they couple with are generically not simple powers of the fields, they can be rather complicated. Let $\mathcal{O}_a$ denote such an operator. This means the interaction terms in the action will look like,

\begin{equation}
    \int\dd[d]{x} g_a \mathcal{O}_a
\end{equation}

In analogy to the Gaussian fixed point, we call $\Delta_{g_a}$ the \textit{scaling dimension} of the coupling, or equivalently (and more commonly) $\Delta_{\mathcal{O}_a}=d-\Delta_{g_a}$ the scaling dimension of the operator. This scaling dimension does not need to coincide with the naive mass dimension of the operator (which is also, somewhat rudely, called the engineering dimension), and the difference between the two is the \textit{anomalous dimension}. We can also use these scaling dimensions to classify the operators into relevant, irrelevant, or marginal:

\begin{itemize}
    \item $\beta_{g_a}>0\Leftrightarrow\Delta_{g_a}<0\Leftrightarrow\Delta_{\mathcal{O}_a}>d$: irrelevant deformation
    \item $\beta_{g_a}<0\Leftrightarrow\Delta_{g_a}>0\Leftrightarrow\Delta_{\mathcal{O}_a}<d$: relevant deformation
    \item $\beta_{g_a}=0\Leftrightarrow\Delta_{g_a}=0\Leftrightarrow\Delta_{\mathcal{O}_a}=d$: marginal deformation
\end{itemize}

So we see that near any fixed point, RG behaves exactly like we were just doing a scaling transformation. The only catch is that the actual, physical scaling dimensions need not coincide with the naive mass dimensions. Indeed, experimentally, these scaling dimensions can be measured through \textit{critical exponents}. Quite famously, these exponents seemed quite bizarre, sometimes even irrational! RG managed to correctly explain them, we were merely measuring the \textit{scaling} dimensions rather than the \textit{mass} dimension. Dimensional analysis is saved by the present of a new dimensionful quantity to play with: the cutoff.

But wait a moment, what about the kinetic operator? We have insisted to have a canically normalised kinetic term, so it naively seems like it doesn't run, by definition. This is kind of true, however, there is some physics left to be extracted. The reason is that the field strength renormalisation $Z_\phi$ depends on the cutoff itself. So, in principle, we can define some sort of beta function for $Z_\phi$. Even though the coupling by itself has little physical meaning, due to the arbitrariness in the overall scale of the field, the beta function will have a meaning, as the anomalous dimension of the field. 

First, what happens to the field if we just make some naive scale transformation $\vb{x}\to\vb{x}'=\frac{\Lambda}{\Lambda_0}\vb{x}$? Well it changes according to the mass dimension, i.e. 

\begin{equation}
    \phi\to\phi'=\qty(\frac{\Lambda}{\Lambda_0})^{-[\phi]}\phi=\qty(\frac{\Lambda}{\Lambda_0})^{\frac{2-d}{2}}\phi
\end{equation}

However, a real change of scale in our theory is given not by that simple transformation, but by an RG flow, and in that case, the transformation is

\begin{equation}
    \phi\to\phi'=\sqrt{Z_\phi}\phi=\qty(\frac{\Lambda}{\Lambda_0})^{\frac{2-d}{2}}\sqrt{Z'_\phi}\phi
\end{equation}
where both $Z_\phi$ and $Z'_\phi$ are functions of $\Lambda$. Expanding $\Lambda$ about $\Lambda_0$ (and noting that $Z'_\phi(\Lambda_0)=1$ by definition) gives

\begin{equation}
    \phi'=\qty(1+\qty(\frac{d-2}{2}-\eval{\frac{1}{2}\Lambda\dv{\log Z_\phi'(\Lambda)}{\Lambda}}_{\Lambda=\Lambda_0})\qty(1-\frac{\Lambda}{\Lambda_0})+\dots)\phi
\end{equation}

We see that we get the naive mass dimension plus an extra contribution which involves the derivative of the logarithm of the field strength renormalisation. This prompts us to the define the field anomalous dimension as 

\begin{equation}
    \gamma_\phi=-\frac{1}{2}\Lambda\dv{\log Z_\phi'(\Lambda)}{\Lambda}
\end{equation}
which gives the interpretation for the flow of the field strength renormalisation. It gives the anomalous dimension for how the field itself scales as we change scale.

And once more, as you are probably sick of hearing at this point, we can and we have measured the field anomalous dimension, and RG agrees with Nature.

\subsection{Perturbative Renormalisation}

In the last section, we have discussed the structure of RG, and calculated explicitly the flow near the Gaussian fixed point. However, we have neglected steps 1 and non-trivial 3, which means we never had to perform any functional integration, it came down to simple scaling and dimensional analysis. In fact, for these calculations, we didn't even need to invoke any partition function.

Now, it's finally time to grow up, take the path integral seriously, and tackle the quantum corrections to the beta functions. But we have a problem, if we start in an arbitrary point in theory space, we are entirely lost. The reason being we have no idea how to actually calculate a generic path integral. As path integrals are concerned, we can only do two things analytically: Gaussian integrals, and perturbation theory. Therefore, to make progress, we will assume we start with a weakly coupled theory, i.e. we assume that $g_{0,n}\ll 1,~n>2$. In this way we can use the techniques developed in Appendix \ref{pathqft} to evaluate the path integral.

Since the behaviour of the flows will depend on the number of dimensions, we shall organise our calculation starting from higher dimensions and progressively lowering the number of dimensions.

\subsubsection{$d>4$}

This case is rather trivial. If we stay at weak coupling we can trust the classical scaling analysis to determine the character of our deformation. Using the results from the last section, we conclude that all interactions are irrelevant. Hence, if we move away from the Gaussian fixed point, we quickly focus on the free theory, all interactions die out in the IR. The theory is always free in the IR and scaling always works.

\subsubsection{$d=4$}

In four dimensions life is more interesting. All interactions are irrelevant except for the quartic one. This means we can drop most interactions and just start with the following action:

\begin{equation}
    S_{\Lambda_0}[\phi]=\int\dd[4]{x}\qty(\frac{1}{2}(\grad{\phi})^2+\frac{1}{2}m_0^2\phi^2+\frac{\lambda_0}{4!}\phi^4)
\end{equation}
where, despite using continuum notation, we implicitly cut off modes with $\abs{\vb{p}}>\Lambda_0$. 

Now we split the fields into high and low energy modes, as we did before, the only difference is that this time the action doesn't factorise neatly. This means we must calculate the full path integral given by

\begin{equation}
    \ee^{-S_\Lambda[\phi^-]}=\int\mathcal{D}\phi^+\ee^{-S_{\Lambda_0}[\phi^++\phi^-]}
\end{equation}

Before we attempt calculating this object let's massage this expression a little bit. Notice that, just as before, when expanding $S_{\Lambda_0}[\phi^++\phi^-]$ in powers of the fields, since the supports in momentum space of the high and low energy fields don't intersect, there is no mixing at the quadratic level. This means we can make the split,

\begin{equation}
    S_{\Lambda_0}[\phi^++\phi^-]=S_\text{free}[\phi^+]+S_\text{free}[\phi^-]+S_\text{int}[\phi^+,\phi^-]
\end{equation}
where
\begin{align}
    S_\text{free}[\phi]&=\int\dd[4]{x}\qty(\frac{1}{2}(\grad{\phi})^2+m_0^2\phi^2)\\
    S_\text{int}[\phi^+,\phi^-]&=\int\dd[4]{x}\frac{\lambda_0}{4!}(\phi^++\phi^-)^4
\end{align}

Now, since $S_\text{free}[\phi^-]$ is independent of $\phi^+$, we can pull it out of the integral. Defining $\Var{S_\Lambda[\phi^-]}=S_\Lambda[\phi^-]-S_\text{free}[\phi^-]$, we get

\begin{equation}
    \ee^{-\Var{S_\Lambda[\phi^-]}}=\int\mathcal{D}\phi^+\ee^{-S_\text{free}[\phi^+]-S_\text{int}[\phi^+,\phi^-]}
\end{equation}

Recalling Appendix \ref{pathqft}, we see that this is exactly like the Wilsonian effective action where the low energy fields serve as sources for the high energy fields, which are the only ones that can propagate. Therefore, we can calculate this object perturbatively by summing over all connected Feynman diagrams with a set number of external $\phi^-$ fields. We shall represent the external $\phi^-$ fields with black lines, and we won't attach propagators to these lines. All internal lines correspond to $\phi^+$ fields being integrated over, they have propagators associated with them, and we shall represent them with blue lines to emphasise the difference.\footnote{Some authors use double line notation, or difference between solid and dotted lines. We prefer different colours since the other notations have other uses, respectively to denote matrix fields, and to distinguish between scalars and spinors}

\begin{itemize}
    \item \textbf{No external fields:}
\end{itemize}

\begin{equation*}
\begin{tikzpicture}[baseline=(a)]
  \begin{feynman}
    \vertex (a);
    \vertex [above=1cm of a] (aux1);
    \vertex [below=1cm of a] (aux2);
    
    \diagram* {
      (a) -- [blue,plain,out=135,in=180] (aux1),
      (aux1) -- [blue,plain,out=0,in=45] (a),
      (a) -- [blue,plain,out=-45,in=0] (aux2),
      (aux2) -- [blue,plain,out=180,in=-135] (a),
    };
  \end{feynman}
  \draw[fill=black] (a) circle (1.5pt);
\end{tikzpicture}
+
\begin{tikzpicture}[baseline=(centre)]
  \begin{feynman}
    \vertex (centre);
    \vertex [above=0.5cm of centre] (a);
    \vertex [below=0.5cm of centre] (b);
    \vertex [above=1cm of a] (aux1);
    \vertex [below=1cm of b] (aux2);
    
    \diagram* {
      (a) -- [blue,plain,out=135,in=180] (aux1),
      (aux1) -- [blue,plain,out=0,in=45] (a),
      (b) -- [blue,plain,out=-45,in=0] (aux2),
      (aux2) -- [blue,plain,out=180,in=-135] (b),
      (a) -- [blue,plain,out=-135,in=135] (b),
      (a) -- [blue,plain,out=-45,in=45] (b),
    };
  \end{feynman}
  \draw[fill=black] (a) circle (1.5pt);
  \draw[fill=black] (b) circle (1.5pt);
\end{tikzpicture}
+
\begin{tikzpicture}[baseline=(centre)]
  \begin{feynman}
    \vertex (centre);
    \vertex [above=0.75cm of centre] (a);
    \vertex [below=0.75cm of centre] (b);
    \vertex [right=0.75cm of centre] (aux1);
    \vertex [right=0.375cm of centre] (aux2);
    \vertex [left=0.75cm of centre] (aux3);
    \vertex [left=0.375cm of centre] (aux4);
    
    \diagram* {
      (b) -- [blue,plain,out=0,in=-90] (aux1),
      (aux1) -- [blue,plain,out=90,in=0] (a),
      (b) -- [blue,plain,out=180,in=-90] (aux3),
      (aux3) -- [blue,plain,out=90,in=180] (a),
      (b) -- [blue,plain,out=45,in=-90] (aux2);
      (aux2) -- [blue,plain,out=90,in=-45] (a);
      (b) -- [blue,plain,out=135,in=-90] (aux4);
      (aux4) -- [blue,plain,out=90,in=-135] (a);
    };
  \end{feynman}
  \draw[fill=black] (a) circle (1.5pt);
  \draw[fill=black] (b) circle (1.5pt);
\end{tikzpicture}
+\cdots
\end{equation*}

These diagrams are known as \textit{vacuum bubbles}. They contribute with a scale dependent, but field \textit{in}dependent, additive constant to the action. In a theory of gravity, these diagrams would have to be accounted for, since gravity couples to everything. And, in this context, this would be known as the \textit{cosmological constant}. The name arises because, in our universe, this constant is very small, and hence, it is only relevant at cosmological scales. Correctly accounting for the smallness of the cosmological constant is one of the big unsolved problems in physics. For our purposes, however, this constant will not contribute to correlation functions, so we can neglect it.

\begin{itemize}
    \item \textbf{Two external fields:}
\end{itemize}

The leading order contribution with two external fields is:

\begin{equation*}
\begin{tikzpicture}[baseline=(b.south)]
  \begin{feynman}
    \vertex (a);
    \vertex [right=1cm of a] (c);
    \vertex [right=1cm of c] (b);
    \vertex [above=1cm of c] (d);
    
    \diagram* {
      (a) -- [plain,out=0,in=180] (c),
      (c) -- [plain,out=0,in=180] (b),
      (c) -- [blue,plain,out=135,in=180,momentum={[arrow style=blue, arrow shorten=0.27] $\vb{k}$}] (d),
      (d) -- [blue,plain,out=0,in=45] (c),
    };
    \draw[fill=black] (c) circle (1.5pt);
  \end{feynman}
\end{tikzpicture}
=-\frac{\lambda_0}{2}\int_\Lambda^{\Lambda_0}\frac{\dd[4]k}{(2\pi)^4}\frac{1}{\vb{k}^2+m_0^2}=-\lambda_0\frac{\mathrm{Vol}(S^{3})}{2(2\pi)^4}\int_\Lambda^{\Lambda_0}\dd{k}\frac{k^{3}}{k^2+m_0^2}=
\end{equation*}
\begin{equation}
=-\frac{\lambda_0}{32\pi^2}\qty(\Lambda_0^2-\Lambda^2+m_0\log(\frac{\Lambda^2+m_0^2}{\Lambda_0^2+m_0^2}))
\label{1-loopmass}
\end{equation}

This means\footnote{There is an annoying minus sign coming from the fact the Wilsonian action is defined as \textit{minus} the logarithm of the partition function, if you look carefully at Appendix \ref{pathqft} you will spot the bastard}

\begin{equation}
    m'^2=m_0^2+\frac{\lambda_0}{32\pi^2}\qty(\Lambda_0^2-\Lambda^2+m_0\log(\frac{\Lambda^2+m_0^2}{\Lambda_0^2+m_0^2}))
\label{mprime}
\end{equation}

Knowing that $g_2=\Lambda^{-2}m'^2$, we could use (\ref{mprime}) directly to calculate the beta function. However, it's a lot easier to go directly from the momentum integral, using the fact that

\begin{equation}
    \Lambda\dv{\Lambda}\int_\Lambda^{\Lambda_0}\dd{k}f(k)=-\Lambda f(\Lambda)
\end{equation}

Either way, the end result is,

\begin{equation}
    \beta_2=\Lambda\dv{g_2}{\Lambda}=\Lambda\dv{\Lambda}(\Lambda^{-2}m'^2)=-2g_2-\frac{\lambda_0}{16\pi^2}\frac{1}{1+\Lambda^{-2}m_0^2}=-2g_2-\frac{1}{16\pi^2}\frac{g_4}{1+g_2}
\end{equation}
where in the final equality we used the fact that, to leading order, we can substitute $g_4=\lambda_0$ and $g_2=\Lambda^{-2}m_0^2$, which is legitimate since our calculations are only valid up to that order anyway.

\begin{itemize}
    \item \textbf{Four external fields:}
\end{itemize}

The leading order contributions with four external fields are (we shall use conventions were the external momenta are all pointing inwards):

\begin{equation*}
\begin{tikzpicture}[baseline=(e)]
  \begin{feynman}
    \vertex (a) {1};
    \vertex [below=1.5cm of a] (b) {2};
    \vertex [right=1.5cm of a] (c) {4};
    \vertex [right=1.5cm of b] (d) {3};
    \vertex [below=0.75cm of a] (f);
    \vertex [right=0.75cm of f] (e);
    
    \diagram* {
      (a) -- [plain,out=-45,in=135] (e),
      (b) -- [plain,out=45,in=-135] (e),
      (e) -- [plain,out=45,in=-135] (c),
      (e) -- [plain,out=-45,in=135] (d),
    };
  \end{feynman}
  \draw[fill=black] (e) circle (1.5pt);
\end{tikzpicture}
+
\begin{tikzpicture}[baseline=(e)]
  \begin{feynman}
    \vertex (a) {1};
    \vertex [below=1.5cm of a] (b) {2};
    \vertex [below=0.75cm of a] (f);
    \vertex [right=2.5cm of a] (c) {4};
    \vertex [right=2.5cm of b] (d) {3};
    \vertex [right=0.75cm of f] (e);
    \vertex [right=1cm of e] (e2);
    \vertex [right=0.5cm of e] (g);
    \vertex [above=0.5cm of g] (g1);
    \vertex [below=0.5cm of g] (g2);
    
    \diagram* {
      (a) -- [plain,out=-45,in=135] (e),
      (b) -- [plain,out=45,in=-135] (e),
      (e) -- [blue,plain,out=90,in=180] (g1),
      (e) -- [blue,plain,out=-90,in=180,rmomentum'={[arrow style=blue, arrow shorten=0.25]$\vb{k}$}] (g2),
      (g1) -- [blue,plain,out=0,in=90] (e2),
      (g2) -- [blue,plain,out=0,in=-90] (e2),
      (e2) -- [plain,out=45,in=-135] (c),
      (e2) -- [plain,out=-45,in=135] (d),
    };
  \end{feynman}
  \draw[fill=black] (e) circle (1.5pt);
  \draw[fill=black] (e2) circle (1.5pt);
\end{tikzpicture}
+
\begin{tikzpicture}[baseline=(h)]
  \begin{feynman}
    \vertex (a) {1};
    \vertex [below=2.5cm of a] (b) {2};
    \vertex [right=1.5cm of a] (c) {4};
    \vertex [right=1.5cm of b] (d) {3};
    \vertex [right=0.75cm of a] (f);
    \vertex [below=0.75cm of f] (e);
    \vertex [below=1cm of e] (e2);
    \vertex [below=0.5cm of e] (h);
    \vertex [left=0.5cm of h] (h1);
    \vertex [right=0.5cm of h] (h2);
    
    \diagram* {
      (a) -- [plain,out=-45,in=135] (e),
      (b) -- [plain,out=45,in=-135] (e2),
      (e) -- [plain,out=45,in=-135] (c),
      (e2) -- [plain,out=-45,in=135] (d),
      (e) -- [blue,plain,out=180,in=90,rmomentum'={[arrow style=blue,arrow shorten=0.25]$\vb{k}$}] (h1),
      (h1) -- [blue,plain,out=-90,in=180] (e2),
      (e) -- [blue,plain,out=0,in=90] (h2),
      (h2) -- [blue,plain,out=-90,in=0] (e2),
    };
  \end{feynman}
  \draw[fill=black] (e) circle (1.5pt);
  \draw[fill=black] (e2) circle (1.5pt);
\end{tikzpicture}
+
\begin{tikzpicture}[baseline=(e)]
  \begin{feynman}
    \vertex (a) {1};
    \vertex [below=1.5cm of a] (b) {2};
    \vertex [below=0.75cm of a] (f);
    \vertex [right=2.5cm of a] (c) {4};
    \vertex [right=2.5cm of b] (d) {3};
    \vertex [right=0.75cm of f] (e);
    \vertex [right=1cm of e] (e2);
    \vertex [right=0.5cm of e] (g);
    \vertex [above=0.5cm of g] (g1);
    \vertex [below=0.5cm of g] (g2);
    
    \diagram* {
      (a) -- [plain,out=-45,in=135] (e),
      (b) -- [plain,out=-90,in=-90] (e2),
      (e) -- [blue,plain,out=90,in=180] (g1),
      (e) -- [blue,plain,out=-90,in=180,rmomentum'={[arrow style=blue, arrow shorten=0.25]$\vb{k}$}] (g2),
      (g1) -- [blue,plain,out=0,in=90] (e2),
      (g2) -- [blue,plain,out=0,in=-90] (e2),
      (e2) -- [plain,out=45,in=-135] (c),
      (e) -- [plain,out=-45,in=-90] (d),
    };
  \end{feynman}
  \draw[fill=black] (e) circle (1.5pt);
  \draw[fill=black] (e2) circle (1.5pt);
\end{tikzpicture}
=
\end{equation*}

\begin{align}
    =-\lambda_0+\frac{\lambda_0^2}{2}\int_\Lambda^{\Lambda_0}\frac{\dd[4]{k}}{(2\pi)^4}\bigg(&\frac{1}{\vb{k}^2+m_0^2}\frac{1}{(\vb{k}+\vb{p}_1+\vb{p}_2)^2+m_0^2}+\nonumber\\
    +&\frac{1}{\vb{k}^2+m_0^2}\frac{1}{(\vb{k}+\vb{p}_1+\vb{p}_4)^2+m_0^2}+\nonumber\\
    +&\frac{1}{\vb{k}^2+m_0^2}\frac{1}{(\vb{k}+\vb{p}_1+\vb{p}_3)^2+m_0^2}\bigg)
\end{align}

Hmm... this time the integral depends on external momenta, what does that mean? Since these integrals are all finite we can do a Taylor series in external momenta. We'll end up with terms that look like $\sim(\tilde{\phi}^-)^2p^2(\tilde{\phi}^-)^2$ which in real space correspond to $\sim(\phi^-)^2\partial^2(\phi^-)^2$. This signals the fact that when we start integrating out degrees of freedom, we will generate all possible terms consistent with the symmetries we impose, our simplification of considering only a handful of terms seems to break down. However, all these interactions are classically irrelevant, which means scaling will come to the rescue and these terms will be completely swamped by the classical contribution, quickly dying off. Therefore, we won't need to consider them.

Looking at the limit when $\vb{p}_a=\vb{0},~a=1,\dots,4$, we get

\begin{align}
    &-\lambda_0+\frac{3\lambda_0^2}{2}\int_\Lambda^{\Lambda_0}\frac{\dd[4]{k}}{(2\pi)^4}\frac{1}{(\vb{k}^2+m_0^2)^2}=-\lambda_0+\lambda_0^2\frac{3\mathrm{Vol}(S^3)}{2(2\pi)^4}\int_\Lambda^{\Lambda_0}\dd{k}\frac{k^3}{(k^2+m_0^2)^2}=\nonumber\\
    =&-\lambda_0+\lambda_0^2\frac{3}{32\pi^2}\qty(\frac{m_0^2}{m_0^2+\Lambda_0^2}-\frac{m_0^2}{m_0^2+\Lambda^2}+\log(\frac{m_0^2+\Lambda_0^2}{m_0^2+\Lambda^2}))
\end{align}
therefore
\begin{equation}
    \lambda'=\lambda_0\qty(1-\lambda_0\frac{3}{32\pi^2}\qty(\frac{m_0^2}{m_0^2+\Lambda_0^2}-\frac{m_0^2}{m_0^2+\Lambda^2}+\log(\frac{m_0^2+\Lambda_0^2}{m_0^2+\Lambda^2})))
\end{equation}

You may be concerned that we are considering a $O(\lambda_0^2)$ correction here, while we only considered until $O(\lambda_0)$ in the mass term. However, the original term here is already $O(\lambda_0)$, the correction still only has an single \textit{extra} power of $\lambda_0$. This feature is completely generic, we should always count loops instead of vertices when calculating a Wilsonian effective action. This can be rigorously shown using elementary graph theory but it's not very relevant for our purposes so we won't delve into it.

The beta function is then

\begin{equation}
    \beta_4=\Lambda\dv{g_4}{\Lambda}=\Lambda\dv{\lambda'}{\Lambda}=\frac{3}{16\pi^2}\frac{\Lambda^4\lambda_0^2}{(\Lambda^2+m_0^2)^2}=\frac{3}{16\pi^2}\frac{g_4^2}{(1+g_2)^2}
\end{equation}
where in the last line we once again approximated $g_4=\lambda_0$ and $g_2=\Lambda^{-2}m_0^2$.

All further interactions are irrelevant and hence are irrelevant for the rest of this section\footnote{Pun intended}. What can we conclude from this analysis?

Looking at the beta function for the quartic coupling, we see that it is positive, this means this coupling is actually \textit{marginally irrelevant} rather than purely marginal. Therefore, $d=4$ behaves similarly to $d>4$ in the sense that all interactions are irrelevant, and in the IR we always flow to a free theory. This phenomenon is known as \textit{triviality}. The only difference between $d=4$ and $d>4$ is that the quartic interaction dies off much slower in $d=4$, it only decays logarithmically, instead of with a power law. We shall have much more to say about these kinds of decays and triviality later on when we discuss the continuum limit, but keep these features in mind.

\begin{center}
    \includegraphics[scale=0.4]{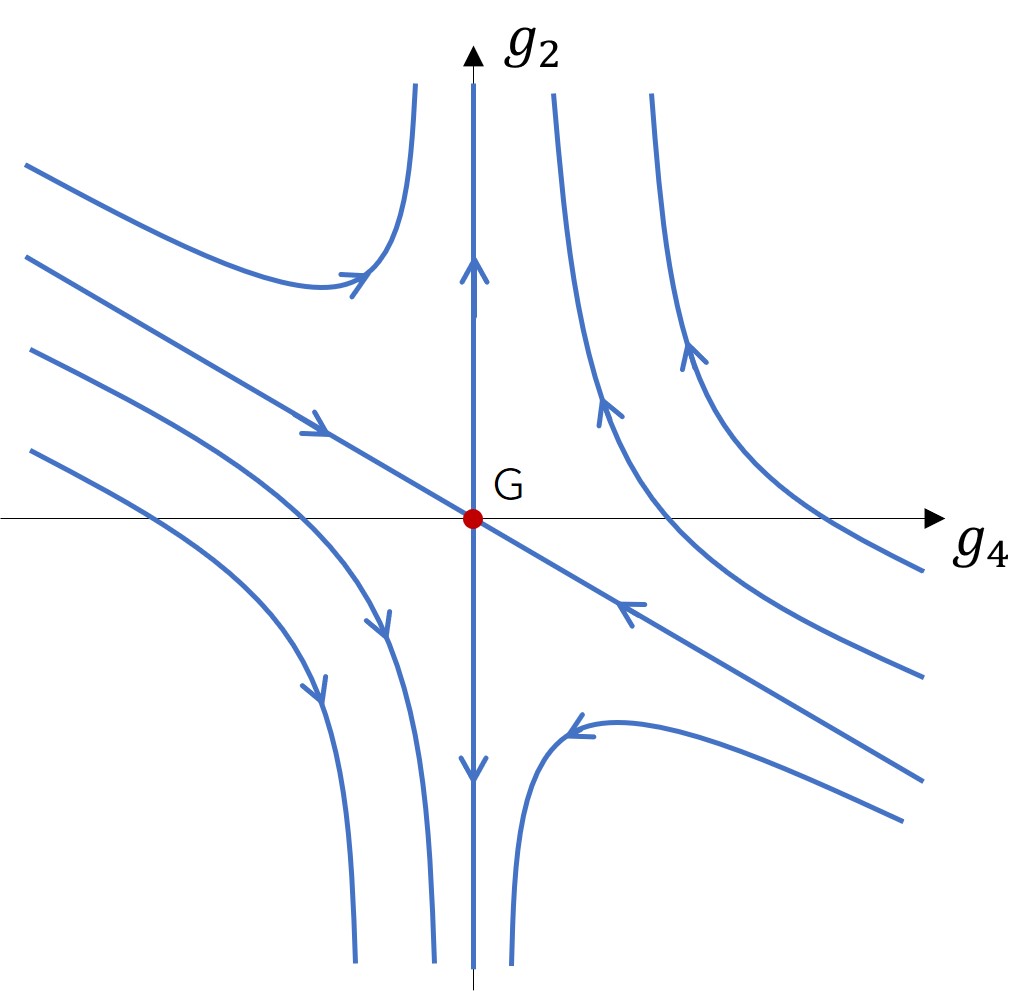}
    \captionsetup{type=figure}
    \caption{RG flows for $d=4$}
\end{center}

Notice that, if you just turn on this coupling, you won't simply flow back to the Gaussian fixed point, there is some mixing between this coupling and the quadratic coupling. We will in fact flow to a free \textit{massive} theory rather than the Gaussian fixed point. In order to hit the Gaussian fixed point we must turn on a particular combination of $g_2$ and $g_4$.

\subsubsection{$d<4$}

We would like to complete our analysis and study the RG flow for $d=1,2,3$. However, perturbation theory is of little help. All these theories feature relevant interactions, which grow as we flow to the IR, rendering our weak coupling approximation useless. 

It turns out that the $d=1,2$ cases are exactly solvable. In $d=1$ we can describe the flows in terms of ordinary quantum mechanics and tunnelling prevents us from reaching a fixed point. In $d=2$ we can use sophisticated conformal field theory techniques to figure out there is actually an infinite number of fixed points. However, these details are beyond the scope of this course. To examine $d=3$ we use a trick called the \textit{epsilon expansion}. It consists of considering $d=4-\epsilon$ and doing perturbation theory for $\epsilon\ll1$. 

And by now you surely think I've gone mad. What the hell to I mean by taking the dimension to be $4-\epsilon$? And haven't I noticed 1 is not actually $\ll1$?

The answer to the first question is that we don't mean considering an action or a path integral in non-integer dimensions. We're just going to \textit{formally} extend the integrals in perturbation theory, and the beta functions, to non-integer dimensions, and then expand in powers of $\epsilon$. The answer to the second question is shhhhhh. Or more explicitly, blah, blah, resummation, blah, blah.

It's straightforward to generalise the beta functions for arbitrary $d$, we get

\begin{align}
    \beta_2&=-2g_2-\frac{\mathrm{Vol}(S^{d-1})}{32\pi^4}\frac{g_4}{1+g_2}\\
    \beta_4&=(d-4)g_4+\frac{3\mathrm{Vol}(S^{d-1})}{32\pi^4}\frac{g_4^2}{(1+g_2)^2}
\end{align}

To leading order, we can just take $\mathrm{Vol}(S^{d-1})=2\pi^2$ to obtain

\begin{align}
    \beta_2&=-2g_2-\frac{1}{16\pi^2}\frac{g_4}{1+g_2}\\
    \beta_4&=-\epsilon g_4+\frac{3}{16\pi^2}\frac{g_4^2}{(1+g_2)^2}
\end{align}

The remarkable thing about these beta functions is that there is an additional fixed point beyond the Gaussian fixed point, this is called the Wilson-Fisher\footnote{Michael Ellis Fisher: born in 1931 in Fyzabad, Trinidad and Tobago; Doctoral Advisor: Cyril Domb} fixed point. To find it, we set the beta functions to zero and solve the resulting system of equations:

\begin{align}
    0&=-2g^*_2-\frac{1}{16\pi^2}\frac{g^*_4}{1+g^*_2}\\
    0&=-\epsilon g^*_4+\frac{3}{16\pi^2}\frac{{g^*_4}^2}{(1+g^*_2)^2}
\end{align}

To leading order in $\epsilon$, the solution for $g^*_2\neq0$ and $g^*_4\neq0$ is

\begin{align}
    g^*_2&=-\frac{1}{6}\epsilon\\
    g^*_4&=\frac{16\pi^2}{3}\epsilon
\end{align}

Since $g^*_4\sim O(\epsilon)$, this fixed point is at weak coupling and our whole analysis is consistent. Now that we've found a fixed point we should analyse the flow around it. To do that we linearise the beta functions around this point, i.e. we set

\begin{align}
    g_2=g^*_2+\var{g_2}\\
    g_4=g^*_4+\var{g_4}
\end{align}
and expand the beta functions up to linear order in $\var{g_2},\var{g_4},\text{ and }\epsilon$. The end result is

\begin{align}
    \Lambda\dv{\Lambda}\mqty(\var{g_2} \\ \var{g_2})=\mqty(-2+\frac{\epsilon}{3} & -\frac{1}{16\pi^2}(1+\frac{\epsilon}{6}) \\ 0 & \epsilon) \mqty(\var{g_2} \\ \var{g_4})
\end{align}

The eigenvalues of this matrix are $\Delta_2=-2+\frac{\epsilon}{3}$ and $\Delta_4=\epsilon$. Since $\epsilon$ is small $\Delta_2<0$ and $\Delta_4>0$ which means we have one relevant and one irrelevant direction. The relevant direction is the mass direction as usual, the irrelevant direction is some combination of $g_2$ and $g_4$ which you can calculate, but it's exact form is not very important. The resulting flow diagram is as follows:

\begin{center}
    \includegraphics[scale=0.4]{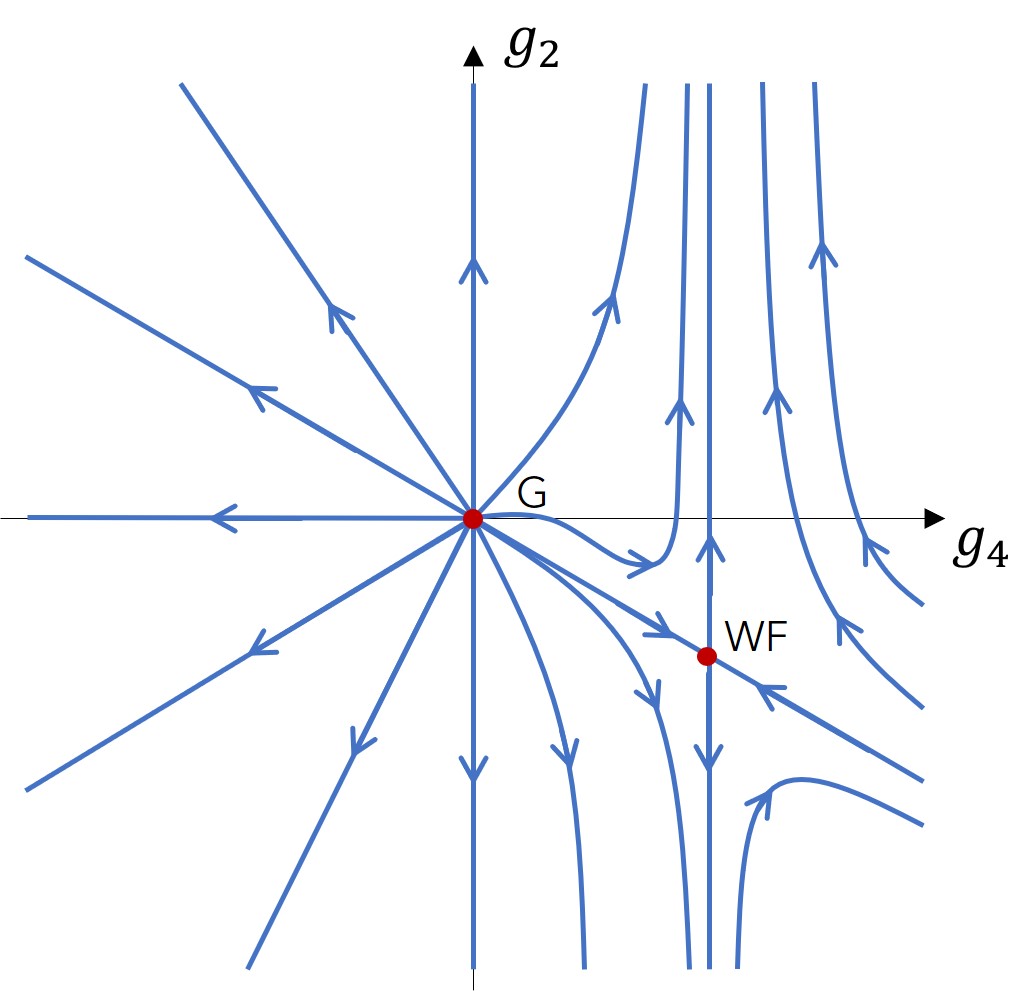}
    \captionsetup{type=figure}
    \caption{RG flows for $d=4-\epsilon$}
\end{center}

This flow diagram is much richer that we had previously, and although the exact details are only valid for small $\epsilon$, we can hope that the broad structure holds all the way down to $d=3$ (and independent calculations show that it does!). In summary, we have two fixed points: the Gaussian fixed point, this is the familiar one, it's non-interacting, and, in this projection to the $g_2-g_4$ plane, it is completely unstable, all deformations lead away from it; and the Wilson-Fisher fixed point, which is an interacting fixed point, and has one stable, and one unstable direction. What can we do with this?

Starting from the Gaussian fixed point we can, for example, flow directly in the $g_2$ direction. In this trajectory, the theory is always free, we say this is \textit{trivial}. Another possibility is that we turn on the deformation that hits exactly the WF fixed point in the IR, now the theory is interacting in the IR. However, it is a free theory in the UV, for that reason, we say that this theory is \textit{asymptotically free}. If we turn on a direction very close to this one, we will miss the WF fixed point. However, we will very rapidly focus on the renormalised trajectory that corresponds to its mass deformation. This is still an asymptotically free direction. If we start on the WF fixed point on the other hand, we can turn on its irrelevant direction which brings us back to the fixed point, and is boring. Alternatively, we can turn on it's relevant direction, $g_2$, this theory is emanating from a fixed point in the UV but it is an interacting fixed point, we call theories such as this \textit{asymptotically safe}. Any other direction will quickly focus on the renormalised trajectory.

These flows are very important for two reasons. First of all this is a flow of a physical theory, the Ising model, or equivalently, gases in the liquid-gas phase transition. Both these systems exhibit second order phase transitions which characteristically are scale invariant, i.e. they correspond to fixed points. And, because the Gaussian fixed point is actually unstable, in the real world, we will most definitely hit the WF fixed point. This has many experimental consequences, and correctly explaining them was one the great successes of RG. 

The second reason is that it illustrates many different features of RG flows. I've been emphasising all these names of triviality, asymptotic freedom and asymptotic safety because they are very important for high energy physics. The reason for this is that, if our world has no intrinsic scale, if spacetime is, in fact, continuous, then if we go up in the UV we must hit some fixed point, otherwise, at some point, the theory breaks down. However, the theories that describe the experiments at low energies are interacting, which means we should avoid theories that are entirely trivial, such as $\phi^4$ in $d=4$. We want either asymptotically free, or asymptotically safe theories. Most remarkably, quantum chromodynamics (QCD), the theory of strong interactions, is asymptotically free, which means it can be valid at arbitrarily high energies. It also means it is more weakly coupled the higher up we go in the energy scale, and it is very strongly coupled in the IR, which causes very severe theoretical challenges such as confinement and hadronisation. It is still an open problem whether the full standard model of particle physics has a fixed point in the UV.

Further, now you are capable of understanding one of the problems with quantum gravity. If we take the action for general relativity and apply the same procedure, we find that all interactions are irrelevant, which makes it very hard to analyse at very high energies and naively the whole theory breaks down. Many approaches to quantum gravity make it have a physical limit above which physics makes no sense to solve this issue. This basically stops the RG flow before it gets too bad. Another possibility is that GR actually comes from a highly non-trivial interacting fixed point in the UV, i.e. it is an asymptotically safe theory, this is the asymptotic safety program for quantum gravity. They try to characterise this fixed point using very sophisticated RG techniques and figure out whether or not we can construct such a theory. So far it is inconclusive which of these leads in the right path, only time will tell.

\subsection{The Local Potential Approximation}\label{lpa}

So far we've been able to do two things: Gaussian integrals and perturbation theory. And while those are really powerful, there remains the question: can we do more? The short answer is: no. The long answer is: kinda. What we can do is use other approximations that are not weak coupling to get some results. However, the trick is to use those approximations to reduce the actual calculations to either Gaussian integrals, or Feynman diagrams. This whole program of trying to examine the RG flows when weak coupling is not a good approximation is called \textit{exact renormalisation group}, and we'll go back to it when we discuss the continuum limit. But for now, let's just look at a simple example, to get a feel for how these things work. We won't extract much more physics than we already did just using perturbation theory, but we'll have a slightly different take on the derivation, which is always useful.

There are two approximations in what follows, the first one is merely for convenience, it can be dropped or changed without affecting the ideas too much, but it will make things easier for us, for the time being. It is the \textit{local potential approximation}, and it amounts to considering that the effective action at the scale $\Lambda$ is given by

\begin{equation}
    S_\Lambda[\phi]=\int\dd[d]{x}\qty(\frac{1}{2}(\grad{\phi})^2+V(\phi))
\end{equation} 
where $V(\phi)$ is a local function of $\phi$, i.e. it doesn't depend on derivatives of $\phi$. This is an approximation, as we've seen, in general, additional derivative interactions will be generated as we flow to the IR. For $d>2$ they are irrelevant, so we can, hopefully, neglect them.

The second approximation is the crucial one, we give up on the idea of integrating out all modes between $\Lambda$ and $\Lambda_0$. Instead, we focus on extracting just the beta functions. That is, we do an infinitesimal lowering of the cutoff $\Lambda=\Lambda_0-\var{\Lambda}$. 

Doing the usual split $\phi=\phi^++\phi^-$ we get

\begin{gather}
    S_{\Lambda_0}[\phi^++\phi^-]=S_{\Lambda_0}[\phi^-]+\nonumber\\
    +\int\dd[d]{x}\qty(\frac{1}{2}(\grad{\phi^+})^2+\phi^+V'(\phi^-)+\frac{1}{2}(\phi^+)^2V''(\phi^-)+\frac{1}{6}(\phi^+)^3 V^{(3)}(\phi^-)+\dots)
    \label{action2}
\end{gather} 

To calculate this it seems we would have to resort to Feynman diagrams again, which wouldn't be too much of a help. But now our second approximation kicks in. Since $\Lambda=\Lambda_0-\var{\Lambda}$, then each high energy loop will come with an integral in momentum similar to

\begin{equation}
    \int_{\Lambda_0-\var{\Lambda}}^{\Lambda_0}\frac{\dd[d]{k}}{(2\pi)^d}\dots\approx\var{\Lambda}\frac{\Lambda_0^{d-1}}{(2\pi^D)}\int_{S^{d-1}}\dd{\Omega}\dots
\end{equation}
where the last integral is over a unit $(d-1)$-sphere. 

This means each loop is of order $\var{\Lambda}$, then, in our approximation, we only need to consider linear order in $\var{\Lambda}$, i.e. we only need to consider 1-loop diagrams\footnote{You may be concerned by the fact that this argument does not seem to rule out tree diagrams. However, with a little bit more thought you can see that they also don't matter. We'll see shortly that, in the local potential approximation, the external momenta should be sent to zero, however, then there is no way any internal momenta can be high energy in a tree diagram, which means they also don't contribute}, which is already a massive simplification. But instead of jumping in and calculating those diagrams, we can do even more progress using some elementary graph theory.

Consider a particular diagram with a number of vertices, $v_i$, with $i$ powers of $\phi^+$ and arbitrary powers of $\phi^-$. Then, if that diagram has $e$ edges and $l$ loops, by Euler's identity, it obeys

\begin{equation}
    e-\sum_i v_i=l-1
\end{equation}

By our reasoning above we know that $l=1$. Further, since all edges must be high energy, and there can be no external $\phi^+$ lines, all edges must end in the the vertices, and then we have 

\begin{equation}
    2e=\sum_i iv_i
\end{equation}

Putting everything together gives

\begin{equation}
    \sum_i \frac{i-2}{2}v_i=0
\end{equation}

This means that the only diagrams that can contribute are those whose vertices have exactly 2 $\phi^+$ lines, therefore we can truncate (\ref{action2}) and just consider

\begin{equation}
    S_{\Lambda_0}[\phi^++\phi^-]=S_\Lambda[\phi^-]+\int\dd[d]{x}\qty(\frac{1}{2}(\grad{\phi^+})^2+\frac{1}{2}(\phi^+)^2V''(\phi^-))
\end{equation}

This is a Gaussian integral! However, $V''(\phi^-)$, which would play the role of mass, is spacetime dependent which complicates everything. It is time for our local potential approximation to come to the rescue. Because the whole point is not that we start with a local potential, that we can always do, there's no approximation there, the real approximation is to impose we end with a local potential. This truly is an approximation, that is, it isn't exactly true, it's just that those extra derivative interactions will be irrelevant for $d>2$, so we're not too far off.

But how do we implement that? Thinking in terms of diagrams again, we want to set all external momenta to zero to get the momentum independent piece. This is equivalent to treating $V''(\phi^-)$ as if it wasn't spacetime dependent. Then we have a usual Gaussian integral, which we know how to do. For convenience let's rewrite (\ref{gauss}).

\begin{equation}
    Z[\te{M}]=\int\dd[n]{\phi}\ee^{-\frac{1}{2}\dotp{\vb{\phi}}{\te{M}\vb{\phi}}}=\sqrt{\frac{(2\pi)^n}{\det \te{M}}}
\end{equation}
using the identity $\log(\det\te{M})=\tr(\log\te{M})$, we can write

\begin{equation}
    Z[\te{M}]\sim\exp(-\frac{1}{2}\tr(\log\te{M}))
\end{equation}

To apply this to our case, we go to momentum space (treating $V''(\phi^-)$ as a constant), then our analogue of $\te{M}$ is

\begin{equation}
    M(\vb{p},\vb{p}')=\delta^{(d)}(\vb{p}+\vb{p}')(\vb{p}^2+V''(\phi^-))
\end{equation}

The log is defined in terms of the eigenvalues of $M$, and the trace just means putting $\vb{p}=-\vb{p}'$ and integrating over $\vb{p}$. Therefore

\begin{equation}
    \tr(\log(M))=\int_{\Lambda_0-\var{\Lambda}}^{\Lambda_0}\frac{\dd[d]p}{(2\pi)^d}\delta^{(d)}(\vb{0})\log(\vb{p}^2+V''(\phi^-))
\end{equation}

Since $\var{\Lambda}\ll\Lambda_0$ the integral is easy to do

\begin{equation}
    \tr(\log(M))=\var{\Lambda}\Lambda_0^{d-1}\frac{\mathrm{Vol}(S^{d-1})}{(2\pi)^d}\delta^{(d)}(\vb{0})\log(\Lambda_0^2+V''(\phi^-))
\end{equation}

The $\delta^{(d)}(\vb{0})$ is also easy to interpret, it is a volume factor, or equivalently an integration over spacetime. In the end we have

\begin{equation}
    \tr(\log(M))=\var{\Lambda}\Lambda_0^{d-1}\frac{\mathrm{Vol}(S^{d-1})}{(2\pi)^d}\int\dd[d]{x}\log(\Lambda_0^2+V''(\phi^-))
\end{equation}

This means we have

\begin{equation}
    S_\Lambda[\phi^-]=\int\dd[d]{x}\qty(\frac{1}{2}(\grad{\phi^-})^2+V(\phi^-)+k\var{\Lambda}\Lambda_0^{d-1}\log(\Lambda_0^2+V''(\phi^-)))
\end{equation}
where
\begin{equation}
    k=\frac{\mathrm{Vol}(S^{d-1})}{2(2\pi)^d}=\frac{1}{(4\pi)^{\frac{d}{2}}\Gamma\qty(\frac{d}{2})}
\end{equation}
To get the beta functions we need to do steps 2 and 3, which give us

\begin{alignat}{2}
    S_\Lambda[\phi']&=\int\dd[d]{x'}\Bigg(&&\frac{1}{2}(\grad'\phi')^2+\qty(\frac{\Lambda_0}{\Lambda})^{d-n_\phi\frac{d-2}{2}}V(\phi')+\nonumber\\
    & &&+k\var{\Lambda}\Lambda_0^{d-1}\qty(\frac{\Lambda_0}{\Lambda})^{d-n_\phi'\frac{d-2}{2}}\log(\Lambda_0^2+V''(\phi^-))\Bigg)=\nonumber\\
    &=\int\dd[d]{x'}\Bigg(&&\frac{1}{2}(\grad'\phi')^2+V(\phi')+\var{\Lambda}\bigg(\qty(d-n_\phi\frac{d-2}{2})\Lambda_0^{-1}V(\phi')+\nonumber\\
    & &&+k\Lambda_0^{d-1}\log(\Lambda_0^2+V''(\phi'))\bigg)+O(\var{\Lambda}^2)\Bigg)
\end{alignat}
where $n_\phi$ counts the number of fields inside $V$, how do we do that? Just think of it as an operator

\begin{equation}
    n_\phi V(\phi)=\phi V'(\phi)
\end{equation}

Keeping in mind the minus sign in $\Lambda=\Lambda_0-\var{\Lambda}$, we have the following expression for the beta function of the potential

\begin{equation}
    \Lambda\dv{V(\phi)}{\Lambda}=-d\cdot V(\phi)+\frac{d-2}{2}\phi V'(\phi)-\Lambda^d\log(\Lambda^2+V''(\phi))
    \label{potential}
\end{equation}

We could also expand $V$ in a Taylor series and track the evolution of the individual couplings. We write

\begin{equation}
    V(\phi)=\sum_{n=1}^\infty\Lambda^{d-n(d-2)}\frac{g_{2n}}{(2n)!}\phi^{2n}
\end{equation}

By expressing $V$ in terms of dimensionless couplings as usual we should not use (\ref{potential}) directly since it encapsulates scaling already. We should instead use the expression just with step 1. Plugging things in you eventually reach the conclusion that

\begin{equation}
    \Lambda\dv{g_{2n}}{\Lambda}=(n(d-2)-d)g_{2n}-k\Lambda^{n(d-2)}\pdv[2n]{\phi}\eval{\log(\Lambda^2+V''(\phi))}_{\phi=0}
    \label{beta}
\end{equation}

Just for concreteness the first few terms in this expansion are (to confirm we haven't done anything too silly, it's also a straightforward exercise to find out which 1-loop diagrams would correspond to each term)

\begin{align}
    \Lambda\dv{g_2}{\Lambda}&=-2g_2-k\frac{g_4}{1+g_2}\nonumber\\
    \Lambda\dv{g_4}{\Lambda}&=(d-4)g_4-k\frac{g_6}{1+g_2}+3k\frac{g_4^2}{(1+g_2)^2}\label{series}\\
    \Lambda\dv{g_6}{\Lambda}&=(2d-6)g_6-k\frac{g_8}{1+g_2}+15k\frac{g_4g_6}{(1+g_2)^2}-30k\frac{g_4^3}{(1+g_2)^3}\nonumber
\end{align}

Before we move on, I must address the elephant in the room. At this point, you may be a bit uneasy. First of all, this whole procedure seems rather ill justified\footnote{Or, as we say in Portugal, \textit{javardo}}, and, to some extent, it is. If we're too far away from the Gaussian fixed point we really can't trust the mass dimensions to tell whether the couplings are relevant or irrelevant. Therefore, strictly speaking, if we have too strong coupling there may be more subtle effects that make the derivative interactions matter. Also, we made all this argument in terms of diagrams, so there may be non-perturbative effects kicking in that we will never capture with diagrams. All this to say this isn't very mathematically rigorous, and we can't really claim these beta functions are the whole story, or even that they are non-perturbative in nature. The strongest statement we can do about these beta functions is that if we start with a theory without derivative interactions the beta functions for the local interactions are given by (\ref{beta}) to all orders in perturbation theory. More geometrically, if you imagine, in theory space, the space of all theories that have no derivative interactions as a plane, we know the projection of the beta functions parallel to this plane at any point in that plane, modulo non-perturbative effects.

Secondly, this doesn't seem like a huge gain to perturbation theory, like, why didn't we just use 1-loop Feynman diagrams to compute (\ref{series}) once we new that just to compute beta functions we could restrict to 1-loop? Well, we did get every term in one go, which is nice. But that's not the real reason, there is hidden power in (\ref{potential}). This equation is really useful, both because it deals directly with the potential and not with the individual couplings, and because it is valid to all orders in perturbation theory, even away from weak coupling (with the caveats from the above paragraph). It is still very hard to solve, however, it is more amenable to plug into a computer. For example, this formulation of the flow has been used to search for fixed points directly in $d=3$, without resorting to the epsilon expansion, and indeed one finds an interacting fixed point whose properties agree precisely with the prediction of the epsilon expansion for the Wilson-Fisher fixed point which gives some faith to both approaches. 

The real conclusion is that, despite all its flaws, it is a take on the RG flow slightly different from standard perturbation theory. And the truth is, computing RG flows is bloody hard, there's no way around it. So, we grab everything we can get our hands on, to get different glimpses into the wilderness of theory space. In the next chapter we will collect a few more perspectives under our belt, more experimentally driven. And, historically, these perspectives actually came first, but they were clouded in misconceptions and confusion, and only with what we learned in this chapter about Wilson's view of the Renormalisation Group can we truly understand what is going on.

\newpage

\section{Counterterms and the continuum limit}

In the last chapters we have delved quite deeply into the role of scale in quantum field theory, how, to even define a theory, we need to introduce some scale into the game. We saw how our theories depend on this scale through the study of renormalisation group flows and even calculated explicitly these flows, in some simple cases, elucidating the structure of scalar field theory, and exemplifying important techniques. But we still have a lot of questions left to answer, and, historically, these were actually the first questions to be asked and answered. The structure we exposed previously took some more decades to fully unfold.

The first question is experimental. Yes, we have showed that the couplings in the Lagrangian change as we flow down to the IR. But is this just some theoretical slight of hand, or can we actually measure RG? And related to that, do we really need this scale or is it just because we're too dumb to write the theory without it? Finally, we shall close our lectures with a long promised proof, how does our theory depend on different choices of regularisation schemes? Are some schemes better than others, or is it all the same in the end?

\subsection{The flow of correlation functions}

We wish to connect with the real world of experiments. To do that, we need to connect RG flows to observable quantities. But what can we observe? Well, in QFT experiments, there are two things we can measure: cross sections and decay rates. These can be reduced to the calculation of scattering amplitudes (S-matrix elements), which, in turn, using the LSZ (Lehmann\footnote{Harry Lehmann: born in 1924 in Güstrow, Germany; died in 1998 in Hamburg, Germany; Doctoral Advisor: Friedrich Hund}-Symanzik\footnote{Kurt Symanzik: born in 1923 in Lyck, Germany; died in 1983 in Hamburg, Germany; Doctoral Advisor: Werner Heisenberg}-Zimmermann\footnote{Wolfhart Zimmermann: born in 1928 in Freiburg im Breisgau, Germany; died in 2016 in Munich, Germany; Doctoral Advisors: Emanuel Sperner, Wilhelm Süss}) reduction formula and Wick rotation, can be reduced to the calculation of Euclidean time correlation functions\footnote{Any QFT textbook will have this in detail}.

All this to say we'll consider correlation function as our observables, knowing there are standard methods to convert these into things experimentalists like. So let's calculate the flow of correlation functions! We'll begin with the simplest, the partition function. But before we jump into it, we need to discuss another thing first. What does the partition function depend on? Well, firstly, it cannot depend on the fields directly since we have integrated over them. What the partition function will depend on, is the many choices we had to make in defining our theory, namely, the various couplings (where mass is counted as a coupling), the spacetime manifold, and the field configuration space. In our case, we'll restrict the spacetime manifold to be Euclidean space with a flat metric. Further, the way we'll see the dependence on the field space is purely in terms of the cutoff (remember that the point of the cutoff is precisely to change which fields we're integrating over). Therefore we write

\begin{equation}
    Z\equiv Z(\Lambda_0;g_{0,a})
\end{equation}

What happens when we change the scale? By construction, nothing can happen. We're only changing the order in which we integrate our fields, starting with the high energy ones. This means

\begin{equation}
    Z(\Lambda;g_n(\Lambda))=Z(\Lambda_0;g_{0,a})
\end{equation}
or, infinitesimally,

\begin{equation}
    \Lambda\dv{Z(\Lambda,g_a(\Lambda))}{\Lambda}=\qty(\Lambda\pdv{\Lambda}+\beta_a\pdv{g_a})Z(\Lambda;g_a(\Lambda))=0
\end{equation}
where we substituted $\beta_a=\Lambda\pdv{g_a}{\Lambda}$.

This is the first example of a Callan\footnote{Curtis Gove Callan Jr.: born in 1942 in North Adams, USA; Doctoral Advisor: Sam Treiman}-Symanzik (CS) equation. It tells us that the couplings in the effective action change so as to precisely balance the change in the field configuration space (cutoff), and, in the end, leave the partition function unchanged.

Now, let's tackle correlation functions. Using the original theory with canonically normalised kinetic terms, and cutoff $\Lambda_0$, the $n$-point correlator is given by

\begin{align}
    \expval{\phi(\vb{x}_1)\dots\phi(\vb{x}_n)}&=\frac{1}{Z}\int\mathcal{D}\phi~\phi(\vb{x}_1)\dots\phi(\vb{x}_n)\ee^{-S_{\Lambda_0}[\phi]}=\nonumber\\
    &=\Gamma_{\Lambda_0}^{(n)}(\vb{x}_1,\dots,\vb{x}_n;g_{0,a})
    \label{lambda0}
\end{align}

If the fields are inserted at positions such that $\abs{\vb{x}_i-\vb{x}_j}\gg\frac{1}{\Lambda}$, then we should be able to use the low energy effective theory to compute this correlator. However, we must be careful with field strength renormalisation, i.e.

\begin{equation}
    \expval{\phi(\vb{x}_1)\dots\phi(\vb{x}_n)}=\frac{1}{Z}\int\mathcal{D}\phi'~\phi(\vb{x}_1)\dots\phi(\vb{x}_n)\ee^{-S_\Lambda[\phi']}
\end{equation}
where, as usual, $\phi'=\sqrt{Z_\phi}\phi^-$ and we're using dimensionless couplings to avoid spatial rescalings. Then

\begin{align}
    \expval{\phi(\vb{x}_1)\dots\phi(\vb{x}_n)}&=\frac{1}{Z}Z_\phi^{-\frac{n}{2}}(\Lambda)\int\mathcal{D}\phi'~\phi'(\vb{x}_1)\dots\phi'(\vb{x}_n)\ee^{-S_{\Lambda_0}[\phi']}=\nonumber\\
    &=Z_\phi^{-\frac{n}{2}}(\Lambda)\Gamma^{(n)}_\Lambda(\vb{x}_1,\dots,\vb{x}_n;g_a(\Lambda))
    \label{lambda}
\end{align}

Equating (\ref{lambda0}) and (\ref{lambda}) gives

\begin{equation}
    \Gamma_{\Lambda_0}^{(n)}(\vb{x}_1,\dots,\vb{x}_n;g_{0,a})=Z_\phi^{-\frac{n}{2}}(\Lambda)\Gamma^{(n)}_\Lambda(\vb{x}_1,\dots,\vb{x}_n;g_a(\Lambda))
\end{equation}
or, infinitesimally,

\begin{align}
    &\Lambda\dv{\Lambda}(Z_\phi^{-\frac{n}{2}}(\Lambda)\Gamma^{(n)}_\Lambda(\vb{x}_1,\dots,\vb{x}_n;g_a(\Lambda)))=0\Leftrightarrow\nonumber\\
    \Leftrightarrow&\qty(\Lambda\pdv{\Lambda}+\beta_a\pdv{g_a}+n\gamma_\phi)\Gamma^{(n)}_\Lambda(\vb{x}_1,\dots,\vb{x}_n;g_a(\Lambda))=0
\end{align}
where we've used the field anomalous dimension, $\gamma_\phi=-\frac{1}{2}\Lambda\pdv{\log Z_\phi(\Lambda)}{\Lambda}$.

This is another CS equation that tells us the couplings and field renormalisation precisely balance the changing scale to leave the observable correlation functions invariant. This makes perfect physical sense, since the theories are equivalent, any observable quantity must agree. It's more a matter of which version of the theory is more convenient to use. 

But hold on a minute! This seems to say RG is a theoretical trick. A procedure to go from one theory to another to choose the most useful one, or to gauge which theories should we bother writing, in which the actual measurable quantities are unchanged. This is completely true, however, somewhat paradoxically, the reason why RG isn't measurable will tell us how to measure RG.

Let's go back a moment and understand how RG isn't measurable. After all, we have calculated that the mass of the field should depend on the scale for example. Can't experimentalists measure masses? Yes, they can perfectly measure masses, they just look for peaks in their histograms. Or, in our language, they measure 2-point functions as a function of the momentum, and then call "mass" the value of the momentum for which the correlator diverges. However, that point doesn't have to be the thing we called $m$. We have been saying mass all along, but really it was just the coupling for the quadratic term in the Lagrangian. We haven't really checked if that $m$ was the actual, physical mass of the field.

Let's go back to theory space. In the last chapter, the point of view was that we have defined the theory at some scale $\Lambda_0$ by whatever means necessary. By defining a theory we specified the field configuration space (which will depend on $\Lambda_0$), the spacetime manifold, and the values of the couplings. Then, we used RG to get an equivalent effective description at some other sale $\Lambda$, which gave us some new values for couplings. This procedure tracks a trajectory in theory space. Now, obviously, we're free to parametrise this flow however we want. Instead of focusing on the UV theory given by divine intervention, we can focus on the IR observables, and use that to parametrise the flow. That is, instead of defining the couplings to compute amplitudes in terms of those couplings, we define the couplings in terms of some amplitudes. Then we can use those values to calculate more amplitudes, which could be, for example, the same amplitude but at a different energy scale.

\begin{center}
    \includegraphics[scale=0.7]{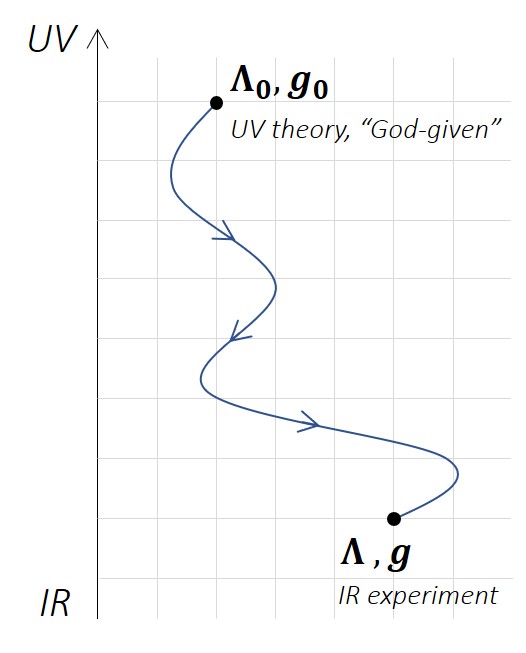}
\end{center}

There is a catch however, we cannot invert RG. Once we've integrated out fields there is no way to integrate them back in. Those modes are gone. This seems like a fatal flaw for high energy physics. Sure you can use experiments at high energies to then calculate low energy amplitudes but that seems very much like our original picture, just with a possible change of basis. And strictly speaking, this is all we can do if we assume nothing more. This is very much the reason why we keep building bigger and bigger accelerators, we don't know what's behind the corner at higher energies.

However, we can be clever about it. Let's assume this theory extends beyond the energies we have so far probed. This may not be true, but let's assume it is. What we can then do is run RG from that hypothetical high energy down to our low energy experiments. With this information, we can then tune the high energy coupling such that the low energy amplitude agrees with experiment. This seems like a silly idea. What have we gained from this? 
First of all, we'll see shortly that this is actually a very practical way to calculate RG flows, and to parametrise the flow in terms of observables. Further, this allows us to leave our comfort zone and ask the question "if this theory were valid up to higher energies, what would it look like?", then we can make predictions about high energy experiments, albeit predictions that can very well not be true, but it gives a guide as to what we're looking for. Finally, this point of view will lend itself quite naturally to the continuum limit. This high energy cutoff is almost begging to be removed at this stage. We'll purposefully delay that removal to the next section to emphasise the relevance and validity of these techniques even when we're not taking the continuum limit.

Now it's time to do some calculations. Although these techniques will be valid in more general circumstances, since we're taking the high energy physics point of view, which is more concerned with 4 dimensions, let's consider 4-dim $\phi^4$ theory. Also, this is the theory where perturbation theory is more reliable, which simplifies our discussion.

We could just use the results from Chapter \ref{wilsonrg} and try to find the correct $m_0$ and $\lambda_0$ such that it reproduces some given amplitude by giving it some $\Lambda_0$ dependence. Although there's nothing wrong with that approach there is a simpler way to phrase the calculations. We split the action into two pieces, usually called the \textit{bare} action, and the \textit{counterterm} action.

\begin{equation}
    S_{\Lambda_0}[\phi]=S_\text{B}[\phi]+S_\text{CT}[\phi]
\end{equation}
where
\begin{align}
    S_\text{B}[\phi]&=\int\dd[4]{x}\qty(\frac{1}{2}(\grad{\phi})^2+\frac{1}{2}m_0^2\phi^2+\frac{\lambda_0}{4!}\phi^4)\\
    S_\text{CT}[\phi]&=\int\dd[4]{x}\qty(\frac{1}{2}\var{Z_\phi}(\grad{\phi})^2+\frac{1}{2}\var{m^2}\phi^2+\frac{\var{\lambda}}{4!}\phi^4)
\end{align}
and we assume that all dependence on the cutoff is in the counterterm action. We shall also assume that each counterterm contribution in a diagram counts as a loop. That is, tree-level contributions from the counterterms come at the same order as 1-loop contributions from the bare action. This may seem odd but there's no physics in this, it is purely a computational choice. After all, we're going to tune the counterterms anyway so they'll come at the order that we wish. You'll see in practice how this help.

There are many different ways we could fix the counterterms and the bare couplings depending on the context we're doing the calculations. These choices are called \textit{renormalisation schemes}. The particular scheme we'll choose in this chapter is called the \textit{on-shell} scheme since it's the one closer to experiment. But it's not the only valid choice, for other contexts there can be more convenient schemes.

We shall now go through both the 2- and 4-point functions in this theory. Both of them will have different annoying subtleties that we'll have to address, but, hopefully, in this simpler theory, we won't have muddled the physics too much. Just note a few details which are common to both. Firstly, we'll be calculating full correlation functions in the high energy theory, we're not integrating out modes, so our propagators will have momentum that goes all the way from 0 to $\Lambda_0$. Secondly, we'll consider only \textit{connected} correlation functions, the reason for this is that the disconnected pieces are fairly trivial. Physically, they correspond to when particles don't "see" each other, it's once again part of the standard bag of techniques to go from connected correlators to physical amplitudes. With those points in mind let's do the calculations.

\subsubsection{2-Point Function}

The irritating technicality with this correlator is the fact that the propagator has the mass in the denominator instead of the numerator. Therefore we need to do some trickery. In general, the diagrams that contribute to the 2-point function, even beyond leading order are (excluding counterterms for the time being)

\begin{equation}
\begin{tikzpicture}[baseline=(b.south)]
  \begin{feynman}
    \vertex (a);
    \vertex [right=1.5cm of a] (b);
    
    \diagram* {
      (a) -- [plain,out=0,in=180] (b),
    };
  \end{feynman}
\end{tikzpicture}
+
\begin{tikzpicture}[baseline=(b.south)]
  \begin{feynman}
    \vertex (a);
    \vertex [right=1cm of a] (c);
    \vertex [right=1cm of c] (b);
    \vertex [above=1cm of c] (d);
    
    \diagram* {
      (a) -- [plain,out=0,in=180] (c),
      (c) -- [plain,out=0,in=180] (b),
      (c) -- [plain,out=135,in=180] (d),
      (d) -- [plain,out=0,in=45] (c),
    };
    \draw[fill=black] (c) circle (1.5pt);
  \end{feynman}
\end{tikzpicture}
+
\begin{tikzpicture}[baseline=(b.south)]
  \begin{feynman}
    \vertex (a);
    \vertex [right=0.5cm of a] (c);
    \vertex [right=0.75cm of c] (d);
    \vertex [right=0.5cm of d] (b);
    \vertex [above=1cm of c] (e);
    \vertex [above=1cm of d] (f);
    
    \diagram* {
      (a) -- [plain,out=0,in=180] (c),
      (c) -- [plain,out=0,in=180] (d),
      (c) -- [plain,out=0,in=180] (b),
      (c) -- [plain,out=135,in=180] (e),
      (e) -- [plain,out=0,in=45] (c),
      (d) -- [plain,out=135,in=180] (f),
      (f) -- [plain,out=0,in=45] (d),
    };
    \draw[fill=black] (c) circle (1.5pt);
    \draw[fill=black] (d) circle (1.5pt);
  \end{feynman}
\end{tikzpicture}
+
\begin{tikzpicture}[baseline=(b.south)]
  \begin{feynman}
    \vertex (a);
    \vertex [right=1cm of a] (c);
    \vertex [right=1cm of c] (b);
    \vertex [above=1cm of c] (d);
    \vertex [above=1cm of d] (e);
    
    \diagram* {
      (a) -- [plain,out=0,in=180] (c),
      (c) -- [plain,out=0,in=180] (b),
      (c) -- [plain,out=135,in=-135] (d),
      (d) -- [plain,out=-45,in=45] (c),
      (d) -- [plain,out=135,in=180] (e),
      (e) -- [plain,out=0,in=45] (d),
    };
    \draw[fill=black] (c) circle (1.5pt);
    \draw[fill=black] (d) circle (1.5pt);
  \end{feynman}
\end{tikzpicture}
+
\begin{tikzpicture}[baseline=(b.base)]
  \begin{feynman}
    \vertex (a);
    \vertex [right=0.5cm of a] (c);
    \vertex [right=1cm of c] (d);
    \vertex [right=0.5cm of d] (b);
    \vertex [right=0.5cm of c] (e);
    \vertex [above=0.5cm of e] (f);
    \vertex [below=0.5cm of e] (g);
    
    \diagram* {
      (a) -- [plain,out=0,in=180] (c),
      (c) -- [plain,out=0,in=180] (d),
      (d) -- [plain,out=0,in=180] (b),
      (c) -- [plain,out=90,in=180] (f),
      (f) -- [plain,out=0,in=90] (d),
      (c) -- [plain,out=-90,in=180] (g),
      (g) -- [plain,out=0,in=-90] (d),
    };
    \draw[fill=black] (c) circle (1.5pt);
    \draw[fill=black] (d) circle (1.5pt);
  \end{feynman}
\end{tikzpicture}
+\ldots
\end{equation}

We see that some of the diagrams are basically copies of each other. For example, the third term looks like two copies of the second. This suggests we should be able to only consider "distinct" diagrams. We say that a diagram is one-particle irreducible (1PI) if it has no bridges, where a bridge is a line that once removed makes the diagram disconnected. For example, the second is 1PI but the third isn't, because, if we remove the middle propagator, we get two disconnected diagrams (two copies of the second term). We can then write a series for the 1PI diagrams with two external legs:

\begin{equation}
\begin{tikzpicture}[baseline=(b.base)]
  \begin{feynman}
    \vertex (a);
    \vertex [blob,right=1cm of a] (c) {1PI};
    \vertex [right=1.25cm of c] (b);

    \diagram* {
      (a) -- [plain,out=0,in=180] (c),
      (c) -- [plain,out=0,in=180] (b),
    };
  \end{feynman}
\end{tikzpicture}
=
\begin{tikzpicture}[baseline=(b.south)]
  \begin{feynman}
    \vertex (a);
    \vertex [right=1cm of a] (c);
    \vertex [right=1cm of c] (b);
    \vertex [above=1cm of c] (d);
    
    \diagram* {
      (a) -- [plain,out=0,in=180] (c),
      (c) -- [plain,out=0,in=180] (b),
      (c) -- [plain,out=135,in=180] (d),
      (d) -- [plain,out=0,in=45] (c),
    };
    \draw[fill=black] (c) circle (1.5pt);
  \end{feynman}
\end{tikzpicture}
+
\begin{tikzpicture}[baseline=(b.south)]
  \begin{feynman}
    \vertex (a);
    \vertex [right=1cm of a] (c);
    \vertex [right=1cm of c] (b);
    \vertex [above=1cm of c] (d);
    \vertex [above=1cm of d] (e);
    
    \diagram* {
      (a) -- [plain,out=0,in=180] (c),
      (c) -- [plain,out=0,in=180] (b),
      (c) -- [plain,out=135,in=-135] (d),
      (d) -- [plain,out=-45,in=45] (c),
      (d) -- [plain,out=135,in=180] (e),
      (e) -- [plain,out=0,in=45] (d),
    };
    \draw[fill=black] (c) circle (1.5pt);
    \draw[fill=black] (d) circle (1.5pt);
  \end{feynman}
\end{tikzpicture}
+
\begin{tikzpicture}[baseline=(b.base)]
  \begin{feynman}
    \vertex (a);
    \vertex [right=0.5cm of a] (c);
    \vertex [right=1cm of c] (d);
    \vertex [right=0.5cm of d] (b);
    \vertex [right=0.5cm of c] (e);
    \vertex [above=0.5cm of e] (f);
    \vertex [below=0.5cm of e] (g);
    
    \diagram* {
      (a) -- [plain,out=0,in=180] (c),
      (c) -- [plain,out=0,in=180] (d),
      (d) -- [plain,out=0,in=180] (b),
      (c) -- [plain,out=90,in=180] (f),
      (f) -- [plain,out=0,in=90] (d),
      (c) -- [plain,out=-90,in=180] (g),
      (g) -- [plain,out=0,in=-90] (d),
    };
    \draw[fill=black] (c) circle (1.5pt);
    \draw[fill=black] (d) circle (1.5pt);
  \end{feynman}
\end{tikzpicture}
+\ldots
\end{equation}

We'll call this expansion $\Pi(\vb{p}^2)$ ($\vb{p}$ is the external momenta, and we have amputated the two external propagators), this is often called the \textit{self-energy} of the field. Using this notation the propagator is then:

\begin{equation*}
\begin{tikzpicture}[baseline=(b.south)]
  \begin{feynman}
    \vertex (a);
    \vertex [right=2cm of a] (b);
    
    \diagram* {
      (a) -- [plain,out=0,in=180] (b),
    };
  \end{feynman}
\end{tikzpicture}
+
\begin{tikzpicture}[baseline=(b.base)]
  \begin{feynman}
    \vertex (a);
    \vertex [blob,right=1cm of a] (c) {1PI};
    \vertex [right=1.25cm of c] (b);

    \diagram* {
      (a) -- [plain,out=0,in=180] (c),
      (c) -- [plain,out=0,in=180] (b),
    };
  \end{feynman}
\end{tikzpicture}
+
\begin{tikzpicture}[baseline=(b.base)]
  \begin{feynman}
    \vertex (a);
    \vertex [blob,right=0.5cm of a] (c) {1PI};
    \vertex [blob,right=1.25cm of c] (d) {1PI};
    \vertex [right=1cm of d] (b);

    \diagram* {
      (a) -- [plain,out=0,in=180] (c),
      (c) -- [plain,out=0,in=180] (d),
      (d) -- [plain,out=0,in=180] (b),
    };
  \end{feynman}
\end{tikzpicture}
+\ldots=
\end{equation*}
\vspace{-4mm}
\begin{align}
    &=\frac{1}{\vb{p}^2+m_0^2}+\frac{1}{\vb{p}^2+m_0^2}\Pi(\vb{p}^2)\frac{1}{\vb{p}^2+m_0^2}+\frac{1}{\vb{p}^2+m_0^2}\Pi(\vb{p}^2)\frac{1}{\vb{p}^2+m_0^2}\Pi(\vb{p}^2)\frac{1}{\vb{p}^2+m^2}+\dots\nonumber\\
    &=\frac{1}{\vb{p}^2+m_0^2-\Pi(\vb{p}^2)}
\end{align}
where in going to the last line we summed the geometric series.

Now all we're left to do is calculate $\Pi(\vb{p}^2)$. To 1-loop order it is given by (including counterterms)

\begin{equation*}
\Pi(\vb{p}^2)=
\begin{tikzpicture}[baseline=(b.south)]
  \begin{feynman}
    \vertex (a);
    \vertex [right=1cm of a] (c);
    \vertex [right=1cm of c] (b);
    \vertex [above=1cm of c] (d);
    
    \diagram* {
      (a) -- [plain,out=0,in=180] (c),
      (c) -- [plain,out=0,in=180] (b),
      (c) -- [plain,out=135,in=180] (d),
      (d) -- [plain,out=0,in=45] (c),
    };
    \draw[fill=black] (c) circle (1.5pt);
  \end{feynman}
\end{tikzpicture}
+
\begin{tikzpicture}[baseline=(b.south)]
  \begin{feynman}
    \vertex (a);
    \vertex [crossed dot,right=0.75cm of a] (c) {};
    \vertex [right=1cm of c] (b);
    \vertex [above=0.45cm of c] (d) {$- \vb{p}^2\var{Z_\phi}$};
    
    \diagram* {
      (a) -- [plain,out=0,in=180] (c),
      (c) -- [plain,out=0,in=180] (b),
    };
  \end{feynman}
\end{tikzpicture}
+
\begin{tikzpicture}[baseline=(b.south)]
  \begin{feynman}
    \vertex (a);
    \vertex [crossed dot,right=0.75cm of a] (c) {};
    \vertex [right=1cm of c] (b);
    \vertex [above=0.45cm of c] (d) {$-\var{m}^2$};
    
    \diagram* {
      (a) -- [plain,out=0,in=180] (c),
      (c) -- [plain,out=0,in=180] (b),
    };
  \end{feynman}
\end{tikzpicture}
=
\end{equation*}

\begin{equation}
    =-\frac{\lambda_0}{32\pi^2}\qty(\Lambda_0^2-m_0^2\log(1+\frac{\Lambda_0^2}{m_0^2}))- \vb{p}^2\var{Z_\phi}-\var{m}^2
\end{equation}
where we've used the results from (\ref{1-loopmass}), setting $\Lambda$ to 0.

It is finally time to go to experiment to fix $\var{Z_\phi}$ and $\var{m^2}$. Experimentally, there is a simple pole when the external momentum is $\vb{p}^2=-m_\text{phys}^2$, further, we require that pole to have unit residue to fix the field strength renormalisation. But wait a moment! We're in Euclidean signature, $\vb{p}^2$ is positive, so it doesn't seem very reliable to set it equal to \textit{minus} the square of a real number. But there's an easy solution to that, you just Wick rotate back to Lorentzian\footnote{Hendrik Antoon Lorentz: born in 1853 in Arnhem, Netherlands; died in 1928 in Haarlem, Netherlands; Doctoral Advisor: Pieter Rijke; Nobel Prize in Physics 1902 together with Pieter Zeeman "in recognition of the extraordinary service they rendered by their researches into the influence of magnetism upon radiation phenomena"} signature, which, if you use the $(-,+,+,+)$ sign convention\footnote{Also known as the correct sign convention.}, means you have to do absolutely nothing! The expression looks exactly the same, except all vectors are now vectors in Minkowski\footnote{Hermann Minkowski: born in 1864 in Aleksotas, Russian Empire; died in 1909 in Göttingen, German Empire; Doctoral Advisor: Ferdinand von Lindemann} space, which just means they'll loose the boldface. And it makes perfect sense that in order to compare to experiment we'd need to go back to the physical Lorentzian world.

In the end, in terms of the self-energy, the two necessary requirements are

\begin{align}
    m_0^2-\Pi(-m_\text{phys}^2)=m_\text{phys}^2\\
    \eval{\pdv{\Pi(p^2)}{p^2}}_{p^2=-m_\text{phys}^2}=0
\end{align}

Plugging things in, we can satisfy these requirements by choosing,

\begin{align}
    \var{Z_\phi}&=0\\
    \var{m^2}&=m_\text{phys}^2-m_0^2-\frac{\lambda_0}{32\pi^2}\qty(\Lambda_0^2-m_0^2\log(1+\frac{\Lambda_0^2}{m_0^2}))
\end{align}
we can further choose $m_0^2=m_\text{phys}^2=m^2$ to get the requirement $\Pi(-m^2)=0$ and

\begin{equation}
    \var{m^2}=-\frac{\lambda_0}{32\pi^2}\qty(\Lambda_0^2-m^2\log(1+\frac{\Lambda_0^2}{m^2}))
\end{equation}

In this case the counterterm is tuned to exactly cancel the 1-loop contributions, leaving a fairly trivial result. This is an accident of the simplicity of this theory. Namely in the fact that the 1-loop diagram did not depend on the external momentum. Generically, we would expect a non-trivial field strength renormalisation, and that the counterterm would only cancel the loop contributions at that particular energy.

Note, however, than we can still extract the mass beta function from this calculation. The reason being that the quadratic coupling is now $m^2+\var{m^2}$, and we have fully specified its dependence on the cutoff such that it yields the physical mass $m^2$. This physical mass is just a result of some measurement, therefore it can't depend on the cutoff $\Lambda_0$, all cutoff dependence is explicit. More concretely, we have, to leading order in $\lambda_0$,

\begin{equation}
    \beta_2=\Lambda_0\dv{g_2}{\Lambda_0}=\Lambda_0\dv{\Lambda_0}[\Lambda_0^{-2}(m^2+\var{m^2})]=-2g_2-\frac{g_4}{16\pi^2}\frac{1}{1+g_2}
\end{equation}
where we take $\dv{m^2}{\Lambda_0}=0$ and $g_4=\lambda_0$.

As a final comment, notice that this calculation hinged on setting $p^2=-m^2$, hence if $m^2$ is higher than the cutoff energy then this step makes no sense, since we don't trust the amplitude up to those energies. In this case, physically, we simply haven't probed high enough energies to be able to measure the mass. And, looking at the propagator, notice that the $k^2$ term is negligible in comparison with the $m^2$ term, if $m^2$ is above the cutoff, which means that the field essentially can't propagate. This is the reasoning behind the common practice of integrating out fields whose masses are well above the energies we can probe. Even though this may seem like a somewhat brutal approximation, it has some merits. For example, by integrating out the W and Z bosons we can go from the standard model to Fermi's theory of weak interactions, which is a very useful approximation and was very important historically.

\subsubsection{4-Point Function}

The diagrams that contribute to this correlator are (up to 1-loop order)

\begin{equation*}
\begin{tikzpicture}[baseline=(e)]
  \begin{feynman}
    \vertex (a) {1};
    \vertex [below=1.5cm of a] (b) {2};
    \vertex [right=1.5cm of a] (c) {4};
    \vertex [right=1.5cm of b] (d) {3};
    \vertex [below=0.75cm of a] (f);
    \vertex [right=0.75cm of f] (e);
    
    \diagram* {
      (a) -- [plain,out=-45,in=135] (e),
      (b) -- [plain,out=45,in=-135] (e),
      (e) -- [plain,out=45,in=-135] (c),
      (e) -- [plain,out=-45,in=135] (d),
    };
  \end{feynman}
  \draw[fill=black] (e) circle (1.5pt);
\end{tikzpicture}
+
\begin{tikzpicture}[baseline=(e)]
  \begin{feynman}
    \vertex (a) {1};
    \vertex [below=1.5cm of a] (b) {2};
    \vertex [right=1.5cm of a] (c) {4};
    \vertex [right=1.5cm of b] (d) {3};
    \vertex [below=0.75cm of a] (f);
    \vertex [crossed dot,right=0.6cm of f] (e) {};
    
    \diagram* {
      (a) -- [plain,out=-45,in=135] (e),
      (b) -- [plain,out=45,in=-135] (e),
      (e) -- [plain,out=45,in=-135] (c),
      (e) -- [plain,out=-45,in=135] (d),
    };
  \end{feynman}
\end{tikzpicture}
+
\begin{tikzpicture}[baseline=(e)]
  \begin{feynman}
    \vertex (a) {1};
    \vertex [below=1.5cm of a] (b) {2};
    \vertex [below=0.75cm of a] (f);
    \vertex [right=2.5cm of a] (c) {4};
    \vertex [right=2.5cm of b] (d) {3};
    \vertex [right=0.75cm of f] (e);
    \vertex [right=1cm of e] (e2);
    \vertex [right=0.5cm of e] (g);
    \vertex [above=0.5cm of g] (g1);
    \vertex [below=0.5cm of g] (g2);
    
    \diagram* {
      (a) -- [plain,out=-45,in=135] (e),
      (b) -- [plain,out=45,in=-135] (e),
      (e) -- [plain,out=90,in=180] (g1),
      (e) -- [plain,out=-90,in=180,rmomentum'={[arrow shorten=0.25]$\vb{k}$}] (g2),
      (g1) -- [plain,out=0,in=90] (e2),
      (g2) -- [plain,out=0,in=-90] (e2),
      (e2) -- [plain,out=45,in=-135] (c),
      (e2) -- [plain,out=-45,in=135] (d),
    };
  \end{feynman}
  \draw[fill=black] (e) circle (1.5pt);
  \draw[fill=black] (e2) circle (1.5pt);
\end{tikzpicture}
+
\begin{tikzpicture}[baseline=(h)]
  \begin{feynman}
    \vertex (a) {1};
    \vertex [below=2.5cm of a] (b) {2};
    \vertex [right=1.5cm of a] (c) {4};
    \vertex [right=1.5cm of b] (d) {3};
    \vertex [right=0.75cm of a] (f);
    \vertex [below=0.75cm of f] (e);
    \vertex [below=1cm of e] (e2);
    \vertex [below=0.5cm of e] (h);
    \vertex [left=0.5cm of h] (h1);
    \vertex [right=0.5cm of h] (h2);
    
    \diagram* {
      (a) -- [plain,out=-45,in=135] (e),
      (b) -- [plain,out=45,in=-135] (e2),
      (e) -- [plain,out=45,in=-135] (c),
      (e2) -- [plain,out=-45,in=135] (d),
      (e) -- [plain,out=180,in=90,rmomentum'={[arrow shorten=0.25]$\vb{k}$}] (h1),
      (h1) -- [plain,out=-90,in=180] (e2),
      (e) -- [plain,out=0,in=90] (h2),
      (h2) -- [plain,out=-90,in=0] (e2),
    };
  \end{feynman}
  \draw[fill=black] (e) circle (1.5pt);
  \draw[fill=black] (e2) circle (1.5pt);
\end{tikzpicture}
+
\begin{tikzpicture}[baseline=(e)]
  \begin{feynman}
    \vertex (a) {1};
    \vertex [below=1.5cm of a] (b) {2};
    \vertex [below=0.75cm of a] (f);
    \vertex [right=2.5cm of a] (c) {4};
    \vertex [right=2.5cm of b] (d) {3};
    \vertex [right=0.75cm of f] (e);
    \vertex [right=1cm of e] (e2);
    \vertex [right=0.5cm of e] (g);
    \vertex [above=0.5cm of g] (g1);
    \vertex [below=0.5cm of g] (g2);
    
    \diagram* {
      (a) -- [plain,out=-45,in=135] (e),
      (b) -- [plain,out=-90,in=-90] (e2),
      (e) -- [plain,out=90,in=180] (g1),
      (e) -- [plain,out=-90,in=180,rmomentum'={[arrow shorten=0.25]$\vb{k}$}] (g2),
      (g1) -- [plain,out=0,in=90] (e2),
      (g2) -- [plain,out=0,in=-90] (e2),
      (e2) -- [plain,out=45,in=-135] (c),
      (e) -- [plain,out=-45,in=-90] (d),
    };
  \end{feynman}
  \draw[fill=black] (e) circle (1.5pt);
  \draw[fill=black] (e2) circle (1.5pt);
\end{tikzpicture}
=
\end{equation*}

\begin{align}
    =-\lambda_0-\var{\lambda}+\frac{\lambda_0^2}{2}\int\frac{\dd[4]{k}}{(2\pi)^4}\bigg(&\frac{1}{\vb{k}^2+m^2}\frac{1}{(\vb{k}+\vb{p}_1+\vb{p}_2)^2+m^2}+\nonumber\\
    +&\frac{1}{\vb{k}^2+m^2}\frac{1}{(\vb{k}+\vb{p}_1+\vb{p}_4)^2+m^2}+\nonumber\\
    +&\frac{1}{\vb{k}^2+m^2}\frac{1}{(\vb{k}+\vb{p}_1+\vb{p}_3)^2+m^2}\bigg)
\end{align}
where we have used the results from the previous section to use the physical mass $m$.

Now we have some dependence on the external momenta, which makes this correlator much more interesting than the previous one. We will at last understand how to measure RG. But first we need to evaluate this integral, keeping the dependence on the external momentum. To do that, we'll need the following trick due to Feynman (which you can check by explicit computation of the RHS):

\begin{equation}
    \frac{1}{AB}=\int_0^1\dd{x}\frac{1}{(xA+(1-x)B)^2}
\end{equation}
which allows us to combine two propagators, e.g. (where we define $\vb{p}_{12}=\vb{p}_1+\vb{p}_2$)

\begin{align}
    \frac{1}{\vb{k}^2+m^2}\frac{1}{(\vb{k}+\vb{p}_{12})^2+m^2}&=\int_0^1\dd{x}\frac{1}{\qty[x\qty(\qty(\vb{k}+\vb{p}_{12})^2+m^2)+(1-x)\qty(\vb{k}^2+m^2)]^2}=\nonumber\\
    &=\int_0^1\dd{x}\frac{1}{\qty[\vb{k}^2+m^2+x\qty(2\dotp{\vb{k}}{\vb{p}_{12}}+\vb{p}_{12}^2)]^2}=\nonumber\\
    &=\int_0^1\dd{x}\frac{1}{\qty[(\vb{k}+x\vb{p}_{12})^2+m^2+x(1-x)\vb{p}_{12}^2]^2}
\end{align}

We then define $\vb{l}=\vb{k}+x\vb{p}_{12}$ to write

\begin{align}
    &\int\frac{\dd[4]{k}}{(2\pi)^4}\frac{1}{\vb{k}^2+m^2}\frac{1}{(\vb{k}+\vb{p}_1+\vb{p}_2)^2+m^2}=\nonumber\\
    =&\int_0^1\dd{x}\int\frac{\dd[4]l}{(2\pi)^4}\frac{1}{\qty[\vb{l}^2+m^2+x(1-x)\vb{p}_{12}^2]^2}
\end{align}

In rigour, we should integrate over $\qty{\vb{k}|~\abs{\vb{k}}<\Lambda_0\land\abs{\vb{k}+\vb{p}_{12}}<\Lambda_0}$, which would be a very complicated $x$ dependent region in terms of $\vb{l}$. However, for $\abs{\vb{p}_{12}}\ll\Lambda_0$ these two regions are almost identical. And anyway, we can only really trust this theory if we probe it at energies much lower than the cutoff. Hence, we will approximate and integrate $\vb{l}$ over the region $\abs{\vb{l}}<\Lambda_0$. This is mainly for theoretical convenience, in practice you could always to the full integral numerically.

Therefore we write

\begin{align}
    &\int_0^1\dd{x}\int^{\Lambda_0}\frac{\dd[4]l}{(2\pi)^4}\frac{1}{\qty[\vb{l}^2+m^2+x(1-x)\vb{p}_{12}^2]^2}=\nonumber\\
    =&\int_0^1\dd{x}\frac{\mathrm{Vol}(S^3)}{(2\pi)^4}\int_0^{\Lambda_0}\dd{l}\frac{l^3}{\qty[l^2+m^2+x(1-x)\vb{p}_{12}^2]^2}=\nonumber\\
    =&\int_0^1\dd{x}\frac{1}{16\pi^2}\qty(\log(\frac{\Lambda_0^2+m^2+x(1-x)\vb{p}_{12}^2}{m^2+x(1-x)\vb{p}_{12}^2})+\frac{m^2+x(1-x)\vb{p}_{12}^2}{\Lambda_0^2+m^2+x(1-x)\vb{p}_{12}^2}-1)
\end{align}

Doing similar manipulations with the other three integrals, and keeping only the terms that don't vanish in the limit $\Lambda_0\to\infty$, since those are the only ones we have reliably computed in our approximation, we get for the full 4 point amplitude:

\begin{align}
    \mathcal{A}(s,t,u)=-\lambda_0-\var{\lambda}+\frac{\lambda_0^2}{32\pi^2}\int_0^1\dd{x}\bigg(&\log(\frac{\Lambda_0^2}{m^2-x(1-x)s})+\log(\frac{\Lambda_0^2}{m^2-x(1-x)t})+\nonumber\\
    &+\log(\frac{\Lambda_0^2}{m^2-x(1-x)u})-3\bigg)
\end{align}
where we've Wick rotated to Lorentzian signature, and used the Mandelstam\footnote{Stanley Mandelstam: born in 1928 in Johannesburg, South Africa; died in 2016 in Berkeley, USA; Doctoral Advisor: Paul Taunton Matthews} variables

\begin{equation}
    s=-p_{12}^2,~~~~~t=-p_{14}^2,~~~~~u=-p_{13}^2
\end{equation}
which, due to Lorentz invariance, completely specify four particle scattering kinematics.

Now, imagine we performed some experiment with kinematics $s_0,~t_0,~u_0$, we can then define the physical coupling at that energy scale to be

\begin{equation}
    \lambda_\text{phys}=-\mathcal{A}(s_0,t_0,u_0)
\end{equation}
which means

\begin{align}
    \var{\lambda}=\lambda_\text{phys}-\lambda_0+\frac{\lambda_0^2}{32\pi^2}\int_0^1\dd{x}\bigg(&\log(\frac{\Lambda_0^2}{m^2-x(1-x)s_0})+\log(\frac{\Lambda_0^2}{m^2-x(1-x)t_0})+\nonumber\\
    &+\log(\frac{\Lambda_0^2}{m^2-x(1-x)u_0})-3\bigg)
\end{align}

Once again we can choose $\lambda_\text{phys}=\lambda_0=\lambda$ for simplicity\footnote{In fact, even though they're conceptually different quantities, in general we can always make this choice of putting the physical coupling as the parameter in the Lagrangian. This choice is sometimes known as \textit{renormalised perturbation theory}, whereas keeping $\lambda_0$ unfixed is known as \textit{bare perturbation theory}. Both names are completely uninformative}. For theoretical purposes it is often convenient to set $s_0=t_0=u_0=0$, even though this point isn't actually experimentally accessible. In this case we are essentially parametrising the RG flow in terms of the asymptotic value of the coupling infinitely far into the IR. 

As we did for the quadratic coupling we can calculate the beta-function.

\begin{equation}
    \beta_4=\Lambda_0\dv{\Lambda_0}(\lambda+\var{\lambda})=\frac{3g_4^2}{16\pi^2}
    \label{beta4}
\end{equation}
where we substituted $g_4=\lambda$.

In this case we don't get the exact same result as we did before, this is due to our approximate way of calculating the integral, that is, we assumed that $\Lambda_0$ was much bigger than any other energy scale, which, given our caveats in the previous section, also means $g_2\ll1$, and taking the $g_2\ll1$ limit on the beta function from Chapter \ref{wilsonrg} gives us (\ref{beta4}). In this limit we still recover the essential physics, namely that this beta function is positive, and therefore this is a marginally irrelevant operator.

Note that this beta function does not depend on the particular point $s_0,~t_0,~u_0$ used to define the coupling. This makes perfect physical sense, we're asking about how the high energy theory coupling depends on the cutoff in order to yield the same IR physics, it shouldn't depend on where exactly are we fixing the IR coupling.

So far this is looking very similar to the 2-point function story, but now comes the cool part. The full amplitude, after plugging in the chosen counterterm is:

\begin{align}
    \mathcal{A}(s,t,u)=-\lambda+\frac{\lambda^2}{32\pi^2}\int_0^1\dd{x}\bigg(&\log(\frac{m^2-x(1-x)s_0}{m^2-x(1-x)s})+\log(\frac{m^2-x(1-x)t_0}{m^2-x(1-x)t})+\nonumber\\
    &+\log(\frac{m^2-x(1-x)u_0}{m^2-x(1-x)u})\bigg)
\end{align}

There are several things we can note about this amplitude. Firstly, we can confidently give this to an experimentalist. Everything is perfectly well defined in terms of experimentally measurable (i.e. physical) quantities. Secondly, the coupling $\lambda$ will depend on the precise energies $(s_0,t_0,u_0)$ we used to define it, however, the amplitude itself can't change, it's a physically measurable quantity, we're just parametrising it differently. This means we can write a kind of Callan-Symanzik equation, but now with respect to the low energy scales. For simplicity, let's set $s_0=t_0=u_0=\mu^2$ (this is not necessary but it makes the expressions neater), then we can write

\begin{equation}
    \mu\dv{\mathcal{A}}{\mu}=\qty(\mu\pdv{\mu}+\mu\pdv{\lambda}{\mu}\pdv{\lambda})\mathcal{A}=0
    \label{CSIR}
\end{equation}
where we can define a sort of experimental/low energy beta function as

\begin{equation}
    \beta^\text{exp}_\lambda=\mu\pdv{\lambda}{\mu}
\end{equation}

Using (\ref{CSIR}) we can calculate this (to leading order in $\lambda$),

\begin{equation}
    \beta^\text{exp}_\lambda=\frac{3\lambda^2}{16\pi^2}\int_0^1\dd{x}\frac{-x(1-x)\mu^2}{m^2-x(1-x)\mu^2}\approx\frac{3\lambda^2}{16\pi^2}
\end{equation}
where in the last step we approximated $\mu^2\gg m^2$ which is true if we're probing energies much higher than the mass of the particle.

We have now computed two beta functions which look rather similar both in shape and in interpretation but they are telling us different things. First, their similarities arise because they both obey a CS equation, and, due to dimensional analysis, there are only so many ways you can play with a dimensionful parameter. They also have very similar interpretation, they are both capturing the variation of the coupling such that the physics stays the same. These similarities are a big source of confusion in the literature, but they are in fact different concepts. So let us delve into it.

The low energy beta function is, in some sense, the more physical one. This one captures the dependence of the amplitude with the energy, which means it can be measured. And in fact, for theories which correspond to physical models such as QCD or QED, these have been measured, verifying the theoretical predictions to an astonishing accuracy! For example, the electron charge (or equivalently the fine structure constant) is the relevant coupling for the interaction between photons and electrons in QED and is predicted to be a marginally irrelevant parameter just like $\lambda\phi^4$. Sure enough this dependence has been measured and the electron charge is smaller at higher energies!

Further, note that you can't use this to play tricks with perturbation theory. Since $\mu^2$ is basically a free parameter, it seems like, in principle, we could set it so that the coupling is very small, helping perturbation theory. However, if then you're probing energies too different from that $\mu^2$, the logarithms that appear inside the integral will grow, ruining everything. So you really need to use physical measurements for this. This doesn't mean we have somehow lost predictability, we can still predict the dependence of the amplitude with the energy scale, and further we can use these amplitudes to calculate higher point amplitudes.

The original high energy beta function is a bit trickier. At first glace it appears to be completely unphysical. It is merely a theoretical construct to compensate for the UV cutoff needed to define the theory. And sure enough, in the final result the cutoff dependence completely disappeared. However, this result was only valid for low energies, far away from the cutoff. If we started to probe energies closer to the cutoff we would most likely start to see it's effects, usually as non-localities in the theory. And at that point we would have to re-examine our beliefs, because it can either be that the theory is the ultimate theory and we're dealing with a discrete spacetime\footnote{This is not suggesting some esoteric quantum gravity theory, for example, it can be a theory to describe some condensed matter system, which usually has an underlying lattice.}, or perhaps there is new physics that comes into play, like a new very massive field. So even though it is not directly measurable it still has a lot of important physics.

\subsection{The continuum limit}

In the last section we have constructed an algorithm to phrase perturbation theory in terms of physically measurable quantities, which involved the use of counterterms to tune the high energy couplings such that it will reproduce the necessary low energy physics. This is all fine and dandy but along the line something miraculous happened. The end result was independent of the UV cutoff. This means that, at the end of the calculation, we can set it to be whatever we want it to be without changing the actual measurable predictions. What if we set it to infinity? There seems to be no problem with this. OK, and what if instead of waiting all this time to phrase things in terms of physically measurable quantities I just try to work with a theory without a UV cutoff? For instance, let's look at the 2-to-2 particle scattering, the integrand goes like

\begin{equation}
    \int\dd[4]{k}\frac{1}{k^4}\sim\int\dd{k}\frac{1}{k}
\end{equation}
which diverges if we have no cutoff! 

This is one example of the famous infinities that seemingly plague quantum field theory. But with all the tools we have developed, it won't be very hard to deal with them. In this case it is obvious what went wrong, even for ordinary QM we can't just take the continuum limit carelessly, we must be very careful in our calculations and only at the end take the appropriate limit. The cool thing is that if we're careful, computing things in a well defined theory, phrase everything in terms of physically measurable quantities, and wait until the very end to take the $\Lambda_0\to\infty$ limit we get a sensible answer. However, it is not at all obvious this behaviour is generic. Sure we can introduce counterterms to parametrise the RG flow, but who is to say the final answer doesn't depend on the cutoff? If the theory is fundamentally discrete, for example, there's no conceptual uneasiness in having physical answers depending on the precise lattice spacing.

It turns out it isn't a generic behaviour. Some theories are "nice", and, after introducing the relevant counterterms, we can remove the cutoff and still get finite answers, and some theories are "less nice" and we we can't completely remove the cutoff dependence. At this point it is helpful to use a slightly less careful nomenclature when discussing these issues. Instead of focusing on the cutoff and talking about whether or not we can remove the cutoff dependence, it is easier to think in terms of divergences. That is, we compute diagrams (including counterterms) with a finite cutoff, and we talk about whether the final amplitude diverges or not as we take the infinite cutoff limit. Then the counterterms play a double role, they are both trying to absorb that infinity and to phrase the answer in terms of physical quantities. Note that this is just a change of words, if you feel uneasy about this mindset of removing infinities, you can always go back to the previous picture of always having a finite cutoff and doing the limit carefully just at the end.

At this point, all QFT textbooks go through a rather involved discussion about how to tell whether or not a diagram will diverge and how bad is the divergence. It involves lots of subtleties with graph theory and the fact that some diagrams appear to converge but then they have a subgraph that is actually divergent. It is all incredibly messy, and the worst thing is, even after all that hard work, most textbooks don't end up using this technology at all! The reason being the actual proof of the theorem is so incredibly fiddly that it is completely not worth the effort. 

In my opinion, that discussion of subgraphs and divergences doesn't really help in understanding the final theorem, so we'll skip it and go straight to the end result. If you want more details you can find them in any self-respecting QFT textbook. The end result is encapsulated in the following theorem due to Bogoliubov\footnote{Nikolay Nikolayevich Bogolyubov: born in 1909 in Nizhny Novgorod, Russian Empire; died in 1992 in Moscow, Russian Federation; Doctoral Advisor: Nikolay Krylov}, Parasiuk\footnote{Ostap Stepanovich Parasyuk; born in 1921 in Belka, Republic of Poland; died in 2007 in Kyiv, Ukraine}, Hepp\footnote{Klaus Hepp: born in 1936 in Kiel, Germany; Doctoral Advisor: Res Jost}, and Zimmermann:

\begin{theorem}[BPHZ]
Quantum Field Theories can be classified into three categories:
\begin{itemize}
    \item \textbf{Super-Renormalisable}: All couplings have positive mass dimension
    \item \textbf{Renormalisable}: All couplings have either positive or vanishing mass dimension
    \item \textbf{Non-Renormalisable}: At least one coupling has negative mass dimension
\end{itemize}
These three categories will have different types of divergences and therefore their behaviour is different:
\begin{itemize}
    \item \textbf{Super-Renormalisable}: There are only a finite number of divergent diagrams and therefore with a finite number of counterterms we can absorb all divergences
    \item \textbf{Renormalisable}: There are an infinite number of divergent diagrams but only a finite number of divergent amplitudes. We can introduce a finite number of counterterms that can be tuned order by order to absorb all divergences. The counterterms at a given order will exactly cancel the divergent sub-diagrams at higher orders
    \item \textbf{Non-Renormalisable}: There are an infinite number of divergent amplitudes, and therefore we need an infinite number of counterterms to absorb all divergences
\end{itemize}
\end{theorem}

This is a crucial result in perturbative QFT, but it is also one of the most controversial (and arguably misinterpreted). There are several claims made about the nature of this theorem, some are a bit misguided and others are just flat out wrong, we'll enumerate a few below to help navigate the literature (especially when older).

\begin{itemize}
    \item \textit{Non-renormalisable theories have infinities which cannot be removed}: This is one the of the "flat out wrong" type, we can perfectly take care of the infinities in non-renormalisable theories, we just need an infinite number of couplings to do it.
    \item \textit{Non-renormalisable theories are not predictive}: The idea behind this is the interpretation that to fix each counterterm we need a relevant experiment, then, since we need an infinite number of counterterms we would need an infinite number of experiments to be able to predict anything. This is technically true but only if you want to take the continuum limit. You can perfectly truncate to just a finite number of terms, keeping the cutoff dependence, then scaling will tell us the higher order terms will matter less and less for lower energies. At that level we can perfectly make predictions using non-renormalisable theories, you just have an additional parameter, the cutoff. Arguably they are even more predictive than renormalisable theories, since renormalisable theories are almost completely insensitive to UV physics, whereas non-renormalisable theories have much sharper bounds on where new physics comes into play, which is around the scale of the cutoff. And this has been widely exploited in the past for the discovery of the W, Z, and the Higgs bosons.
    \item \textit{Non-renormalisable theories are inconsistent}: This was a belief that was held for a long time but is now rather out dated. Usually people complain that either locality or unitarity is broken in these theories. And to some extent this is true. However, it is not because of some fundamental inconsistency, it is merely the statement that these theories appear to not have a continuum limit. But this is perfectly fine so long as we have some fundamental reason behind it (like a lattice) or if we just take them as effective field theories at low energy. It is a bit hubris to think our puny theory is valid up to arbitrarily large scales, it is perfectly fine to deal with theories that don't have a continuum limit
    \item \textit{The BPHZ theorem determines which theories have a sensible continuum limit}: This one is not entirely wrong, it's just a bit too quick of a conclusion. First of all, it is fundamentally perturbative in nature. Secondly, it deals with the consistency of a particular algorithm. All it states is for which theories will the counterterm algorithm work and remove the cutoff dependence, and for which theories it doesn't work. It doesn't mean the theory itself makes or doesn't make sense in the continuum. Essentially it is only asking how sensitive your theory is to the UV physics and classifying whether theories are less sensitive (renormalisable) or more sensitive (non-renormalisable). We'll see below how a theory can be renormalisable, but at the same time, even at the perturbative level appear not to have a sensible continuum limit.
\end{itemize}

A final note before moving on. Not everyone agrees with the classification stated above. Some people call what we've been describing as renormalisability the \textit{power counting} notion of renormalisability, preferring the term renormalisable to mean "theories with a continuum limit". This is partially to address the misconceptions stated above, and is a perfectly fine distinction, one just has to be careful when navigating different conventions. 

Although the BPHZ theorem is a good start when we want to ask the question of the continuum limit, as we stated above, it isn't the whole story. The real world is far more complex than that. So then, which theories do have a continuum limit? To be entirely honest, the answer is not known in its full generality. Mostly what we have is very strong evidence pointing one or the other way but very rarely do we have an actual loophole-free proof.\footnote{In fact, one of the famous Clay Mathematics Institute's Millenium Prize Problems (worth $10^6$ \$) is to prove Yang-Mills theories, i.e. non-Abelian gauge theories, have a continuum limit.} However, we can still make some progress and understand which class of theories would have a continuum limit, or said another way, how to spot if a theory has a continuum limit. To do that, let's bring back our old picture of the RG flows in theory space.

\begin{center}
    \includegraphics[scale=0.5]{fixedpoint.jpg}
    \captionsetup{type=figure}
    \caption{RG flows near a fixed point}
\end{center}

How to probe the continuum limit using this? Technically, the RG flow only goes one way, from high energies to low energies, but let's imagine we can run it backwards, fixing a point in the IR and then increasing the cutoff as we go along. This is indeed the difficulty, we don't really know how to do this is practice, but we can always use our imagination. What we want to know is which theories give sensible answers as we send the energy scale to infinity.

The first example is the trivial one. If we sit at a fixed point, then RG flow does absolutely nothing to the theory, so obviously we can also go backwards and we stay in the same theory. Therefore, conformal field theories have a sensible continuum limit. Like I said, this is a very trivial example, they don't depend on the energy scale in the first place so nothing interesting happens.

Now imagine we sit at the critical surface, these are theories that have a fixed point in the IR. Then, as we flow to higher energies they get further away from the fixed point.  There are two additional possibilities: either they come from another UV fixed point and are really a renormalised trajectory with respect to that CFT; or they go all the way to infinity. In the latter case, we have a theory with infinite couplings in the high energy limit, which is obviously bonkers. These theories don't have a continuum limit.

Finally, we could have a theory that sits in the renormalised trajectory of some UV CFT. These theories have a well defined limit as we go to higher and higher energies, the asymptotic fixed point, hence, they make sense at arbitrarily high energy scales and they have a continuum limit. Usually, these theories are further classified by whether or not they are interacting, a classification we have alluded to in Chapter \ref{wilsonrg}. If there are no interactions everywhere along the flow it is said to be trivial, if there are interactions along the flow but the asymptotic CFT is the Gaussian fixed point it is called asymptotically free, and if the CFT itself is also interacting we call it asymptotically safe.

Now we can truly see how the BPHZ classification can fool us. For instance, $\phi^4$ theory in 4 dimensions has a marginally irrelevant operator, therefore, strictly speaking it appears to not have a continuum limit\footnote{Of course this is a statement about a perturbative calculation up to 1-loop, so we haven't really proved a continuum limit doesn't exist, only that it is very unlikely.}. However the BPHZ theorem classified this theory as "nice", and indeed we were able to remove the cutoff dependence entirely. What is going on?

The reason for this is that the quartic coupling isn't irrelevant but merely marginally irrelevant. This means its divergence is very soft, in fact it is logarithmic. To see this let's recall the beta function (in the massless limit)

\begin{equation}
    \beta_4=\frac{3g_4^2}{16\pi^2}
\end{equation}
which can be solved to yield

\begin{equation}
    g_4(\Lambda)=-\frac{16\pi^2}{16\pi^2 C+3\log\Lambda}
    \label{solutiong4}
\end{equation}
where $C$ is some integration constant. Naively, we would have to parametrise this in terms of some value for $g_4$ at a particular scale. However, there is a natural scale which we can use to parametrise the flow, the scale at which the coupling diverges. This is given by

\begin{equation}
    16\pi^2 C+3\log\Lambda_{\phi^4}=0
\end{equation}
substituting back into (\ref{solutiong4}) gives

\begin{equation}
    g_4(\Lambda)=\frac{16\pi^2}{3\log(\frac{\Lambda_{\phi^4}}{\Lambda})}
\end{equation}

This phenomenon when the coupling appears to diverge at a finite energy scale is called a Landau\footnote{Lev Davidovich Landau: born in 1908 in Baku, Russian Empire; died in 1968 in Moscow, Soviet Union; Academic Advisor: Niels Bohr; Nobel Prize in Physics 1962 "for his pioneering theories for condensed matter, especially liquid helium"} pole. In our case, since this is a 1-loop computation we shouldn't really interpret this as true divergence but more as a signal that perturbation theory breaks down since the theory becomes strongly coupled. Also, the fact that we managed to trade a dimensionless coupling with an energy scale is known as \textit{dimensional transmutation} and used to be a big mystery.

This gives us the clue we need to interpret the BPHZ result. The divergence is so incredibly soft that, if we start in the weak coupling regime, we need to probe energies exponentially higher than that to see any flow whatsoever. Therefore, when perturbation theory is valid, which is also when the counterterm algorithm makes sense, this coupling barely runs at all and it will appear that we can confidently remove this cutoff. It is purely an artefact of the algorithm.

In the real world we have examples for all these kinds of flows. Fermi's theory for the weak interactions was non-renormalisable, which signalled the presence of the W and Z bosons. QCD, the theory of strong interactions, appears to be asymptotically free so it can indeed be valid at arbitrarily high energy scales. QED on the other hand behaves closer in spirit with $\phi^4$ theory, it appears to not have a continuum limit, however, the energy scale at which it would break is $\Lambda_\text{QED}\sim10^{286}$ GeV, which is so ludicrously high it is completely unimportant phenomenologically. The standard model of particle physics which is our best theory to describe the fundamental interactions is renormalisable, however, it is still an open problem whether or not it has a continuum limit. Nowadays, it is generally believed it is but an effective theory, and people are working hard to search for signals of irrelevant operators which could signal new physics, all searches so far have not proved successful. And finally, general relativity is non-renormalisable so it naively appears to not have a continuum limit, there are many approaches to try to circumvent this, by having a discrete theory (e.g. loop quantum gravity), by having new physics at the Planck\footnote{Karl Ernst Ludwig Marx Planck: born in 1858 in Kiel, Duchy of Holstein; died in 1947 in Göttingen, Germany; Doctoral Advisor: 	Alexander von Brill, Gustav Kirchhoff, Hermann von Helmholtz; Nobel Prize in Physics 1918 "in recognition of the services he rendered to the advancement of Physics by his discovery of energy quanta"} scale (e.g. string theory), or by saying it actually comes from a highly non-trivial UV CFT (e.g. asymptotic safety). So far, there is no compelling experimental evidence to confirm or disprove any of these alternatives

\subsection{Scheme (in)dependence}\label{scheme}

Throughout these lectures, we made a bunch of choices, namely, the precise regulator (hard momentum cutoff), and the renormalisation scheme (on-shell) with the promise that the overall physics did not depend on those choices. It is finally the time to address such concerns. Some of them are quite easy to deal with, so we'll start with that.

First, the renormalisation scheme. We chose a physical scheme such that the couplings would relate to actual experimentally measurable quantities. However, we could have chosen other schemes, that is, we could have fixed the counterterms in a different way. They are, after all, arbitrary. If we want to do old school subtracting infinities renormalisation, we would still need the infinite cutoff limit to make sense, but the finite part of the counterterm is always free to choose. What do different choices mean? Well, it's very easy to see they are essentially a reparametrisation of the RG trajectory with possibly a change of basis in theory space. That will clearly not change the overall shape of the flow, e.g. the existence of fixed points, or if we're flowing towards or away from a fixed point. So we can sleep quietly with regards to this one. The on-shell scheme is a good way to keep the physics more explicit, but, in practical calculations, other choices may be more convenient.

Secondly, imagine we are \textit{not} taking the continuum limit, we are keeping the regulator there to define our theory. Then, physical answers must obviously depend on the precise nature of the regulator. For renormalisable theories and for low enough energies, that dependence will go away. However, for non-renormalisable theories, or if we're probing energies near the regulator scale, that dependence will be there and it is absolutely physical. We should take care to use a regulator that corresponds to the actual physical system we're studying.

Finally, imagine we are indeed dealing with a theory with a well defined continuum limit. Will the physics change if we use a different style of regulator? Or said another way, will the answer to a physical question depend on precisely how we take the limit? We would like the answer to those two questions to be a sound \textit{no}, but we should check. And indeed we will prove that, with the added bonus of showcasing some possibly useful techniques along the way.

First of all, we need to be able to encode a more generic regularisation procedure. So, instead of explicitly turning off high momentum modes, we can simply dampen their contribution to the path integral. If we introduce sufficient damping, there will be no UV divergences to worry about. To that effect, we change the kinetic term to

\begin{equation}
    S_{\Lambda_0}[\phi]=\int\frac{\dd[d]p}{(2\pi)^d}\frac{1}{2}\Tilde{\phi}(-\vb{p})\frac{\vb{p}^2+m^2}{K\qty(\frac{\vb{p}}{\Lambda_0})}\Tilde{\phi}(\vb{p})+S_{\text{int,}\Lambda_0}[\phi]
\end{equation}
where $K$ is the function that will do the necessary damping. Ideally, we would demand the following properties:
\begin{enumerate}
    \item $K\qty(\frac{\vb{p}}{\Lambda_0})$ is a smooth non-increasing positive function of $\frac{\vb{p}^2}{\Lambda_0^2}$
    \item $K\qty(\frac{\vb{p}}{\Lambda_0})=1$ for $\vb{p}^2<\Lambda_0^2$
    \item $K\qty(\frac{\vb{p}}{\Lambda_0})\to0$ sufficiently fast as $\frac{\vb{p}^2}{\Lambda_0^2}\to\infty$ 
\end{enumerate}

\begin{center}
    \includegraphics[scale=0.5]{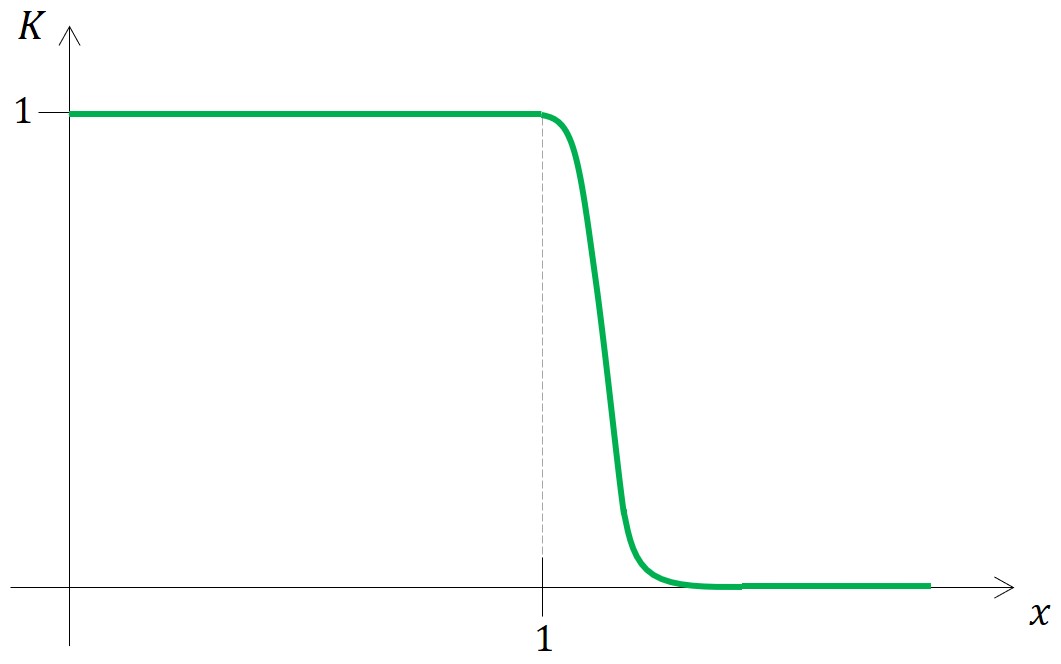}
    \captionsetup{type=figure}
    \caption{Example of an appropriate $K(x)$}
\end{center}

We say ideally because some of them can be tweaked without loosing the main picture, for example, we can give up smoothness (using a hard momentum cutoff does so), or we could instead demand the function to be very close to 1 rather than exactly 1 for $\vb{p}^2<\Lambda_0^2$. The third one however is completely non-negotiable, that's where the physics lies.\footnote{This style of regulator captures every possible regulator that actually presents a definition for the path integral. However, some commonly used regulators don't really do that. Instead, they are fundamentally perturbative and only take care of making the Feynman integrals finite. They work well for renormalisable theories, if we're interested in the continuum limit (or for low enough energies), because, in that case, we won't care about the UV beta function, and then any renormalisation scheme will make the IR flow physical. An example of this is dimensional regularisation, which involves analytically continuing integrals to non-integer dimensions and then removing the poles as $d\to4$. These regulators are often the more convenient choice since they can more easily preserve the symmetries of the theory, e.g. gauge invariance. However, they are far less rigorous (and far less physical) and I won't deal with them in these notes.}

Now using the identity

\begin{align}
    &\int\mathcal{D}\phi_1~\mathcal{D}\phi_2~\exp\Bigg[-\frac{1}{2}\int\frac{\dd[d]p}{(2\pi)^d}\qty[\frac{1}{A(\vb{p})}\Tilde{\phi}_1(\vb{p})\Tilde{\phi}_2(-\vb{p})+\frac{1}{B(\vb{p})}\Tilde{\phi}_2(\vb{p})\Tilde{\phi}_2(-\vb{p})]+S_\text{int}[\phi_1+\phi_2]\Bigg]=\nonumber\\
    =&\int\mathcal{D}\phi~\exp[-\frac{1}{2}\int\frac{\dd[d]p}{(2\pi)^d}\frac{1}{A(\vb{p})+B(\vb{p})}\Tilde{\phi}(\vb{p})\Tilde{\phi}(-\vb{p})-S_\text{int}[\phi]]\nonumber\cdot\\
    &\cdot\int\mathcal{D}\phi'~\exp[\frac{1}{2}\int\frac{\dd[d]p}{(2\pi)^d}\qty(\frac{1}{A(\vb{p})}+\frac{1}{B(\vb{p})})\Tilde{\phi}'(\vb{p})\Tilde{\phi}'(-\vb{p})]
\end{align}
where

\begin{align}
    \Tilde{\phi}&=\Tilde{\phi}_1+\Tilde{\phi}_2\\
    \Tilde{\phi}'&=-\frac{B}{A+B}\Tilde{\phi}_1+\frac{A}{A+B}\Tilde{\phi}_2
\end{align}

We can write (by appropriately choosing $A$ and $B$, and neglecting the $\phi'$ integral since it only contributes with a field independent constant)

\begin{align}
    \int\mathcal{D}\phi~\ee^{-S_{\Lambda_0}[\phi]}=&\int\mathcal{D}\phi^-\mathcal{D}\phi^+\exp\Bigg[-\frac{1}{2}\int\frac{\dd[d]p}{(2\pi)^d}\frac{\vb{p}^2+m^2}{K\qty(\frac{\vb{p}}{\Lambda})}\Tilde{\phi}^-(\vb{p})\Tilde{\phi}^-(-\vb{p})-\nonumber\\
    &-\frac{1}{2}\int\frac{\dd[d]p}{(2\pi)^d}\frac{\vb{p}^2+m^2}{K\qty(\frac{\vb{p}}{\Lambda_0})-K\qty(\frac{\vb{p}}{\Lambda})}\Tilde{\phi}^+(\vb{p})\Tilde{\phi}^+(-\vb{p})-S_{\text{int},\Lambda_0}[\phi^++\phi^-]\Bigg]
    \label{rot}
\end{align}

Then, using the shorthand notation $K(\vb{p})\equiv K\qty(\frac{\vb{p}}{\Lambda})$ and $K_0(\vb{p})\equiv K\qty(\frac{\vb{p}}{\Lambda_0})$, we define the interaction part of the effective action as,

\begin{equation}
    \ee^{-S_{\text{int,}\Lambda}[\phi^-]}=\int\mathcal{D}\phi^+\exp[-\frac{1}{2}\int\frac{\dd[d]p}{(2\pi)^d}\frac{\vb{p}^2+m^2}{K_0-K}\Tilde{\phi}^+(\vb{p})\Tilde{\phi}^+(-\vb{p})-S_{\text{int},\Lambda_0}[\phi^++\phi^-]]
    \label{rgeq}
\end{equation}
and therefore the full action,

\begin{equation}
    S_\Lambda[\phi^-]=\frac{1}{2}\int\frac{\dd[d]p}{(2\pi)^d}\frac{\vb{p}^2+m^2}{K}\Tilde{\phi}^-(\vb{p})\Tilde{\phi}^-(-\vb{p})+S_{\text{int,}\Lambda}[\phi^-]
\end{equation}

All this together then establishes how to perform the RG procedure with a smooth regulator. It perfectly agrees with our intuition about how RG should work. We managed to have a simple splitting between high and low energy modes. The low energy mode looks like the original one but with a different cutoff scale, and the high energy mode only has support in the interval $[\Lambda,\Lambda_0]$. Note however a few differences, in this time we are not changing the field configuration space, the transformations were mere shifts, all the physics is in the quadratic damping. Therefore, there is no implicit dependence on the scale in the spacetime integration or in the field definition. This doesn't mean we can't perform the second and third steps of RG, we can recover both scaling and the field anomalous dimension. However, they don't play any role in what follows so we will ignore them. Finally, we chose to regulate the full quadratic term rather than just the kinetic term, this simplifies our story when we go to perturbative Feynman diagrams calculations, however, it is just a choice, we could have chosen otherwise without affecting the main points.

Given the above considerations about what depends and doesn't depend on the cutoff, it's a straightforward exercise to derive the differential form of the RG equation (\ref{rgeq}). This is known as the Polchinski\footnote{Joseph Gerard Polchinski Jr.: born in 1954 in White Plains, USA; died in 2018 in Santa Barbara, USA; Doctoral Advisor: Stanley Mandelstam} equation:

\begin{equation}
    \Lambda\pdv{\Lambda}S_{\text{int,}\Lambda}[\phi]=\int\frac{\dd[d]p}{(2\pi)^d}\frac{1}{2}\frac{\Delta\qty(\frac{\vb{p}}{\Lambda})}{\vb{p}^2+m^2}\qty{\fdv{S_{\text{int,}\Lambda}}{\tilde{\phi}(-\vb{p})}\fdv{S_{\text{int,}\Lambda}}{\tilde{\phi}(\vb{p})}-\frac{\var{^2S_{\text{int,}\Lambda}}}{\var{\tilde{\phi}(-\vb{p})}\var{\tilde{\phi}(\vb{p})}}}
\end{equation}
where $\Delta\qty(\frac{\vb{p}}{\Lambda})=\Lambda\pdv{\Lambda}K\qty(\frac{\vb{p}}{\Lambda})$.

This equation is entirely equivalent to (\ref{rgeq}), it's just the usual transition between integral and differential forms. This means we can calculate RG flows entirely based on the Polchinski equation, rather than trying to compute Feynman integrals. To a large extent, what we did in Section \ref{lpa} was precisely that, just dressed in a slightly different language.

Now, to proceed, we need to be able to compare (connected) correlation functions computed with both actions. To that effect we define the Wilsonian effective action with a source term, i.e. the generating functional of connected correlation functions:

\begin{equation}
    \ee^{-W_\Lambda[J]}=\int\mathcal{D}\phi~\exp[-S_{\Lambda}[\phi]+\int\frac{\dd[d]p}{(2\pi)^d}\tilde{J}(-\vb{p})\tilde{\phi}(\vb{p})]=\int\mathcal{D}\phi~\ee^{-S_\Lambda[\phi;J]}
\end{equation}

Using (\ref{rgeq}), we can relate the Wilsonian effective actions calculated at different energies (we shall omit the argument of the various functions appearing for brevity and because we're only to write the quadratic piece explicitly, in which case one of the fields/sources is evaluated at $\vb{p}$ and the other at $-\vb{p}$).

\begin{align}
    \ee^{-W_{\Lambda_0}[J]}=&\int\mathcal{D}\phi~\ee^{-S_{\Lambda_0}[\phi;J]}=\nonumber\\
    =&\int\mathcal{D}\phi~\exp\bigg[-\int\frac{\dd[d]p}{(2\pi)^d}\qty(\frac{1}{2}\frac{\vb{p}^2+m^2}{K_0}\tilde{\phi}\tilde{\phi}-\tilde{J}\tilde{\phi})-S_{\text{int,}\Lambda_0}[\phi]\bigg]=\nonumber\\
\intertext{Doing the rotation as in (\ref{rot}), where the source term behaves as an interaction,}
    =&\int\mathcal{D}\phi^-\mathcal{D}\phi^+~\exp\bigg[-\int\frac{\dd[d]p}{(2\pi)^d}\bigg(\frac{1}{2}\frac{\vb{p}^2+m^2}{K}\tilde{\phi}^-\tilde{\phi}^-+\frac{1}{2}\frac{\vb{p}^2+m^2}{K_0-K}\tilde{\phi}^+\tilde{\phi}^+-\nonumber\\
    &-\tilde{J}\qty(\tilde{\phi}^-+\tilde{\phi}^+)\bigg)-S_{\text{int,}\Lambda_0}[\phi^-+\phi^+]\bigg]=\nonumber\\
\intertext{Completing the square}
    =&\int\mathcal{D}\phi^-\mathcal{D}\phi^+~\exp\bigg[-\int\frac{\dd[d]p}{(2\pi)^d}\bigg(\frac{1}{2}\frac{\vb{p}^2+m^2}{K}\tilde{\phi}^-\tilde{\phi}^-+\nonumber\\
    &+\frac{1}{2}\frac{\vb{p}^2+m^2}{K_0-K}\qty(\tilde{\phi}^+-\frac{K_0-K}{\vb{p}^2+m^2}\tilde{J})\qty(\tilde{\phi}^+-\frac{K_0-K}{\vb{p}^2+m^2}\tilde{J})-\tilde{J}\tilde{\phi}^--\frac{1}{2}\frac{K_0-K}{\vb{p}^2+m^2}\tilde{J}\tilde{J}\bigg)-\nonumber\\
    &-S_{\text{int,}\Lambda_0}[\phi^-+\phi^+]\bigg]=\nonumber\\
\intertext{Defining $\tilde{\phi}'=\tilde{\phi}^+-\frac{K_0-K}{\vb{p}^2+m^2}\tilde{J}$}
    =&\int\mathcal{D}\phi^-\mathcal{D}\phi'\exp\bigg[-\int\frac{\dd[d]p}{(2\pi)^d}\bigg(\frac{1}{2}\frac{\vb{p}^2+m^2}{K}\tilde{\phi}^-\tilde{\phi}^-+\frac{1}{2}\frac{\vb{p}^2+m^2}{K_0-K}\tilde{\phi}'\tilde{\phi}'-\tilde{J}\tilde{\phi}^--\nonumber\\
    &-\frac{1}{2}\frac{K_0-K}{\vb{p}^2+m^2}\tilde{J}\tilde{J}\bigg)-S_{\text{int,}\Lambda_0}\qty[\tilde{\phi}^-+\tilde{\phi}'+\frac{K_0-K}{\vb{p}^2+m^2}\tilde{J}]\bigg]=\nonumber\\
\intertext{Using (\ref{rgeq}) to perform the $\phi'$ integration}
    =&\int\mathcal{D}\phi^-\exp\bigg[-\int\frac{\dd[d]p}{(2\pi)^d}\bigg(\frac{1}{2}\frac{\vb{p}^2+m^2}{K}\tilde{\phi}^-\tilde{\phi}^--\tilde{J}\tilde{\phi}^--\frac{1}{2}\frac{K_0-K}{\vb{p}^2+m^2}\tilde{J}\tilde{J}\bigg)-\nonumber\\
    &-S_{\text{int,}\Lambda}\qty[\tilde{\phi}^-+\frac{K_0-K}{\vb{p}^2+m^2}\tilde{J}]\bigg]\nonumber\\
\intertext{Defining $\tilde{\phi}''=\tilde{\phi}^-+\frac{K_0-K}{\vb{p}^2+m^2}\tilde{J}$}
    =&\int\mathcal{D}\phi''\exp\bigg[-\int\frac{\dd[d]p}{(2\pi)^d}\bigg(\frac{1}{2}\frac{\vb{p}^2+m^2}{K}\tilde{\phi}''\tilde{\phi}''-\frac{K_0}{K}\tilde{J}\tilde{\phi}''+\frac{1}{2}\frac{K_0-K}{\vb{p}^2+m^2}\frac{K_0}{K}\tilde{J}\tilde{J}\bigg)-S_{\text{int,}\Lambda}[\phi'']\bigg]=\nonumber\\
    =&\int\mathcal{D}\phi''\exp[-S_\Lambda[\phi'']+\int\frac{\dd[d]p}{(2\pi)^d}\frac{K_0}{K}\qty(\tilde{J}\tilde{\phi}''-\frac{1}{2}\frac{K_0-K}{\vb{p}^2+m^2}\tilde{J}\tilde{J})]=\nonumber\\
    =&\exp[-W_\Lambda\qty[\frac{K_0}{K}\tilde{J}]-\frac{1}{2}\int\frac{\dd[d]p}{(2\pi)^d}\frac{K_0}{K}\frac{K_0-K}{\vb{p}^2+m^2}\tilde{J}\tilde{J}]
\end{align}

After all this, we get our desired relation between the Wilsonian effective actions calculated in both theories:

\begin{equation}
    W_{\Lambda_0}[J]=W_\Lambda\qty[\frac{K_0}{K}\tilde{J}]+\frac{1}{2}\int\frac{\dd[d]p}{(2\pi)^d}\frac{K_0}{K}\frac{K_0-K}{\vb{p}^2+m^2}\tilde{J}(\vb{p})\tilde{J}(-\vb{p})
\end{equation}

From this, we can use (\ref{correlator}) to compare connected correlation functions in both theories:

\begin{align}
    \expval{\tilde{\phi}(\vb{p}_1)\tilde{\phi}(\vb{p}_2)}_{\Lambda_0}&=\frac{K_0^2}{K^2}\expval{\tilde{\phi}(\vb{p}_1)\tilde{\phi}(\vb{p}_2)}_\Lambda-\frac{K_0-K}{\vb{p}_1^2+m^2}\frac{K_0}{K}\delta^{(d)}(\vb{p}_1+\vb{p}_2)\\
    \expval{\tilde{\phi}(\vb{p}_1)\cdots\tilde{\phi}(\vb{p}_n)}_{\Lambda_0}&=\expval{\tilde{\phi}(\vb{p}_1)\cdots\tilde{\phi}(\vb{p}_n)}_\Lambda\prod_{i=1}^n\frac{K_0(\vb{p}_i)}{K(\vb{p}_i)}
    \label{rgcorr}
\end{align}
where $n>2$, and in the first line $K$ and $K_0$ can be evaluated at either $\vb{p}_1$ or $\vb{p}_2$, the distinction is irrelevant.

Note that, if we use a smooth regulator that doesn't vanish at a finite value of the momentum, these relations are perfectly invertible. We are not loosing any information. It seems like we could use correlation functions at the scale $\Lambda$ to calculate correlation functions at the scale $\Lambda_0$, this would be true if were actually able to calculate the correlators exactly, however, since we only have access to perturbation theory it is not that straightforward to invert them.

We now have everything we need under our belts. We take the limit $\Lambda_0\to\infty$, which means $K_0\to1$. Then imagine we deform $K$ to $K+\var{K}$, where $\var{K}\ll 1$, and that we have an action $S_\Lambda$ with that deformed regulator. Then (\ref{rgcorr}) gives, to leading order in $\var{K}$,

\begin{align}
    \expval{\tilde{\phi}(\vb{p}_1)\tilde{\phi}(\vb{p}_2)}_\infty=&\frac{1}{K^2}\expval{\qty(1-\frac{\var{K}}{K})\tilde{\phi}(\vb{p}_1)\qty(1-\frac{\var{K}}{K})\tilde{\phi}(\vb{p}_2)}_\Lambda+\nonumber\\
    &+\qty(1-\frac{1}{K}+\frac{\var{K}}{K^2})\frac{1}{\vb{p}_1^2+m^2}\delta^{(d)}(\vb{p}_1+\vb{p}_2)\\
    \expval{\tilde{\phi}(\vb{p}_1)\cdots\tilde{\phi}(\vb{p}_n)}_\infty=&\expval{\qty(1-\frac{\var{K}}{K})\tilde{\phi}(\vb{p}_1)\cdots\qty(1-\frac{\var{K}}{K})\tilde{\phi}(\vb{p}_n)}_\Lambda\prod_{i=1}^n\frac{1}{K(\vb{p}_i)}
\end{align}

From this action we can construct a new one by

\begin{equation}
    S'_\Lambda=S_\Lambda+\int\frac{\dd[d]{p}}{(2\pi)^d}\qty(\frac{\var{K}}{K}\tilde{\phi}(\vb{p})\fdv{S_\Lambda}{\tilde{\phi}(-\vb{p})}+\frac{1}{2}\frac{\var{K}}{\vb{p}^2+m^2}\qty(\fdv{S_\Lambda}{\tilde{\phi}(\vb{p})}\fdv{S_\Lambda}{\tilde{\phi}(-\vb{p})}+\frac{\var{^2S_\Lambda}}{\var{\tilde{\phi}(-\vb{p})}\var{\tilde{\phi}(\vb{p})}}))
    \label{vark}
\end{equation}

This new action will have the following correlation functions, to leading order in $\var{K}$\footnote{To see this you should write the correlation function in terms of a path integral, expand the exponential to leading order in $\var{K}$, then use $\displaystyle\fdv{S_\Lambda}{\tilde{\phi}}\ee^{-S_\Lambda}=-\fdv{\tilde{\phi}}\ee^{-S_\Lambda}$ and similarly for the second derivative, and finally integrate by parts wrt $\tilde{\phi}$. The 2-point function should be straightforward, the second term in (\ref{vark}) only gives a disconnected contribution to higher-point correlators which does not contribute to the connected correlation function}:

\begin{align}
    \expval{\tilde{\phi}(\vb{p}_1)\tilde{\phi}(\vb{p}_2)}'_\Lambda=&\expval{\qty(1-\frac{\var{K}}{K})\tilde{\phi}(\vb{p}_1)\qty(1-\frac{\var{K}}{K})\tilde{\phi}(\vb{p}_2)}_\Lambda+\nonumber\\
    &+\frac{\var{K}}{\vb{p}_1^2+m^2}\delta^{(d)}(\vb{p}_1+\vb{p}_2)\\
    \expval{\tilde{\phi}(\vb{p}_1)\cdots\tilde{\phi}(\vb{p}_n)}'_\Lambda=&\expval{\qty(1-\frac{\var{K}}{K})\tilde{\phi}(\vb{p}_1)\cdots\qty(1-\frac{\var{K}}{K})\tilde{\phi}(\vb{p}_n)}_\Lambda
\end{align}

Therefore

\begin{align}
    \expval{\tilde{\phi}(\vb{p}_1)\tilde{\phi}(\vb{p}_2)}_\infty&=\frac{1}{K^2}\expval{\tilde{\phi}(\vb{p}_1)\tilde{\phi}(\vb{p}_2)}'_\Lambda+\frac{1-\frac{1}{K}}{\vb{p}_1^2+m^2}\delta^{(d)}(\vb{p}_1+\vb{p}_2)\\
    \expval{\tilde{\phi}(\vb{p}_1)\cdots\tilde{\phi}(\vb{p}_n)}_\infty&=\expval{\tilde{\phi}(\vb{p}_1)\cdots\tilde{\phi}(\vb{p}_n)}'_\Lambda\prod_{i=1}^n\frac{1}{K(\vb{p}_i)}
\end{align}

This result means $S'_\Lambda$ satisfies the RG equation (\ref{rgeq}) (or equivalently the Polchinski equation) with regulator $K$ and at the same time gives exactly the same continuum limit as $S_\Lambda$. Therefore regulating with $K$ or $K+\var{K}$ gives exactly the same continuum limit. Alas, the physics does not depend on our choice of regulator, as we wanted in the first place.

\newpage

\section{Historical Perspective and Concluding Remarks}

During these lectures, we have done a path that is almost the exact opposite to the way renormalisation was unveiled historically. Physicists were quick to develop the basic framework of QFT after the discovery of quantum mechanics. Much of the underlying structure was known by the end of the 1930s, however, they were almost at the point of giving up on the whole endeavour! The reason being they were getting infinity for any calculation beyond tree-level. Dirac\footnote{Paul Adrien Maurice Dirac: born in 1902 in Bristol, England; died in 1984 in Tallahasee, USA; Doctoral Advisor: Ralph Fowler; Nobel Prize in Physics 1933 together with Erwin Schrödinger "for the discovery of new productive forms of atomic theory"} even said in 1937 "Because of its extreme complexity, most physicists will be glad to see the end of QED".

Later, after the war, a new generation of physicists, such as Feynman, Schwinger\footnote{Julian Seymour Schwinger: born in 1918 in New York City, USA; died in 1994 in Los Angeles, USA; Doctoral Advisor: Isidor Isaac Rabi; Nobel Prize in Physics 1965 together with Sin-Itiro Tomonaga and Richard P. Feynman "for their fundamental work in quantum electrodynamics, with deep-ploughing consequences for the physics of elementary particles"}, Tomonaga\footnote{Sin-Itiro Tomonaga: born in 1906 in Tokyo, Japan; died in 1979 in Tokyo, Japan; Academic Advisor: Yoshio Nishina; Nobel Prize in Physics 1965 together with Julian Schwinger and Richard P. Feynman "for their fundamental work in quantum electrodynamics, with deep-ploughing consequences for the physics of elementary particles"}, and Dyson\footnote{Freeman John Dyson: born in 1923 in Crowthorne, England; Academic Advisor: Hans Bethe} took matters into their own hands and realised there was some structure behind the infinities. They eventually managed to develop a systematic procedure to sweep away the infinities under a very large rug, earning Feynman, Schwinger, and Tomonaga the 1965 Nobel\footnote{Alfred Bernhard Nobel: born in 1833 in Stockholm, Sweden; died in 1896 in Sanremo, Italy} prize. In our language, they developed the idea of counterterms, and phrasing perturbation theory in terms of physical observables.

However, everyone was rather displeased with this procedure, Feynman, the one who first came up with it, famously called it a "dippy process". Further studies by Gell-Mann\footnote{Murray Gell-Mann: born in 1929 in Manhattan, USA; died in 2019 in Santa Fe, USA; Doctoral Advisor: Victor Weisskopf; Nobel Prize in Physics 1969 "for his contributions and discoveries concerning the classification of elementary particles and their interactions"} and Low\footnote{Francis Eugene Low: born in 1924 in New York City, USA; died in 2007 in Haverford, USA; Doctoral Advisor: Hans Bethe} in 1954 revealed some hints of a group action, and how one could define an effective electron charge at a certain energy scale. And later the Callan-Symanzik equations were written down, long before any interpretation of integrating out degrees of freedom was available.

Simultaneously, in the condensed matter community, there was a lot of confusion surrounding second order phase transitions. Namely the phenomena of universality, and anomalous dimensions were a huge mystery. Later on, Leo Kadanoff\footnote{Leo Philip Kadanoff: born in 1937 in New York City, USA; died in 2015 in Chicago, USA; Doctoral Advisor: Paul Martin} introduced the idea of blocking, which is essentially RG in real space.

The final picture, which we presented in Chapter \ref{wilsonrg}, only came along with Kenneth Wilson in the 1970s, it was him that unveiled the full picture of RG flows, theory space, and integrating out high energy modes. It was also him, together with Michael Fisher that first used the epsilon expansion\footnote{The paper in question has, arguably, the best title ever: "Critical Exponents in 3.99 Dimensions"}. The Polchinski equation and the exact RG formalism we presented in Sections \ref{lpa} and \ref{scheme} came along much later, in 1980s and 1990s.

This was a very complex history (of which I have only grazed the very surface). It took almost fifty years since the infinities first appeared to when they were completely understood, and even longer before they started appearing in standard textbooks. And for a good reason! This is an incredibly subtle subject. Even now, when many more techniques are available, and we have a much better grasp on the wilderness of theory space, there is a lot of confusion surrounding renormalisation. 

I hope these lectures have been useful to dispel some of that confusion. Using the most basic example, the one for which all these methods were originally designed to deal with, we can get a clear picture of what is going on. So let us summarise the broad structure of what has been covered:

\begin{itemize}
    \item To define a quantum field theory we need to introduce some kind of regulator that has an intrinsic scale associated with it (e.g. the cutoff), otherwise we get non-sensical answers.
    \item This gives us a way to define an action whereby we change that scale to some lower scale by integrating out degrees of freedom, ending up with an effective description at lower energies. This is the Renormalisation Group. 
    \item What RG boils down to is changing the couplings of our theory as we change the energy scale such that the observable low energy dynamics are unchanged. This gives us a flow in theory space that let's us ask (and possibly answer) questions about the structure of theories themselves and their relations
    \item Using counterterms, we can phrase all this in terms of physically measurable quantities. However, we're forced to introduce a second scale, the scale at which we do experiments. These two scales, the cutoff and the experimental, behave in very similar manners but are conceptually distinct. Most notably, the experimental scale is always observable, whereas the cutoff may not be.
    \item Finally, using all these tools, we can probe how sensible theories are to the UV data, and further, for which theories can we sensibly take the continuum limit. It is here the only place where infinities may appear.
\end{itemize}

But this is not the whole story! We have left out one of the most crucial parts of field theories, symmetries. The relationship between scale and symmetry is deep and fruitful. There is a beautiful story about how symmetries dictate which phase we're in, or which universality class. And this is relevant far away from condensed matter, particle physicists must also worry about this to be able to describe, for example, the quark-gluon plasma. There is also an intricate connection between renormalisation and anomalies, that is, symmetries present in the classical theory but not in the quantum one.

And finally, the deepest connections are with the two kinds of symmetries that aren't really symmetries: gauge redundancies and dualities. Gauge theories are the pinnacle of our understanding of physics, underlying both the standard model of particle physics and general relativity, but they are bizarre in the sense they are defined in terms of things we can't observe, even in principle. Gauge symmetries, should really be thought of as redundancies in the way we write the theory, not as physical symmetries. There is a lot to learn from the interplay between gauge invariance and RG flow, and a lot of it is still very much in the realm of ignorance. 

Dualities are quite a different beast. These are symmetries not within a given theory but that relate seemingly different theories. And I do mean different, for example, the AdS/CFT correspondence is a duality between a strongly coupled quantum field theory and classical gravity! There are a lot of examples out there and it is a very active field of research to understand whether there's some deep mystery of nature or just a mathematical accident. But in every example, scale and renormalisation play a crucial role in our understanding of Nature.

\newpage
\appendix
\section{Notation and conventions}\label{notation}

We use natural units where $c=\hbar=k_\text{B}=1$. This means we only have one dimension left which can be either mass, energy, length, time, or temperature. We'll prefer the use of energy and more specifically electron volt\footnote{Alessandro Giuseppe Antonio Anastasio Volta: born in 1754 in Como, Duchy of Milan; died in 1827 in Como, Lombardy-Venetia} (eV). They are related by 
\begin{equation}
    [\text{mass}]=[\text{energy}]=[\text{length}]^{-1}=[\text{time}]^{-1}=[\text{temperature}]
\end{equation}

We'll denote Euclidean signature vectors with boldface, e.g. $\vb{x}=(x^1,x^2,x^3,\dots)$, and the spatial indices by Latin letters starting from the middle of the alphabet $i,j,k,\dots=1,2,3,\dots$. Lorentzian signature vectors will have no special notation, e.g. $x = (x^0,\vb{x})$, where we use the "mostly plus" convention: $\eta=\mathrm{diag}(-1,1,1,1,\dots)$, the spacetime indices will be Greek letters starting from the middle of the alphabet $\mu,\nu,\dots=0,1,2,3,\dots$. Any other kind of index such as internal indices will be denoted by Latin letters starting from the beginning of the alphabet. Einstein's\footnote{Albert Einstein: born in 1879 in Ulm, German Empire; died in 1955 in Princeton, USA; Doctoral Advisor: Alfred Kleiner; Nobel Prize in Physics 1921 "for his services to Theoretical Physics, and especially for his discovery of the law of the photoelectric effect"} summation convention for repeated indices will be used unless explicitly stated.

For Fourier\footnote{Jean-Baptiste Joseph Fourier: born in 1768 in Auxerre, Kingdom of France; died in 1830 in Paris, Kingdom of France; Academic Advisor: Joseph-Louis Lagrange} transforms we use the following conventions:

\begin{align}
    \tilde{f}(k)&=\int\dd{x}f(x)\ee^{-\ii kx}\\
    f(x)&=\int\frac{\dd{k}}{2\pi}\tilde{f}(k)\ee^{\ii kx}
\end{align}

Putting it simply, all factors of $2\pi$ appear in $k$ (or $p$) integrals. In particular this gives the following normalisation for the momentum states

\begin{align}
    \bra{\vb{p}}\ket{\vb{p}'}&=(2\pi)^d\delta^{(d)}(\vb{p}-\vb{p}')\\
    \int\frac{\dd[d]p}{(2\pi)^d}\dyad{\vb{p}}{\vb{p}}&=\mathbb{1}\\
    \bra{\vb{x}}\ket{\vb{p}}&=\ee^{-\ii\dotp{\vb{p}}{\vb{x}}}
\end{align}

Further, throughout the text (but mostly in Chapter \ref{pathqm}) we'll merrily swap between Heisenberg\footnote{Werner Karl Heisenberg: born in 1901 in Würzburg, German Empire; died in 1976 in Munich, West Germany; Doctoral Advisor: Arnold Sommerfeld; Nobel Prize in Physics 1932 "for the creation of quantum mechanics, the application of which has, inter alia, led to the discovery of the allotropic forms of hydrogen"} and Schrödinger pictures. To simplify our notation, instead of keeping track of H and S subscripts we'll simply make the time dependence of operators/states explicit, i.e.

\begin{align}
    \hat{O}(t)&=\ee^{\ii\hat{H}t}\hat{O}(0)\ee^{-\ii\hat{H}t},~~~~\hat{O}\equiv\hat{O}(0)\\
    \ket{\psi,t}&=\ee^{-\ii\hat{H}t}\ket{\psi,0},~~~~\ket{\psi}\equiv\ket{\psi,0}
\end{align}

\newpage
\section{Path integrals in quantum field theory}\label{pathqft}

This appendix is intended for those who are not very familiar with the path integral in the context of quantum field theory, especially how to write the path integral in terms of Feynman diagrams. We assume that we have Wick rotated\footnote{Refer to Chapter \ref{pathqm} for motivation and details} to Euclidean signature therefore $\vb{x}=(x^1,x^2,x^3,\dots,x^d)$, where $x^d=\tau$ is the Euclidean time.

\subsection{The partition function}

In quantum field theory, the fundamental degrees of freedom are functions of both space and time, i.e. fields $\phi_a(\vb{x})$, where $a$ is some internal index. We can then define actions for these fields and write the transition amplitude in terms of a path integral, as we did for ordinary quantum mechanics. However, we'll be most interested in amplitudes where $t_\text{i}\to-\infty$ and $t_\text{f}\to\infty$, therefore, inside the action, we integrate over the entirety of spacetime. This new transition amplitude is called \textit{partition function} due to the connection with statistical field theory, and we'll write it as

\begin{equation}
    Z=\int\mathcal{D}\phi_a~\ee^{-S[\phi_a]}
\end{equation}

As we learned in Chapter \ref{pathqm}, this is not a very well defined object if we allow the fields to vary continuously over the spacetime variables. We also saw we could solve this by discretising our spacetime, effectively making the number of degrees of freedom finite, and turning the functional integration into a finite number of ordinary integrations. Because of this, henceforth, I shall be lazy and use finite dimensional notation to perform the necessary integrations, at the end there will be a section highlighting the main results using continuum notation.\footnote{This also has the side effect of making my manipulations more well defined in the eyes of a pure mathematician, but that's not the main motivation. Laziness is definitely a bigger factor.} 

This means fields will be denoted by $\vb{\phi}\in\mathbb{R}^n$, such that all complicated index structure and spacetime dependence is summarised into one index. I'll restrict to real fields since complex fields can be decomposed in terms of real fields\footnote{This is a bit of a lie since there are way more convenient ways to deal with complex fields than to reduce them to real fields, but it won't be very important for the main text}, and moreover, as in the rest of the notes, I won't deal with fermions.

\subsubsection{Free Theory}

Let's start with the simplest case, a free theory, i.e. an action at most quadratic in the fields. Consider a positive definite symmetric matrix $\te{M}$, and the following partition function

\begin{equation}
    Z[\te{M}]=\int\dd[n]{\phi}\ee^{-\frac{1}{2}\dotp{\vb{\phi}}{\te{M}\vb{\phi}}}
\end{equation}

To evaluate this integral note that, because $\te{M}$ is symmetric, there is an orthogonal matrix $\te{O}$ such that $\te{O}^\TT\te{M}\te{O}=\te{D}=\text{diag}(\lambda_1,\dots,\lambda_n)$ is diagonal. Then we make the change of variables $\vb{\phi}'=\te{O}\vb{\phi}$, which leaves the measure invariant since $\abs{\det \te{O}}=1$, to get

\begin{equation}
    Z[\te{M}]=\int\dd[n]{\phi'}\ee^{-\frac{1}{2}\dotp{\vb{\phi}'}{\te{D}\vb{\phi}'}}=\prod_{i=1}^n\int\dd{\phi'_i}\ee^{-\frac{1}{2}{\phi'_i}^2\lambda_i}=\prod_{i=1}^n\sqrt{\frac{2\pi}{\lambda_i}}=\sqrt{\frac{(2\pi)^n}{\det \te{M}}}
    \label{gauss}
\end{equation}

It is completely straightforward to generalise this result by adding a linear term, also called a \textit{source term},

\begin{equation}
    Z[\te{M},\vb{J}]=\int\dd[n]{\phi}\ee^{-\frac{1}{2}\dotp{\vb{\phi}}{\te{M}\vb{\phi}}+\dotp{\vb{\phi}}{\vb{J}}}
\end{equation}

We can then complete the square by performing the transformation $\vb{\phi}'=\vb{\phi}-\te{M}^{-1}\vb{J}$, which is a simple shift, so no Jacobian\footnote{Carl Gustav Jacob Jacobi: born in 1804 in Potsdam, Kingdom of Prussia; died in 1851 in Berlin, Kingdom of Prussia; Doctoral Advisor: Enno Dirksen} penalty,
\begin{equation}
    Z[\te{M},\vb{J}]=\int\dd[n]{\phi'}\ee^{-\frac{1}{2}\dotp{\vb{\phi}'}{\te{M}\vb{\phi}'}+\frac{1}{2}\dotp{\vb{J}}{\te{M}^{-1}\vb{J}}}=\ee^{\frac{1}{2}\dotp{\vb{J}}{\te{M}^{-1}\vb{J}}}\sqrt{\frac{(2\pi)^n}{\det \te{M}}}
\end{equation}

This gives us a neat trick to calculate expectation values of functions of $\vb{\phi}$,

\begin{equation}
    \expval{f(\vb{\phi})}=\frac{1}{Z[\te{M}]}\int\dd[n]{\phi}f(\vb{\phi})\ee^{-\frac{1}{2}\dotp{\vb{\phi}}{\te{M}\vb{\phi}}}
\end{equation}

Using the trick,

\begin{equation}
    f(\vb{\phi})=f\qty(\pdv{\vb{J}})\eval{\ee^{\dotp{\vb{\phi}}{\vb{J}}}}_{\vb{J}=\vb{0}}
    \label{trick}
\end{equation}
we write
\begin{align}
    \expval{f(\vb{\phi})}&=\frac{1}{Z[\te{M}]}\int\dd[n]{\phi}f\qty(\pdv{\vb{J}})\eval{\ee^{-\frac{1}{2}\dotp{\vb{\phi}}{\te{M}\vb{\phi}}+\dotp{\vb{\phi}}{\vb{J}}}}_{\vb{J}=\vb{0}}=\frac{1}{Z[\te{M}]}f\qty(\pdv{\vb{J}})\eval{Z[\te{M},\vb{J}]}_{\vb{J}=\vb{0}}=\nonumber\\
    &=f\qty(\pdv{\vb{J}})\eval{\ee^{\frac{1}{2}\dotp{\vb{J}}{\te{M}^{-1}\vb{J}}}}_{\vb{J}=\vb{0}}\label{free}
\end{align}

This means we only need to know the partition function with a source term to get every possible expectation value. For that reason the partition function with a source term is often called the \textit{generating functional}. To get a feel for this expression let's consider some examples. You should try to apply the formula (\ref{free}) carrying all the derivatives yourself to make sure you understand where this all comes from,

\begin{itemize}
    \item $\expval{\phi_i}=0$
    \item $\expval{\phi_i\phi_j}=M_{ij}^{-1}$
    \item $\expval{\phi_i \phi_j \phi_k \phi_l}=M_{ij}^{-1}M_{kl}^{-1}+M_{ik}^{-1}M_{jl}^{-1}+M_{il}^{-1}M_{jk}^{-1}$
\end{itemize}

The latter two have a diagrammatic representation. We represent a factor of $M_{ij}^{-1}$ by a line from $i$ to $j$:

\begin{equation*}
\begin{tikzpicture}[baseline=(j.south)]
  \begin{feynman}
    \vertex (i) {$i$};
    \vertex [right=2cm of i] (j) {$j$};
    
    \diagram* {
      (i) -- (j),
    };
  \end{feynman}
\end{tikzpicture}
\end{equation*}

Then to calculate a generic correlation function you add an external point for each field inserted and connect them, summing all the possible alternatives. For example, for the 4-point function above we can draw

\begin{equation*}
\begin{tikzpicture}[baseline=(aux2)]
  \begin{feynman}
    \vertex (i) {$i$};
    \vertex [right=2cm of i] (j) {$j$};
    \vertex [below=2cm of i] (k) {$k$};
    \vertex [right=2cm of k] (l) {$l$};
    \vertex [below=1cm of i] (aux1);
    \vertex [right=1cm of aux1] (aux2);
    
    \diagram* {
      (i) -- (j),
      (k) -- (l),
    };
  \end{feynman}
\end{tikzpicture}
+
\begin{tikzpicture}[baseline=(aux2)]
  \begin{feynman}
    \vertex (i) {$i$};
    \vertex [right=2cm of i] (j) {$j$};
    \vertex [below=2cm of i] (k) {$k$};
    \vertex [right=2cm of k] (l) {$l$};
    \vertex [below=1cm of i] (aux1);
    \vertex [right=1cm of aux1] (aux2);
    
    \diagram* {
      (i) -- (k),
      (j) -- (l),
    };
  \end{feynman}
\end{tikzpicture}
+
\begin{tikzpicture}[baseline=(aux2)]
  \begin{feynman}
    \vertex (i) {$i$};
    \vertex [right=2cm of i] (j) {$j$};
    \vertex [below=2cm of i] (k) {$k$};
    \vertex [right=2cm of k] (l) {$l$};
    \vertex [below=1cm of i] (aux1);
    \vertex [right=1cm of aux1] (aux2);
    
    \diagram* {
      (i) -- (l),
      (j) -- (k),
    };
  \end{feynman}
\end{tikzpicture}
\end{equation*}

It's not very hard to convince yourself that this procedure will work for all correlation functions. The diagrams we just introduced are the famous \textit{Feynman diagrams}, and this result for the free theory is called \textit{Wick's theorem}.

\subsubsection{Interacting Theory}

But these are boring Feynman diagrams, what we really want to see is what happens when we turn on interactions. Turning on interactions is equivalent to adding a potential term to the action,

\begin{equation}
    Z[\te{M},\vb{J},V]=\int\dd[n]{\phi}\ee^{-\frac{1}{2}\dotp{\vb{\phi}}{\te{M}\vb{\phi}}+\dotp{\vb{\phi}}{\vb{J}}-V(\vb{\phi})}
\end{equation}
where we have included a source term both for generality, and because we know it may come in handy in the future. Also, we assume $V(\vb{0})=0$, $\eval{\pdv{V}{\phi_i}}_{\phi_i=0}=0$, and $\eval{\pdv{V}{\phi_i}{\phi_j}}_{\phi_i=0}=0$, which is tantamount to ignoring a field independent constant which doesn't affect the dynamics, absorbing the linear term into the source, and absorbing the quadratic term into $\te{M}$. These assumptions could be dropped, but, in most setups, they are equivalent to expanding around the correct vacuum, which is a physical requirement.

For an arbitrary potential, we do not know how to calculate this integral.\footnote{We could plug it into a computer but, as Eugene Wigner once said "It is nice to know that the computer understands the problem. But I would like to understand it too."} What we can do is approximate, namely we approximate that $V$ is, in some sense, a "small" perturbation. It is hard to be more specific than this without being given an actual expression for $V$, but, in general, there will be coefficients inside the function and those coefficients are the small parameters. The details will depend on the case but, assuming such a statement can be made, we can expand in powers of $V$. And, although we can do that straight away, we shall be clever about it.

First, use the trick in (\ref{trick}) to write

\begin{equation}
    Z[\te{M},\vb{J},V]=\ee^{-V\qty(\pdv{\vb{J}})}\int\dd[n]{\phi}\ee^{-\frac{1}{2}\dotp{\vb{\phi}}{\te{M}\vb{\phi}}+\dotp{\vb{\phi}}{\vb{J}}}=Z[\te{M}]\ee^{-V\qty(\pdv{\vb{J}})}\ee^{\frac{1}{2}\dotp{\vb{J}}{\te{M}^{-1}\vb{J}}}
    \label{preexpansion}
\end{equation}

The $Z[\te{M}]$ factor is not important, it will disappear in actual correlation function calculations. The physics is in the two exponentials, we could expand them right away but that still is too complicated. We can be even more clever about it by using this identity.

\begin{claim}
\begin{equation}
    G\qty(\pdv{\vb{J}})F(\vb{J})=\eval{F\qty(\pdv{\vb{\phi}})G(\vb{\phi})\ee^{\dotp{\vb{\phi}}{\vb{J}}}}_{\vb{\phi}=\vb{0}}
    \label{dirtytrick}
\end{equation}
\end{claim}

\begin{proof}
As always in physics, when in doubt, you Fourier transform and see what happens.\footnote{If that fails resort to waving hands, works best in seminars.} Let $\tilde{G}(\vb{k})$ and $\tilde{F}(\vb{k})$ denote the Fourier transforms of $G$ and $F$ respectively, then the LHS of (\ref{dirtytrick}) can be rewritten as

\begin{align}
    &\int\frac{\dd[n]k_1}{(2\pi)^n}\tilde{G}(\vb{k}_1)\ee^{\ii\dotp{\pdv{\vb{J}}}{\vb{k}_1}}\int\frac{\dd[n]k_2}{(2\pi)^n}\tilde{F}(\vb{k}_2)\ee^{\ii\dotp{\vb{J}}{\vb{k}_2}}=\nonumber\\
    =&\int\frac{\dd[n]k_1}{(2\pi)^n}\frac{\dd[n]k_2}{(2\pi)^n}\tilde{G}(\vb{k}_1)\tilde{F}(\vb{k}_2)\ee^{\ii\dotp{\vb{J}}{\vb{k}_2}}\ee^{-\dotp{\vb{k}_1}{\vb{k}_2}}
    \label{final}
\end{align}

Similarly, the RHS becomes
\begin{align}
    &\eval{\int\frac{\dd[n]k_2}{(2\pi)^n}\tilde{F}(\vb{k}_2)\ee^{\ii\dotp{\pdv{\phi}}{\vb{k}_2}}\int\frac{\dd[n]k_1}{(2\pi)^n}\tilde{G}(\vb{k}_1)\ee^{\ii\dotp{\vb{\phi}}{\vb{k}_1}}\ee^{\dotp{\vb{\phi}}{\vb{J}}}}_{\vb{\phi}=\vb{0}}=\nonumber\\
    =&\eval{\int\frac{\dd[n]k_1}{(2\pi)^n}\frac{\dd[n]k_2}{(2\pi)^n}\tilde{G}(\vb{k}_1)\tilde{F}(\vb{k}_2)\ee^{\dotp{\vb{\phi}}{(\ii\vb{k}_1+\vb{J})}}\ee^{-\dotp{\vb{k}_1}{\vb{k}_2}}\ee^{\ii\dotp{\vb{J}}{\vb{k}_2}}}_{\vb{\phi}=\vb{0}}=\nonumber\\
    =&\int\frac{\dd[n]k_1}{(2\pi)^n}\frac{\dd[n]k_2}{(2\pi)^n}\tilde{G}(\vb{k}_1)\tilde{F}(\vb{k}_2)\ee^{\ii\dotp{\vb{J}}{\vb{k}_2}}\ee^{-\dotp{\vb{k}_1}{\vb{k}_2}}
\end{align}

Which is equal to (\ref{final}), finishing out proof.
\end{proof}

Now we apply this new result to (\ref{preexpansion}) to get
\begin{equation}
    Z[\te{M},\vb{J},V]=Z[\te{M}]\eval{\ee^{\frac{1}{2}\dotp{\pdv{\vb{\phi}}}{\te{M}^{-1}\pdv{\vb{\phi}}}}\ee^{-V\qty(\vb{\phi})+\dotp{\vb{J}}{\vb{\phi}}}}_{\vb{\phi}=\vb{0}}
\end{equation}

Finally we expand both exponentials to get a perturbative expansion. Let's first analyse the case $\vb{J}=\vb{0}$. You can think of this as just calculating the actual partition function. We'll use the notation,

\begin{equation}
    V_{i_1\dots i_k}=\eval{\pdv{\phi_{i_1}}\dots\pdv{\phi_{i_k}}V(\vb{\phi})}_{\vb{\phi}=\vb{0}}
\end{equation}

The first few terms in the expansion are (Check!):

\begin{align}
    \frac{Z[\te{M},V]}{Z[\te{M}]}&=1-\frac{1}{8}M^{-1}_{ij}M^{-1}_{kl}V_{ijkl}-\frac{1}{48}M^{-1}_{ij}M^{-1}_{kl}M^{-1}_{mn}V_{ijklmn}+\frac{1}{8}M^{-1}_{ij}M^{-1}_{kl}M^{-1}_{mn}V_{ijk}V_{lmn}+\nonumber\\
    &+\frac{1}{12}M^{-1}_{ij}M^{-1}_{kl}M^{-1}_{mn}V_{ikm}V_{jln}+\frac{1}{384}M^{-1}_{ij}M^{-1}_{kl}M^{-1}_{mn}M^{-1}_{pq}V_{ijklmnpq}+\dots
    \label{terms}
\end{align}

You might be worried about expanding in the propagator exponential as well as the potential since we made no assumption about propagator, only about the potential. For an arbitrary potential you'd be right, our procedure isn't very legitimate, however, most potentials we deal with have a finite (and small) number of terms. Therefore, eventually, the derivatives will simply vanish, terminating the series and rendering the expansion consistent. The reason why we only consider such simple potentials is explained in Chapter \ref{wilsonrg}.

This is progress but we are not happy yet, we want to rephrase this in terms of Feynman diagrams, like we did for the free theory. We'll represent a factor of $M_{ij}^{-1}$ by a line joining two points, the \textit{propagator}

\begin{equation*}
\begin{tikzpicture}[baseline=(j.south)]
  \begin{feynman}
    \vertex (i) {$i$};
    \vertex [right=2cm of i] (j) {$j$};
    
    \diagram* {
      (i) -- (j),
    };
  \end{feynman}
\end{tikzpicture}
\end{equation*}

Since the end result is a scalar, all indices must be contracted. From the first exponential the $\te{M}$ indices are contracted with derivatives that will act on $V$. This means we have a second kind of term which joins up several lines through derivatives of the potential, we call it a \textit{vertex}. It comes with a factor of $-V_{i_1\dots i_k}$ and we draw it as (the dots denote the other lines)

\begin{equation*}
\begin{tikzpicture}[baseline=(a)]
  \begin{feynman}
    \vertex (a);
    \vertex [left=1.8cm of a] (i2) {$i_2$};
    \vertex [right=1.8cm of a] (i5) {$i_{k-1}$};
    \vertex [above=1.6cm of a] (aux1);
    \vertex [below=1.6cm of a] (aux2);
    \vertex [right=0.8cm of aux1] (i6) {$i_k$};
    \vertex [left=0.8cm of aux1] (i1) {$i_1$};
    \vertex [below=0.1cm of a] (aux3);
    \vertex [right=1.0cm of aux3] (aux4);
    \vertex [left=1.0cm of aux3] (aux5);
    \vertex [below=0.9cm of aux3] (aux6);
    
    \diagram* {
      (a) -- (i1),
      (a) -- (i2),
      (a) -- (i5),
      (a) -- (i6),
      (aux4) -- [ghost,out=-90,in=0] (aux6),
      (aux6) -- [ghost,out=180,in=-90] (aux5),
    };
  \end{feynman}
  \draw[fill=black] (a) circle (1.5pt);
\end{tikzpicture}
\end{equation*}

Hence, the terms in (\ref{terms}) become the following diagrams,

\begin{equation*}
1+
\begin{tikzpicture}[baseline=(a)]
  \begin{feynman}
    \vertex (a);
    \vertex [above=1cm of a] (aux1);
    \vertex [below=1cm of a] (aux2);
    
    \diagram* {
      (a) -- [plain,out=135,in=180] (aux1),
      (aux1) -- [plain,out=0,in=45] (a),
      (a) -- [plain,out=-45,in=0] (aux2),
      (aux2) -- [plain,out=180,in=-135] (a),
    };
  \end{feynman}
  \draw[fill=black] (a) circle (1.5pt);
\end{tikzpicture}
+
\begin{tikzpicture}[baseline=(a)]
  \begin{feynman}
    \vertex (a);
    \vertex [left=1cm of a] (aux1);
    \vertex [above=1cm of a] (aux2);
    \vertex [right=1cm of a] (aux3);
    
    \diagram* {
      (a) -- [plain,out=-135,in=-90] (aux1),
      (aux1) -- [plain,out=90,in=135] (a),
      (a) -- [plain,out=135,in=180] (aux2),
      (aux2) -- [plain,out=0,in=45] (a),
      (a) -- [plain,out=45,in=90] (aux3),
      (aux3) -- [plain,out=-90,in=-45] (a),
    };
  \end{feynman}
  \draw[fill=black] (a) circle (1.5pt);
\end{tikzpicture}
+
\begin{tikzpicture}[baseline=(aux3)]
  \begin{feynman}
    \vertex (a);
    \vertex [above=1cm of a] (aux1);
    \vertex [below=1cm of a] (b);
    \vertex [below=1cm of b] (aux2);
    \vertex [below=0.5cm of a] (aux3);
    
    \diagram* {
      (a) -- [plain,out=135,in=180] (aux1),
      (aux1) -- [plain,out=0,in=45] (a),
      (b) -- [plain,out=-45,in=0] (aux2),
      (aux2) -- [plain,out=180,in=-135] (b),
      (a) -- (b),
    };
  \end{feynman}
  \draw[fill=black] (a) circle (1.5pt);
  \draw[fill=black] (b) circle (1.5pt);
\end{tikzpicture}
+
\begin{tikzpicture}[baseline=(a)]
  \begin{feynman}
    \vertex (a);
    \vertex [right=1cm of a] (b);
    \vertex [right=0.5cm of a] (aux1);
    \vertex [above=0.5cm of aux1] (aux2);
    \vertex [below=0.5cm of aux1] (aux3);

    \diagram* {
    (a) -- (b),
    (a) -- [plain,out=90,in=180] (aux2),
    (aux2) -- [plain,out=0,in=90] (b),
    (a) -- [plain,out=-90,in=180] (aux3),
    (aux3) -- [plain,out=0,in=-90] (b),
    };
  \end{feynman}
  \draw[fill=black] (a) circle (1.5pt);
  \draw[fill=black] (b) circle (1.5pt);
\end{tikzpicture}
+
\begin{tikzpicture}[baseline=(a)]
  \begin{feynman}
    \vertex (a);
    \vertex [left=1cm of a] (aux1);
    \vertex [above=1cm of a] (aux2);
    \vertex [right=1cm of a] (aux3);
    \vertex [below=1cm of a] (aux4);
    
    \diagram* {
      (a) -- [plain,out=-135,in=-90] (aux1),
      (aux1) -- [plain,out=90,in=135] (a),
      (a) -- [plain,out=135,in=180] (aux2),
      (aux2) -- [plain,out=0,in=45] (a),
      (a) -- [plain,out=45,in=90] (aux3),
      (aux3) -- [plain,out=-90,in=-45] (a),
      (a) -- [plain,out=-45,in=0] (aux4),
      (aux4) -- [plain,out=180,in=-135] (a),
    };
  \end{feynman}
  \draw[fill=black] (a) circle (1.5pt);
\end{tikzpicture}
+\dots
\end{equation*}

But we're still missing a piece, how to get those factors out front? For that we need a bit of group theory. Let's analyse a generic term in the double expansion with $n$ powers of the potential and $p$ propagators. Schematically, it looks like $\sim\frac{1}{2^p p!}(M^{-1})^p\partial^{2p}\frac{1}{n!}V^n$. Since $V(\vb{0})=0$, we'll need to act with the derivatives on all factors of $V$ which will cancel the $1/n!$ term. Further, due to the Leibniz\footnote{Gottfried Wilhelm (von) Leibniz: born in 1646 in Hanover, Holy Roman Empire; died in 1716 in Hanover, Holy Roman Empire; Doctoral Advisor: Bartholomäus Leonhard von Schwendendörffer; Academic Advisors: Erhard Weigel, Jakob Thomasius, Christiaan Huygens} rule, the derivatives will simply generate all possible terms with that number of $V$s and that number of derivatives. Thinking in terms of diagrams, the action of the derivatives on the potential will generate all possible diagrams with $n$ vertices and $2p$ lines, where we must label the lines and vertices uniquely. The propagators will then pair up these lines.

Now comes the complication, all these diagrams will be distinct if we take into account their labelling. However, some of them will be completely identical up to relabelling, we say they are \textit{topologically} identical, and, since in the end the indices are contracted, the labelling doesn't matter and they will give exactly the same contribution. The question is how many diagrams are actually topologically identical. 

What made the diagrams identical was the contraction with the propagators, which are symmetric and are all identical. This means that, after contraction, we can freely swap two indices of the same propagator, which corresponds to flipping the propagator's ends in the diagram, or exchanging two propagators. The terms/diagrams that give the same contribution are the ones who are mapped to one another by this kind of action.

It's easy to see that this action forms a discrete group $G$ with size $\abs{G}=2^p p!$. The factor of $2^p$ comes from the flipping of the propagators' ends, and the factor of $p!$ comes from the exchange. This is the original factor in front of this term! We're almost there. Take one diagram and call it $x$. The set of diagrams that can be reached via an element of $G$ from that diagram is called the \textit{orbit} of the group $O_x$. This is the set of diagrams that give the same contribution. Then the factor you need to include is $\abs{O_x}/\abs{G}$.

Although we are technically done, this is not what is usually taught in textbooks on quantum field theory. To get to the most common phrasing we need to use the \textit{orbit-stabiliser theorem}. We just need one more definition, the stabiliser $S_x$ of the diagram $x$ is the set of elements of $G$ that leave $x$ unchanged as a labelled diagram. Now let's prove this theorem\footnote{I present the proof mainly for completion and because it's really short. If you don't follow it don't worry, nothing else depends on this, or on similar reasoning, alternatively there's a very nice explanation on \url{https://gowers.wordpress.com/2011/11/09/group-actions-ii-the-orbit-stabilizer-theorem/} (which is where I found this statement of the proof)}

\begin{theorem}
Let $G$ be a group acting on a set $X$, let $x$ be an element of $X$, let $O_x$ be the orbit of $x$ and let $S_x$ be the stabiliser of $x$. Then $\abs{O_x}\abs{S_x}=\abs{G}$
\end{theorem}
\begin{proof}
Let $y$ be an arbitrary element of $O_x$, and let $S_{xy}=\{g\in G:gx=y\}$. Pick $h\in S_{xy}$ and define a map $\phi:S_x\to S_{xy}$ by $\phi:g\to hg$. The map $\psi:S_{xy}\to S_x$ defined by $\psi:u\to h^{-1}u$ inverts $\phi$, so $|S_x|=|S_{xy}|$ for every $y\in O_x$. But the sets $S_{xy}$ with $y\in O_x$ form a partition of $G$. It follows that $|G|=|S_x||O_x|$.
\end{proof}

This means we can write the factor $\abs{O_x}/\abs{G}$ as $1/\abs{S_x}=1/S$. This is what physicists usually call the \textit{symmetry factor} of the diagram. To calculate it you need to find the number of actions you can perform on a labelled diagram that leave it completely unchanged (including the labelling), where you can either flip the propagator's ends or exchange it with another propagator. Check you can reproduce the symmetry factors of the diagrams shown above.\footnote{If you're having difficulty reproducing them, here's a hint: the factors shown (in order) are equal to: $\frac{1}{2^3}$, $\frac{1}{2^3 3!}$, $\frac{1}{2^3}$, $\frac{1}{2 \cdot3!}$, $\frac{1}{2^4 4!}$}

There's only one thing remaining, sources. But this is really easy, sources don't depend of $\vb{\phi}$ hence exactly one derivative needs to act on it. Diagrammatically, we represent this by

\begin{equation*}
\begin{tikzpicture}[baseline=(i.south)]
  \begin{feynman}
    \vertex (i);
    \vertex [right=2cm of i] (j);
    \vertex [left=0.05cm of i] (aux) {$J_i$};
    
    \diagram* {
      (i) -- (j),
    };
  \end{feynman}
  \draw[fill=red] (i) circle (1.5pt);
\end{tikzpicture}
\end{equation*}

To calculate actual correlation functions you just need to use the trick in (\ref{trick}). This means you calculate the diagrams in the presence of sources, and then differentiate with respect to the sources. For example, for the 4-point function, you'll need to differentiate exactly 4 times wrt the sources, hence, you need to calculate all diagrams with exactly 4 sources. Usually, since we just want to correlation functions of simple powers of the fields, we skip the sources bit and add external points to our diagrams. For the 4-point function we'd add 4 external $\phi_i$ points.

To summarise, here's how to associate a QFT calculation with a diagram:

\begin{enumerate}
    \item For every field inside the correlation function you associate an external point.
    \item Draw every possible diagram (up to the order required) connecting the external points.
    \item Associate a number to each diagram using the following rules:
    \begin{itemize}
        \item A line joining two points, called a \textit{propagator}, carries a factor of $M_{ij}^{-1}$:
        \begin{tikzpicture}[baseline=(j)]
            \begin{feynman}
                \vertex (i) {$i$};
                \vertex [right=1.475cm of i] (j) {$j$};
    
                \diagram* {
                    (i) -- (j),
                };
            \end{feynman}
        \end{tikzpicture}
        \item You join several internal lines in a \textit{vertex}. A vertex with $k$ lines carries a factor of $-V_{i_1\dots i_k}(\vb{0})$:
        \begin{tikzpicture}[baseline=(a)]
          \begin{feynman}
            \vertex (a);
            \vertex [left=0.9cm of a] (i2) {$i_2$};
            \vertex [right=0.9cm of a] (i5) {$i_{k-1}$};
            \vertex [above=0.8cm of a] (aux1);
            \vertex [below=0.8cm of a] (aux2);
            \vertex [right=0.4cm of aux1] (i6) {$i_k$};
            \vertex [left=0.4cm of aux1] (i1) {$i_1$};
            \vertex [below=0.05cm of a] (aux3);
            \vertex [right=0.5cm of aux3] (aux4);
            \vertex [left=0.5cm of aux3] (aux5);
            \vertex [below=0.45cm of aux3] (aux6);
            
            \diagram* {
              (a) -- (i1),
              (a) -- (i2),
              (a) -- (i5),
              (a) -- (i6),
              (aux4) -- [ghost,out=-90,in=0] (aux6),
              (aux6) -- [ghost,out=180,in=-90] (aux5),
            };
          \end{feynman}
          \draw[fill=black] (a) circle (1.5pt);
        \end{tikzpicture}
        \item Contract all indices
        \item Divide by the symmetry factor, which is the size of the stabiliser, i.e. the number of actions you can perform on a labelled diagram that leave it completely unchanged (including the labelling), where you can either flip a propagator's ends or exchange it with another propagator.
    \end{itemize}
\end{enumerate}

\subsection{The Wilsonian effective action} 

To calculate the partition function we needed to calculate all diagrams, even those who were disconnected. For example, even though we didn't draw it, the following diagram would also contribute:

\begin{equation*}
\begin{tikzpicture}[baseline=(a)]
  \begin{feynman}
    \vertex (a);
    \vertex [above=1cm of a] (aux1);
    \vertex [below=1cm of a] (aux2);
    \vertex [right=1cm of a] (b);
    \vertex [above=1cm of b] (aux3);
    \vertex [below=1cm of b] (aux4);
    
    \diagram* {
      (a) -- [plain,out=135,in=180] (aux1),
      (aux1) -- [plain,out=0,in=45] (a),
      (a) -- [plain,out=-45,in=0] (aux2),
      (aux2) -- [plain,out=180,in=-135] (a),
      (b) -- [plain,out=135,in=180] (aux3),
      (aux3) -- [plain,out=0,in=45] (b),
      (b) -- [plain,out=-45,in=0] (aux4),
      (aux4) -- [plain,out=180,in=-135] (b),
    };
  \end{feynman}
  \draw[fill=black] (a) circle (1.5pt);
  \draw[fill=black] (b) circle (1.5pt);
\end{tikzpicture}
\end{equation*}

This is clearly not very efficient since it is just two copies of the same diagram. It would be really nice if we had a way of just considering connected diagrams. It turns out we do, it's the \textit{Wilsonian effective action}:

\begin{equation}
    W[\vb{J}]=-\log Z[\vb{J}]
    \label{wilsoneff}
\end{equation}
where from now on we define $Z[\vb{J}]\equiv Z[\te{M},\vb{J},V]$ and $Z_0\equiv Z[\te{M}]$ for convenience.

This object is related to the free energy of statistical mechanics, and will be very useful in the main text. But for now, let's see how to phrase it in terms of connected Feynman diagrams. Let the set of all connected diagrams be denoted by $\qty{\Gamma_a}$, and the symmetry factor of the diagram $\Gamma_a$ be denoted $S_{\Gamma_a}\equiv S_a$. We interpret the product of two diagrams $\Gamma_a\Gamma_b$ to be the disjoint union of the two, i.e. the disconnected diagram consisting of one copy of $\Gamma_a$ and one copy of $\Gamma_b$. Then any disconnected diagram is defined by a set of numbers $\qty{n_a}$, with $n_a\in\mathbb{N}_0$, which specify how many copies of the diagram $\Gamma_a$ it contains. The symmetry factor of the disconnected diagram $\Gamma=\prod_a\Gamma_a^{n_a}$ is

\begin{equation}
    S_\Gamma=\prod_a n_a! S_a^{n_a}
\end{equation}
This is just a product of all the symmetry factors, times a factor of $n_a!$ to account for the exchange of identical diagrams. Let $F(\Gamma)$ be the factor coming from the Feynman rules associated with a given diagram $\Gamma$ then

\begin{equation}
    F(\Gamma)=\prod_a F(\Gamma_a)^{n_a}
\end{equation}

Then we can write the partition function as

\begin{align}
    \frac{Z[\vb{J}]}{Z_0}&=\sum_{\Gamma\in\text{disconnected}}\frac{F(\Gamma)}{S_\Gamma}=\sum_{\qty{n_a}}\prod_a\frac{F(\Gamma_a)^{n_a}}{n_a! S_a^{n_a}}=\prod_a\sum_{n_a=0}^\infty \frac{1}{n_a!}\qty(\frac{F(\Gamma_a)}{S_a})^{n_a}=\nonumber\\
    &=\prod_a\exp(\frac{F(\Gamma_a)}{S_a})=\exp(\sum_a\frac{F(\Gamma_a)}{S_a})=\exp(\sum_{\Gamma\in\text{connected}}\frac{F(\Gamma)}{S_\Gamma})
\end{align}

Therefore

\begin{equation}
    W[\vb{J}]=W_0-\sum_{\Gamma\in\text{connected}}\frac{F(\Gamma)}{S_\Gamma}
\end{equation}
where $W_0=-\log Z_0$.

As promised the Wilsonian effective action can be built out of connected Feynman diagrams. Consequently, we can use this to calculate \textit{connected} $n$-point correlation functions as

\begin{equation}
    \expval{\phi_{i_1}\dots\phi_{i_n}}_{\text{connected}}=-\eval{\pdv{J_{i_1}}\dots\pdv{J_{i_n}}W[\vb{J}]}_{\vb{J}=\vb{0}}
    \label{correlator}
\end{equation}
for this reason the Wilsonian effective action is also called the generating functional for correlation functions.

There is one further generating functional we could build, which would generate 1PI correlation functions. We won't need it for the main text hence I won't cover it. Just for reference it is called the \textit{quantum effective action} and is related to the Gibbs\footnote{Josiah Willard Gibbs: born in 1839 in New Haven, USA; died in 1903 in New Haven, USA; Doctoral Advisor: Hubert Anson Newton} free energy of statistical mechanics.

\subsection{An example: $\phi^4$ theory}

Just to illustrate how this all works in practice, we'll consider the following very simple theory: $\phi^4$ with a single real scalar field in $d$ dimensions. The action for this theory is

\begin{equation}
    S[\phi]=\int\dd[d]x\qty(\frac{1}{2}(\grad{\phi})^2+\frac{1}{2}m^2\phi^2+\frac{\lambda}{4!}\phi^4)
\end{equation}

The partition function with a source term is

\begin{equation}
    Z[J]=\int\mathcal{D}\phi~\ee^{-S[\phi]+\int\dd[d]{x}J(\vb{x})\phi(\vb{x})}
\end{equation}

Correlation functions can be calculated via

\begin{equation}
    \expval{\phi(\vb{x}_1)\dots\phi(\vb{x}_n)}=\fdv{J(\vb{x}_1)}\dots\fdv{J(\vb{x}_n)}Z[J]
\end{equation}
where $\fdv{J}$ denotes the functional derivative defined by the essential rule

\begin{equation}
    \fdv{J(\vb{x})}{J(\vb{y})}=\delta^{(d)}(\vb{x}-\vb{y})
\end{equation}

The quadratic term can be integrated by parts to give 

\begin{equation}
    \int\dd[d]{x}\frac{1}{2}\qty((\grad{\phi})^2+m^2\phi^2)=\int\dd[d]{x}\frac{1}{2}\phi(-\laplacian+m^2)\phi
\end{equation}

In the continuum, instead of a matrix we have an operator, which you can think of as an infinite-dimensional matrix. It's inverse is the Green's function,

\begin{equation}
    (-\laplacian+m^2)\Delta_\text{F}(\vb{x},\vb{y})=\delta^{(d)}(\vb{x}-\vb{y})
\end{equation}
which we can solve easily by taking a Fourier transform,

\begin{equation}
    (\vb{p}^2+m^2)\tilde{\Delta}_\text{F}(\vb{p})=1\Leftrightarrow\tilde{\Delta}_\text{F}(\vb{p})=\frac{1}{\vb{p}^2+m^2}
\end{equation}

Also, the potential is just $V(\phi)=\frac{\lambda}{4!}\phi^4$, hence all it's derivatives will vanish except for the $4^\text{th}$ one. This means our Feynman rules will be\footnote{Usually scalars are represented with dotted lines, however, since we will only be dealing with scalars we can represent them by full lines with no chance of confusion with fermions},

\begin{itemize}
    \item Propagator:
    \begin{equation}
        \begin{tikzpicture}[baseline=(j)]
            \begin{feynman}
                \vertex (i) {$\vb{x}$};
                \vertex [right=2cm of i] (j) {$\vb{y}$};
    
                \diagram* {
                    (i) -- (j),
                };
            \end{feynman}
        \end{tikzpicture}
        =\Delta_\text{F}(\vb{x}-\vb{y})
    \end{equation}
    \item Vertex:
    \begin{equation}
        \begin{tikzpicture}[baseline=(a)]
            \begin{feynman}
                \vertex (a);
                \vertex [above=0.75cm of a] (aux1);
                \vertex [below=0.75cm of a] (aux2);
                \vertex [left=0.75cm of aux1] (b);
                \vertex [right=0.75cm of aux1] (c);
                \vertex [left=0.75cm of aux2] (d);
                \vertex [right=0.75cm of aux2] (e);
    
                \diagram* {
                    (a) -- (b),
                    (a) -- (c),
                    (a) -- (d),
                    (a) -- (e),
                };
            \end{feynman}
            \draw[fill=black] (a) circle (1.5pt);
        \end{tikzpicture}
        =-\lambda
    \end{equation}
    \item Contracting indices means integrating over the position of internal vertices
\end{itemize}

Alternatively we can write this in momentum space, the action becomes,

\begin{equation}
    S[\tilde{\phi}]=\int\frac{\dd[d]p}{(2\pi)^d}\frac{1}{2}\tilde{\phi}(-\vb{p})\qty(\vb{p}^2+m^2)\tilde{\phi}(\vb{p})+\frac{\lambda}{4!}\int\prod_{i=1}^4\qty(\frac{\dd[d]p_i}{(2\pi)^d}\tilde{\phi}(\vb{p}_i))\delta^{(d)}\qty(\sum_{i=1}^4 \vb{p}_i)
\end{equation}

The structure of the delta functions means we should impose momentum conservation along a line and at each vertex. The Feynman rules then become:

\begin{itemize}
    \item Propagator:
    \begin{equation}
        \begin{tikzpicture}[baseline=(j)]
            \begin{feynman}
                \vertex (i);
                \vertex [right=2cm of i] (j);
    
                \diagram* {
                    (i) -- [momentum=$\vb{p}$] (j),
                };
            \end{feynman}
        \end{tikzpicture}
        =\frac{1}{\vb{p}^2+m^2}
    \end{equation}
    \item Vertex:
    \begin{equation}
        \begin{tikzpicture}[baseline=(a)]
            \begin{feynman}
                \vertex (a);
                \vertex [above=0.75cm of a] (aux1);
                \vertex [below=0.75cm of a] (aux2);
                \vertex [left=0.75cm of aux1] (b);
                \vertex [right=0.75cm of aux1] (c);
                \vertex [left=0.75cm of aux2] (d);
                \vertex [right=0.75cm of aux2] (e);
    
                \diagram* {
                    (a) -- (b),
                    (a) -- (c),
                    (a) -- (d),
                    (a) -- (e),
                };
            \end{feynman}
            \draw[fill=black] (a) circle (1.5pt);
        \end{tikzpicture}
        =-\lambda
    \end{equation}
    \item After imposing conservation of momenta at each vertex there may be undetermined momenta in a closed loop. You should integrate over that momenta.
    \item External sources in momentum space are usually denoted not by points but by external lines with a momentum label, there should be no propagator associated to those lines.
\end{itemize}

\newpage

\end{document}